\documentclass{aa}  
\usepackage{graphicx}
\usepackage[T1]{fontenc}
\usepackage{amsmath}    % Advanced maths commands
\usepackage{longtable}
\usepackage{lipsum}
\usepackage{hyperref}
\usepackage{tabularx}
\usepackage{multirow}
\usepackage{rotating}
\usepackage{lscape}  
\usepackage[dvipsnames]{xcolor}
\usepackage{txfonts}
\usepackage{ulem}

\newcommand{\mariac}{}
\newcommand{\maria}{}
\newcommand{\emi}{}
\newcommand{\varsha}{}

\begin{document}

   \title{Supernova search with active learning in ZTF DR3}

%   \subtitle{I. Overviewing the $\kappa$-mechanism}

   \author{M.~V.~Pruzhinskaya\inst{1,2}\thanks{\email{pruzhinskaya@gmail.com}},
           E.~E.~O.~Ishida\inst{1},
           A.~K.~Novinskaya\inst{2},
           E.~Russeil\inst{1},
           A.~A.~Volnova\inst{3},
           K.~L.~Malanchev\inst{4,2},
           M.~V.~Kornilov\inst{2,5},
           P.~D.~Aleo\inst{4,6},
           V.~S.~Korolev\inst{7},
           V.~V.~Krushinsky\inst{8},
           S.~Sreejith\inst{9}
          \and
          E.~Gangler\inst{1} (The SNAD team)}

   \authorrunning{Pruzhinskaya et al.} 
   
   \institute{Universit\'e Clermont Auvergne, CNRS/IN2P3, LPC, F-63000 Clermont-Ferrand, France
        \and
        Lomonosov Moscow State University, Sternberg Astronomical Institute, Universitetsky pr.~13, Moscow, 119234, Russia
        \and
        Space Research Institute of the Russian Academy of Sciences (IKI), 84/32 Profsoyuznaya Street, Moscow, 117997, Russia
        \and
        Department of Astronomy, University of Illinois at Urbana-Champaign, 1002 West Green Street, Urbana, IL 61801, USA
        \and
        National Research University Higher School of Economics, 21/4 Staraya Basmannaya Ulitsa, Moscow, 105066, Russia
        \and
        Center for Astrophysical Surveys, National Center for Supercomputing Applications, Urbana, IL, 61801, USA
        \and
        Independent researcher
        \and
        Laboratory of Astrochemical Research, Ural Federal University, Ekaterinburg, Russia, ul. Mira d. 19, Yekaterinburg, 620002, Russia
        \and
         Physics Department, Brookhaven National Laboratory, Upton, NY 11973}

   \date{Received September 15, 1996; accepted March 16, 1997}

% \abstract{}{}{}{}{} 
% 5 {} token are mandatory
 
  \abstract
   {We provide the first results from the complete
SNAD adaptive learning pipeline in the context of a broad scope of data from
large-scale astronomical surveys.}
   {The main goal of this work is to explore the \mariac{potential} of adaptive learning techniques \mariac{in application} to big data sets.}
   {Our SNAD team used Active Anomaly Discovery (AAD) as a tool to search for new supernova (SN) candidates in the photometric data from the first 9.4 months of the Zwicky Transient Facility (ZTF) survey, namely, between March 17 and December 31 2018 (58194 $\leq$ MJD $\leq$ 58483). We analysed 70 ZTF fields at a high galactic latitude and visually inspected 2100 outliers.}
   {This resulted in 104 \maria{SN}-like objects \varsha{being} found, 57 of \varsha{which} were reported to the Transient Name Server for the first time and with 47 having previously been mentioned in other catalogues, either as \maria{SNe} with known types or as \maria{SN} candidates. We visually inspected the \mariac{multi-colour} light curves of the non-catalogued transients and performed  fittings with different supernova models to assign it to a \maria{probable photometric} class: Ia, Ib/c, IIP, IIL, or IIn. Moreover, we also identified unreported slow-evolving transients that are good superluminous SN candidates, along with a few other non-catalogued objects, such as red dwarf flares and active galactic nuclei.}
   {Beyond confirming the effectiveness of human-machine integration underlying the AAD strategy, our results shed light on potential leaks in currently available pipelines. These findings can help avoid similar losses in future large-scale astronomical surveys. Furthermore, the algorithm enables direct searches of any type of data and based on any definition of an anomaly set by the expert.}


   \maketitle
%
%-------------------------------------------------------------------

\section{Introduction}
The advent of modern astronomical surveys, initiated by the Sloan Digital Sky Survey \citep[SDSS, ][]{Blanton2017} and further propelled by the Zwicky Transient Facility \citep[ZTF, ][]{Bellm2019}, has popularised the use of automated machine learning methods \citep{baron2019}. This shift towards a data-driven approach to astronomical research has been developing swiftly for supervised learning tasks in the areas of classification \citep[see e.g. ][ and references therein]{carleo2019, Ishida2019, malik2022} and regression \citep[e.g.][]{martins2014, Pasquet2019, Cabayol2021, Henghes2021, chen2022}. 

Nevertheless, thanks to the availability of continuous scans of the sky with instruments that are capable of achieving unprecedented resolution, it is natural to expect that new and interesting astrophysical sources will continue to be detected. The challenge then becomes \emi{developing} automated unsupervised learning strategies that can successfully identify such sources among large and complex data sets. The astronomical community has devoted significant efforts to this direction. For example, \citet{Pruzhinskaya2019} applied the isolation forest \citep[IF, ][]{Liu2008} algorithm to identify contaminants in the Open Supernova Catalog~\citep{guillochon2017}. \citet{Malanchev2021} used four different anomaly detection (AD) algorithms and a comprehensive feature extraction process to identify unusual light curves in \maria{the third ZTF data release (DR)}. In searching for changing-state \maria{active galactic nuclei (AGNs)}, \cite{Sanchez2021} identified 75 promising candidates by combining dimensionality reduction via deep learning with IF. \citet{fisher2021} applied a Wasserstein generative adversarial network on nearly one million optical galaxy images in the Hyper Suprime-Cam survey. \citet{martinez2021} combined tree-based \maria{AD} and manifold learning to identify sets of unusual light curves in Kepler data. \citet{chan2022} applied a similar strategy to identify anomalous periodic variables in ZTF data. \citet{sarkar2022} used the Earth as an anomaly example in order to estimate the habitability of exoplanets using a multi-stage memetic algorithm. \citet{kovacevic2022} used self-organising maps to analyse  temporal-only parameters computed from $\gtrapprox 10^5$ sources from the Exploring the X-ray Transient and variable Sky catalogue and \citet{aleo2022} used simulated light curves to search for counterparts in ZTF DR4, identifying 11 non-catalogued transients.

Despite such promising results, all \maria{AD} studies need to deal with the discrepancy between the statistical definition of an outlier (which directly affects the output from traditional machine learning models) and astrophysically interesting anomalies\footnote{Nomenclature from \citet{Malanchev2021}.} (unforeseen or yet to be confirmed events generated by unusual astrophysical phenomena). In large data sets, outliers tend to dominate the set of objects with high anomaly scores \citep{Malanchev2021}. Adaptive learning techniques are aimed at sequentially incorporating expert knowledge in machine learning models \citep[see e.g. ][]{Ishida2021, lochner2021}. The SNAD team\footnote{\url{https://snad.space/}} has been consistently improving and testing such an adaptive learning strategy, whereby at each iteration, a binary reply from the expert is incorporated into the weight calculation of an IF model, producing updated anomaly scores. The active anomaly discovery~\citep[AAD, ][]{Das2017} algorithm has proven to be effective in its first application to real data~\citep{Ishida2021}. In this work, we stress test the effectiveness of this strategy by applying it to light curves from ZTF DR3. Considering \mariac{as anomalies} any light curves that resemble those of\maria{  supernovae (SNe)}, 
 our experts scanned 70 ZTF fields \emi{searching} for uncatalogued or anomalous transients. 

This paper is organised as follows. Section \ref{sec:search} describes the data selection process (\ref{subsec:ZTF}), learning algorithm (\ref{subsec:AAD}), and a summary of the results (\ref{subsec:results}). In Section \ref{sec:models}, we present the results of our light-curve modelling for a subset of the newly reported transients. Section \ref{sec:discussion} presents an  in-depth discussion on superluminous supernova (SLSN) candidates (\ref{subsec:slsn}), along with a complete set of labels within the SNAD viewer knowledge database (\ref{subsec:labels}) and a description of other non-catalogued objects found during our search (\ref{subsec:other}). We present our conclusions in Section \ref{sec:conclusions}. Additionally, the complete SNAD catalogue of discovered transients is shown in Appendix~\ref{ap:catalogue}. Appendix~\ref{ap:fit_results} shows light curves and corresponding fit models for SNAD objects. Appendix~\ref{ap:viewer} gives a glimpse of the domain knowledge database within the SNAD viewer\footnote{\url{https://ztf.snad.space/}}.

%--------------------------------------------------------------------

\section{Supernova search}
\label{sec:search}

\subsection{ZTF data and field selection}
\label{subsec:ZTF}

We analysed photometric data from the first 9.4 months of the ZTF survey, between 2018 March 17 and December 31 (58194 $\leq$ MJD $\leq$ 58483). This period includes data from the ZTF private survey, thus offering a better cadence than the rest of DR3\footnote{\url{https://www.ztf.caltech.edu/ztf-public-releases.html}}. However, 
the expert analysis of discovered \maria{SNe} (see Section~\ref{sec:models})  
used more complete light curves from \maria{ZTF DR8}.

\begin{figure*}
\includegraphics[width=\textwidth]{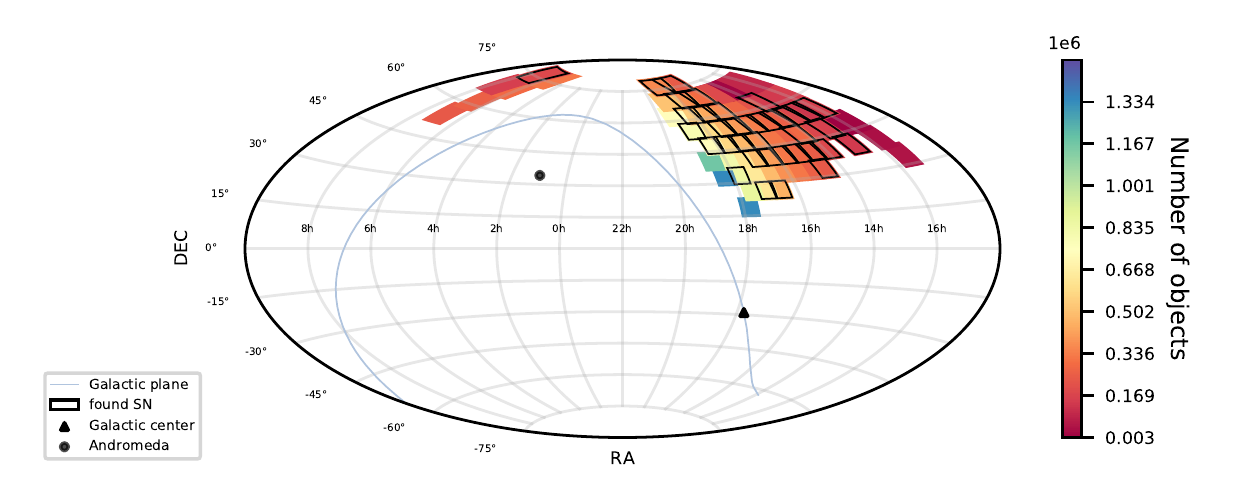}
\caption{Sky map in equatorial coordinates with plotted positions of ZTF fields analysed in this work, the colour bar shows the number of objects in each field. Fields with detected supernova candidates are highlighted with bold black boundaries. The blue curve denotes the galactic plane. The black triangle marks the galactic centre and the black circle corresponds to the position of the Andromeda galaxy.}
\label{fields}
\end{figure*}

Given the higher probability of finding SNe in low extinction regions, we analysed only those fields with centres at $>20^\circ$ above the galactic plane. The distribution of the 70 fields considered in this work is given in Fig.~\ref{fields}.

Each one of the selected fields contains from a few thousand to a little more than a million objects with at least 100 photometric points in $zr$-band (\texttt{catflags $=0$}), thus comprising $\sim$26.5 million light curves in total. Each object is characterised by ZTF \maria{Object ID} (OID). This identifier is unique only within each field and each band, therefore the same source observed in different fields and in different bands can have several OIDs. 

Per each OID, we extracted 42 $zr$-band light curve features including magnitude amplitude, Stetson $K$~coefficient \citep{Stetson1996}, standard deviation of Lomb--Scargle periodogram \citep{1976Ap&SS..39..447L,1982ApJ...263..835S}, and others. A  full description of all features used is given in~\citet{Malanchev2021} \mariac{and} \citet{2021ascl.soft07001M}.
% \LEt{ Please use "and"\ instead of the semicolon.}
%-----------------------------------------------------------------

\subsection{Active anomaly discovery}
\label{subsec:AAD}

Recommendation systems are automatic algorithms whose goal is to minimise the cost of labelling tasks and, at the same time, to optimise classification or anomaly detection results. In this work, we use the AAD algorithm proposed by \citet{Das2017}. It  starts with a traditional \maria{IF} and sequentially presents the object with highest anomaly score to the expert. If the expert judges a particular outlier not to be interesting,  the weights of each decision path is changed to accommodate this new information and the data is passed through the slightly modified forest. The process is repeated until a certain budget has been reached. This framework was first applied to \varsha{a} simulated as well as a small real data set by~\cite{Ishida2021}. 

Here, we present the first application of AAD to a significantly larger data set of real observations ($\sim$26.5 million light curves). 
Since the algorithm can adapt to the expert's opinion, it can be used for a targeted search of transients of a certain type (e.g. SNe). Therefore, in this analysis, a human expert considered only SN-like candidates as anomalies; all other objects proposed by the algorithm are rejected by the expert as 'uninteresting' (i.e. `yes' and `no' in the AAD interface). For each field, the expert has gone through a total budget of 30 objects.

\begin{figure}
    \centering
    \includegraphics[scale=0.35]{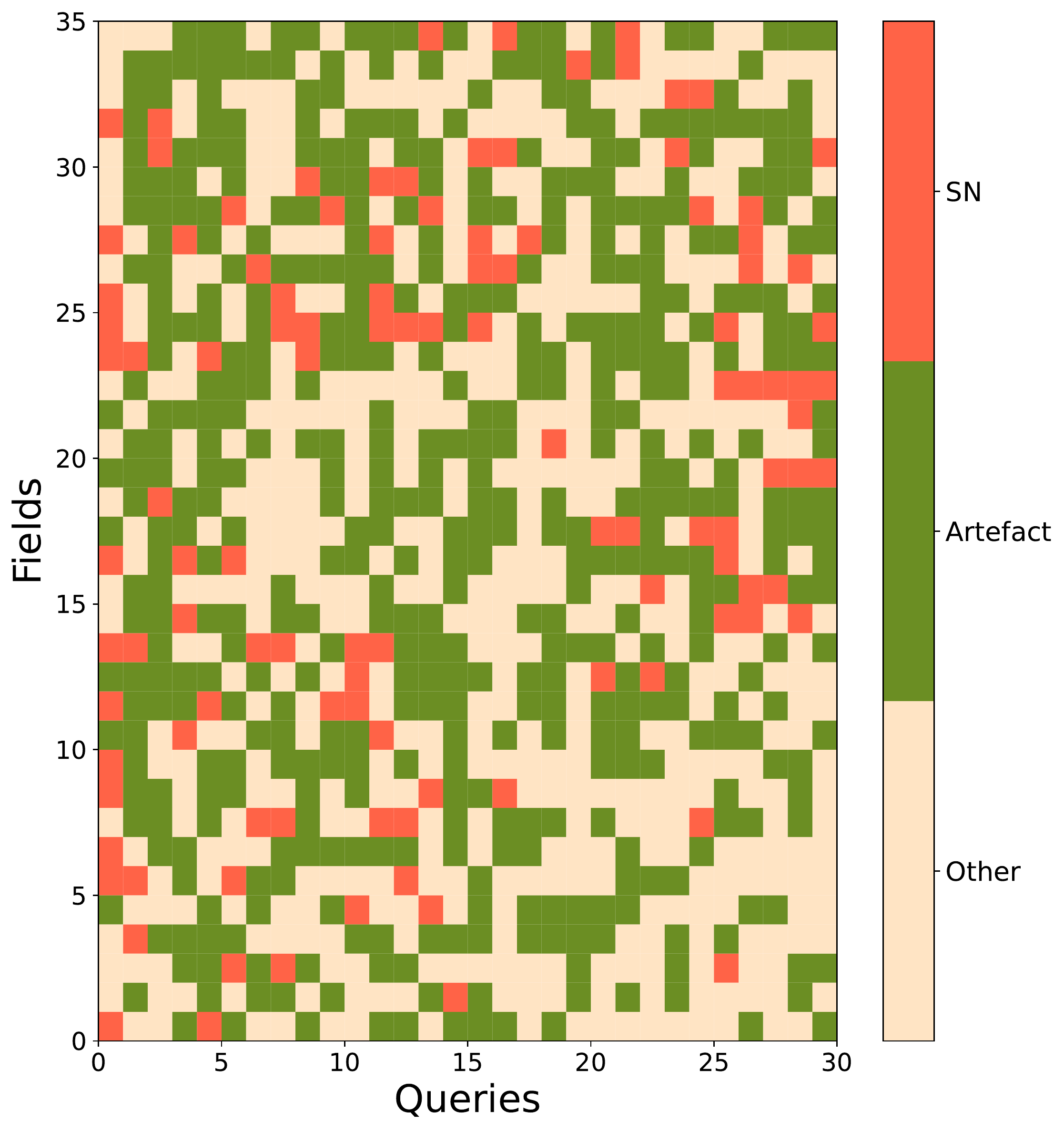}
    \caption{Distribution of objects by type for each of \varsha{the} 35 fields containing supernova or supernova candidate. Each line represents one AAD run with 30 queries. Red, green, and beige colours denote the supernova candidate, artefact or other type of object, respectively. Fields are ordered by \emi{the} number of objects \maria{in them}, from 135681 (bottom line) to 856453 (top line).
    }
    \label{fig:aad_queries}
\end{figure}

In order to enable a smooth interaction between our experts and the AAD algorithm when dealing with such a large data set, we developed the SNAD knowledge database \citep{2022arXiv221107605M}, a framework used by our experts to log their input as one entry in a tailored set of labels (see further details in Section \ref{subsec:labels}). For each one of the ZTF fields, our experts went through 30 objects registering their feedback as a binary answer. The distribution of objects by type for each of the 35 fields containing SNe or SN candidate is given in Fig. \ref{fig:aad_queries}. Each line represents one AAD run with 30 queries in the order of appearance to the expert. The colour denotes the assigned tag, namely, whether it is a supernova, artefact, or other type of object.

\emi{In what follows, we further investigate the most interesting objects we encountered.}
The source code is publicly available as a part of \texttt{zwad}~\citep{2021ascl.soft06033M} GitHub repository\footnote{\url{https://github.com/snad-space/zwad}}.

%-----------------------------------------------------------------

\subsection{Results}
\label{subsec:results}

We visually inspected  2100 ($70\times30$) outliers. Among them, we found 104 SN-like objects, 57 of which were reported 
for the first time and 47 were previously mentioned in other catalogues, either as \maria{SNe} of known types or as \maria{SN} candidates (see Section~\ref{subsec:labels} for other type of objects found). Sources which were not previously mentioned in the Transient Name Server\footnote{\url{https://www.wis-tns.org/}} (TNS) received an internal SNAD name, were added to the SNAD catalogue\footnote{\url{https://snad.space/catalog/} \label{foot:cat}}, and reported to TNS. 
The full list of transients found by the AAD algorithm is given in Table~\ref{tab:sn_list}. Column 1 contains the internal SNAD names of the non-catalogued SN-like candidates. The equatorial coordinates are given in columns 2 and 3. In  column 4, we list the ZTF OIDs. The suggested transient type is defined in column 5. If the object also exists in ZTF alerts, the corresponding target alert name is given in column 6. Column 7 contains the TNS name. Column 8 reports OIDs output by the pipeline that corresponds to the same astrophysical source.

Figure~\ref{fields} shows the distribution of inspected fields on the sky in equatorial coordinates, along with the corresponding number of objects. There are 35 fields with detected SN candidates that are  outlined in black. 
Naively, we would expect that fields with supernovae should be concentrated at the regions further away from the galactic plane and galactic centre. However, we observe that they are located in the middle galactic longitude and latitude. This can be explained by the smaller number of observations in more extragalactic regions. Moreover, the number of objects in different fields varies from a few thousand to more than a million, and the fact that we did not detect any SN in regions with more than a million objects (only three \mariac{regions}), which are also very close to the Milky Way centre, may indicate that the budget of 30 objects was not enough for the AAD to adapt and ideally should be scaled according to  the number of objects in the field.

Among the previously reported supernovae candidates, there are 14 SNe~Ia, 13 possible SNe, 7 SNe~II, 3 SNe~Ic, 2 SNe~IIP, and 1 SN Ib; the remaining 7 catalogued \maria{SNe} belong to the rare supernova classes considered as anomalies in~\citet{Pruzhinskaya2019, Ishida2021}, namely, 2 SNe~IIb, 1 SN~Ia~Pec, 1 SN~Ia-91bg, 1 SN~Ic~BL, 1 SN~IIn, and 1 SLSN-I. To compare the efficiency of the AAD algorithm in searching for more rare  and therefore potentially interesting objects, we recorded the number of spectroscopically confirmed \maria{SNe}  found in this work and discovered by different groups, in the ZTF data, according to TNS for the same period of time (58194 $\leq$ MJD $\leq$ 58483), as shown in Table~\ref{tab:tns}. The fraction of rare SN types among the total number is $\sim$21\%\ for AAD discoveries and $\sim$10\%\  for general TNS findings.

\begin{table}
\centering
\caption{Sub-populations of spectroscopically confirmed supernovae, found in this work (AAD) and total reported in TNS (TNS) for the same time period. TNS numbers report findings within ZTF data only.}
\begin{tabular}{lcc}
\hline
\hline
Type    &       AAD (\%)        & TNS (\%) \\
\hline
SN~Ia   &       14 (41) & 591 (69)\\
SN~II   &       7 (20) & 124 (14) \\
SN~IIP  &       2 (6) & 25 (3)\\
SN~Ib   &       1 (3) & 16 (2) \\
SN~Ic   &       3 (9) & 20 (2) \\
% Total:        &       27      & 776 \\
\hline
\multicolumn{3}{c}{Rare SN types}\\
\hline
SN~Ia~Pec       &       1 (3)   &  11 (1)\\
SN~Ia-91bg      &       1 (3)  &  7 (1)\\
SN~IIb      &   2 (6)   &  13 (2)\\
SN~Ic~BL        &       1 (3)   &  12 (1)\\
SN~IIn      &   1 (3)   &  25 (3)\\
SLSN-I      &   1 (3)   &  15 (2)\\
% Total:        &       7       & 83 \\
\hline
Total:  &       34      (100) & 859 (100) \\
\hline
\end{tabular}
\label{tab:tns}\\
\end{table}

Non-catalogued SN-like objects are listed in the beginning of Table~\ref{tab:sn_list}. 
We note that 15 SNAD possible supernovae (PSNe) are missing in the official ZTF alert stream (Table~\ref{tab:sn_list}, column 6). Missed transients have peak $zr$ magnitude $\sim$19.5--20 mag, which is indeed quite low, but still compatible with those of some other SNAD transients detected by \varsha{the} alert system. Furthermore, some of our candidates (e.g. {\tt SNAD128}, {\tt SNAD165}) have well-sampled early light curves which is of interest for surveys such as the Young Supernova Experiment~\citep{Jones2021}. 

%-----------------------------------------------------------------

\section{Supernova modelling}
\label{sec:models}

We used the \textsc{Python} library \textsc{sncosmo}\footnote{\url{https://sncosmo.readthedocs.io/en/stable/}} to obtain a preliminary \maria{photometric} classification for SNAD objects. Their light curves were fitted with Peter Nugent's supernova models,\footnote{\url{https://c3.lbl.gov/nugent/nugent_templates.html}} which cover the main \maria{SN} types (Ia, Ib/c, IIP, IIL, IIn). Nugent's models are simple spectral time series that can be scaled up and down. The model parameters are the redshift, $z$, the observer-frame time corresponding to the source's zero phase, $t_0$, and the amplitude. The zero phase is defined relative to the explosion moment and the observed time, $t,$ is related to phase via $t = t_0 + {\rm phase} \times (1 + z)$.

In order to perform a preliminary fit, we used only \varsha{the} $zr$-band from DR8. We subtracted the reference magnitude from ZTF light curves, thus roughly accounting for the host galaxy contamination. \maria{The reference magnitude was retrieved from ZTF archival data\footnote{\url{https://irsa.ipac.caltech.edu/Missions/ztf.html}} and listed in the SNAD catalogue\footnote{\url{https://snad.space/}}. We also corrected for a line-of-sight reddening in the Milky Way galaxy using \cite{2011ApJ...737..103S} estimates.} For sources holding SDSS DR16~\citep{2020ApJS..249....3A} photometric redshift of a host  galaxy at the source position, we fixed the redshift to this value.  If this was not available, we adopted $[-15; -22]$ as an acceptable \varsha{range} for the supernova absolute magnitude \citep{2014AJ....147..118R} and then, using the maximum apparent magnitude, roughly transformed it to the corresponding redshift range. We applied a $\chi^2$ criterion to choose the best-fit model for each SNAD object. Results of the light curve fit are given in Appendix~\ref{ap:fit_results}, the best-fit model for each SNAD transient is listed in column 5 of Table~\ref{tab:sn_list}. 

\maria{It should be noted that we did not intend to make a detailed fit, but, rather, to show that the candidate light curves, selected initially by eye, can be satisfactorily fitted by different supernova models. That is why only one band ($zr$) has been used in the fit. Also, we did not take into account the possible extinction in host galaxies of the candidates, therefore, our fit is %not very reliable 
\emi{less accurate} 
for highly reddened objects. Moreover, the redshift we assigned to some host galaxies is photometric, which is another source of uncertainty. Finally, the model itself is rather simple and limited in wavelength
and time range. As a result of these conscious simplifications and assumptions, the obtained absolute magnitude for some of the objects is not typical for normal \maria{SNe} (e.g. {\tt SNAD122}, $M_r\rm{(IIP)}\simeq-22.6$~mag) and we cannot trust the classification in those cases. However, this simple fit is enough to show that a few transients have anomalously wide light curves when compared to normal SNe, making them candidates to the SLSN class (Section~\ref{subsec:slsn}).} 

\begin{figure}
    \centering
    \includegraphics[width=0.5\textwidth]{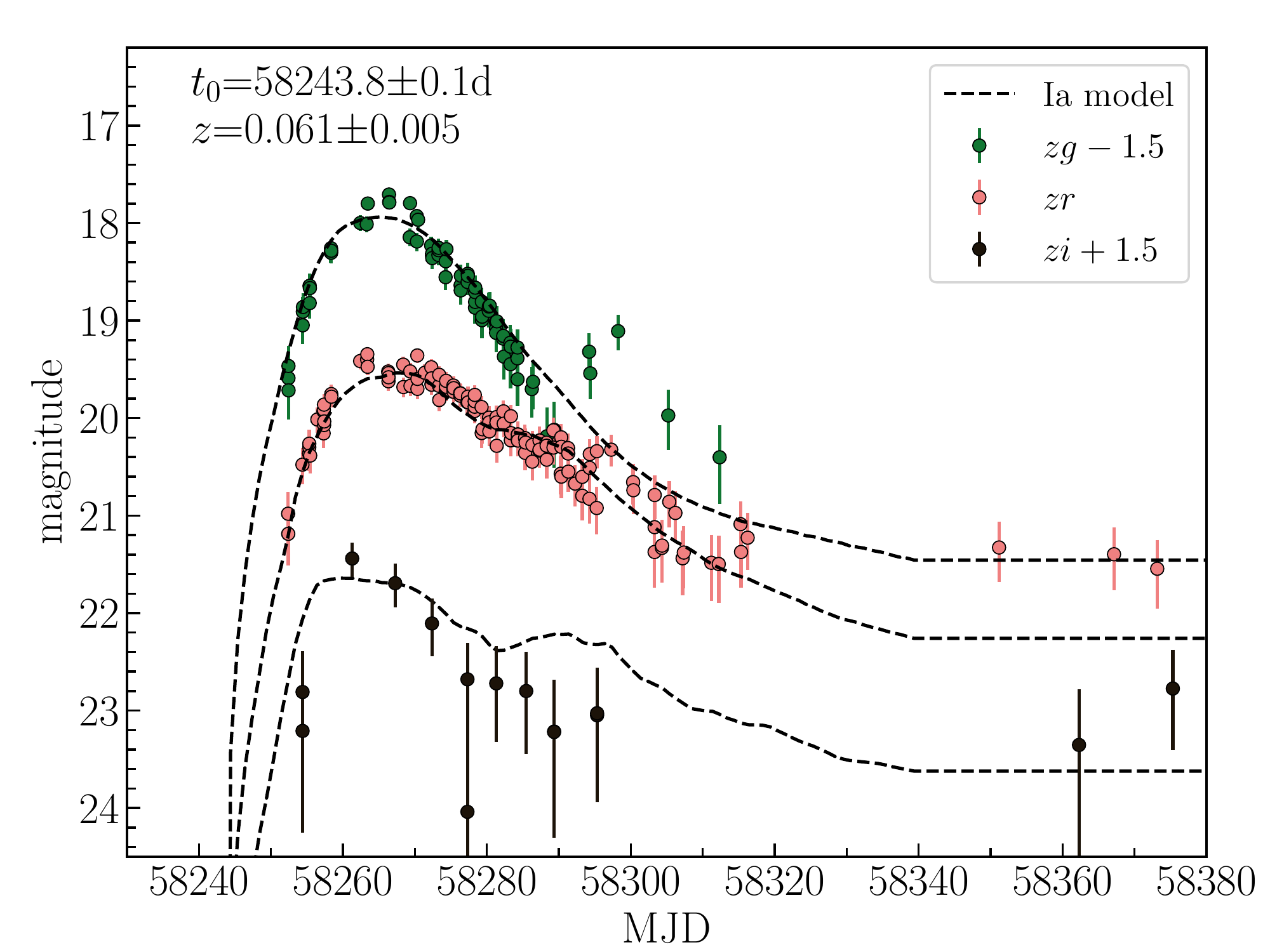}
    \caption{Light curve fit of {\tt SNAD112} by Nugent's Type Ia supernova model. Observational data correspond to OIDs: {\tt 796101400003999} ($zg$), {\tt 796201400007564} ($zr$), {\tt 796301400021875} ($zi$), and {\tt 797304300009092} ($zi$).}
    \label{snad_112_fit}
\end{figure}

\begin{figure}
    \centering
    \includegraphics[width=0.5\textwidth]{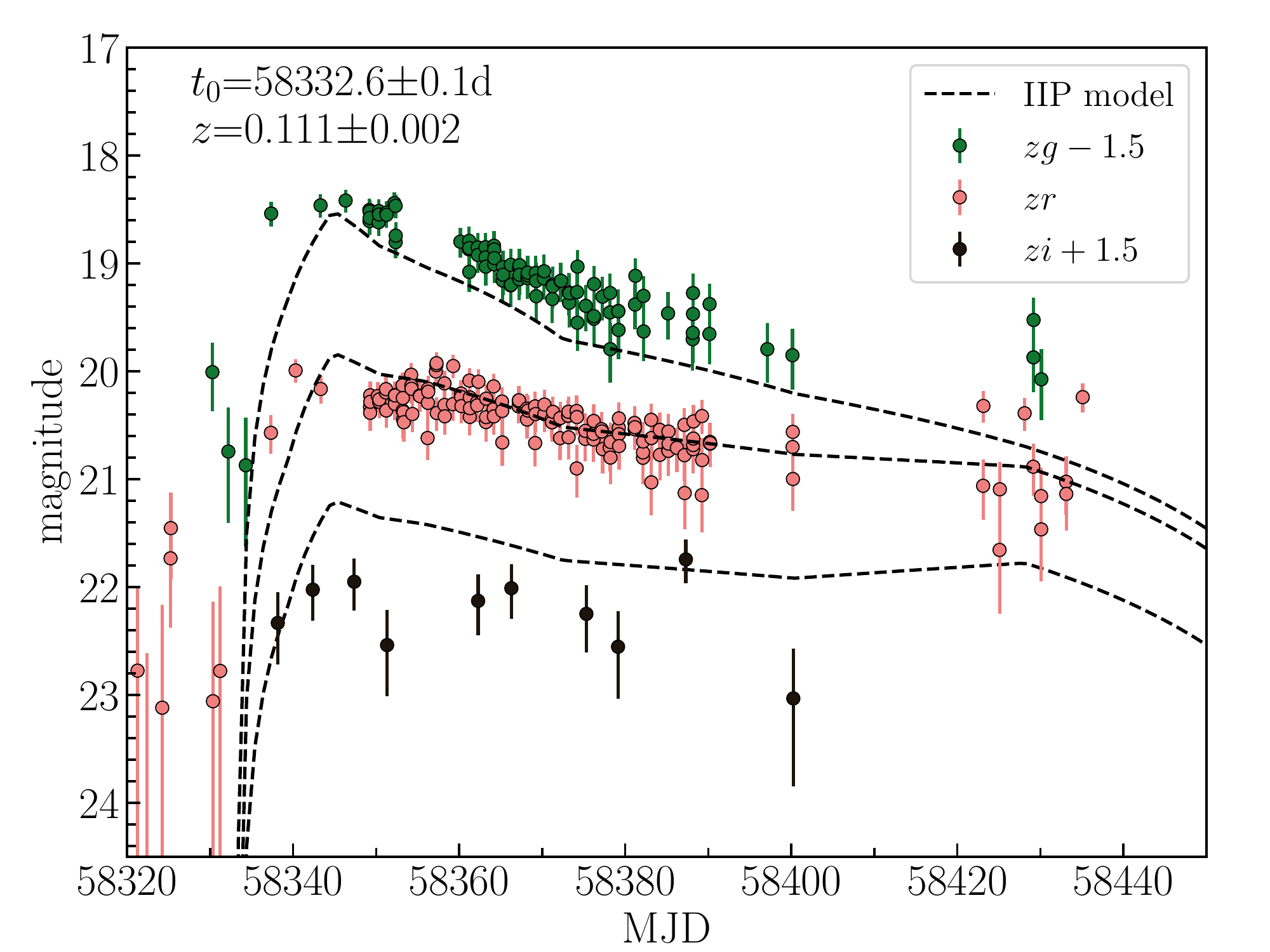}
    \caption{Light curve fit of {\tt SNAD142} by Nugent's Type IIP supernova model. Observational data correspond to OIDs: {\tt 826102200028756} ($zg$), {\tt 826202200030732} ($zr$), and {\tt 826302200021568} ($zi$).}
    \label{snad_142_fit}
\end{figure}

\begin{figure}
    \centering
    \includegraphics[width=0.5\textwidth]{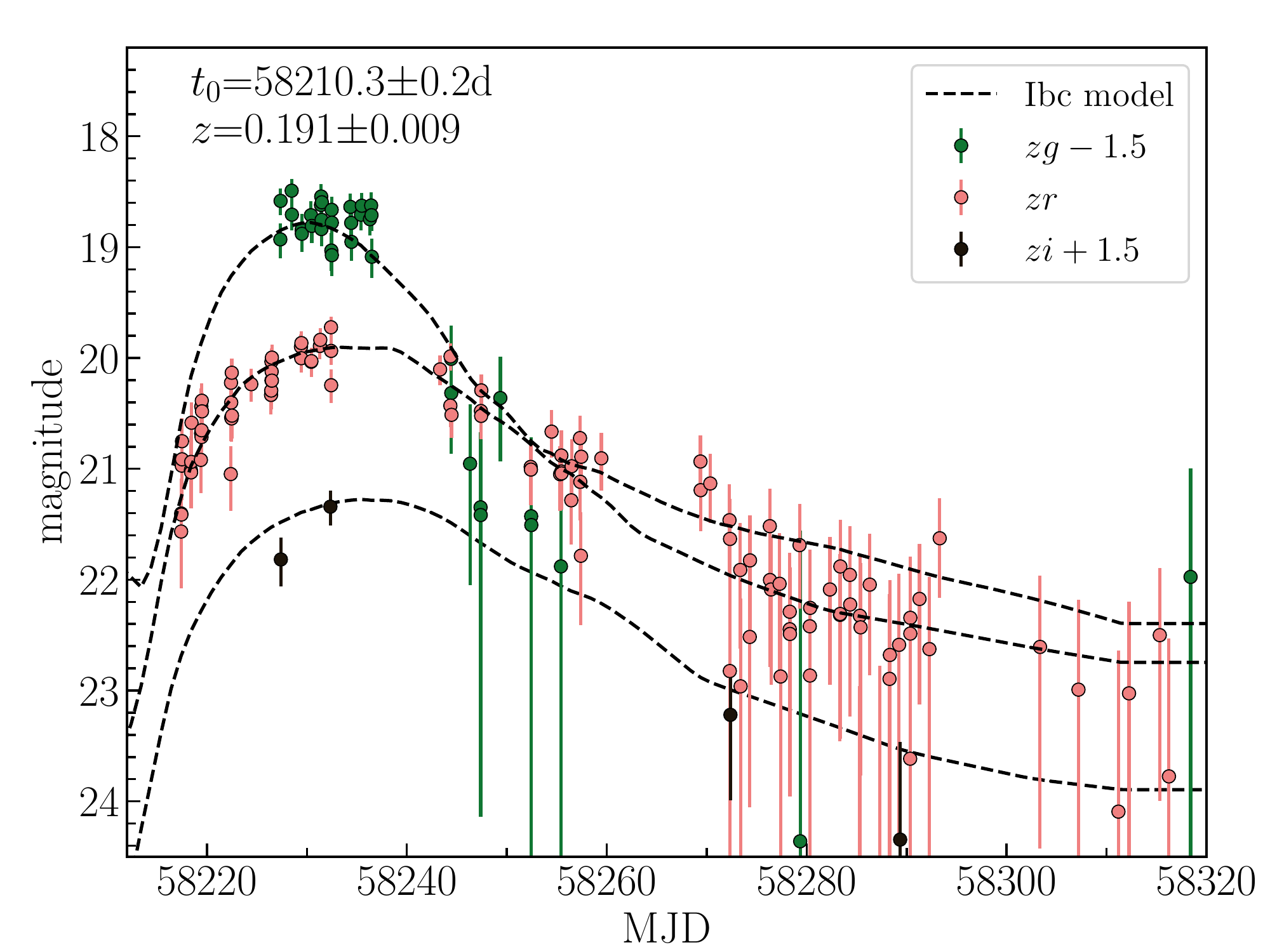}
    \caption{Light curve fit of {\tt SNAD165} by Nugent's Type Ibc supernova model. Observational data correspond to OIDs: {\tt 763104300002058} ($zg$), {\tt 763204300004087} ($zr$), and {\tt 763304300014301} ($zi$).}
    \label{snad_165_fit}
\end{figure}

\begin{figure}
    \centering
    \includegraphics[width=0.5\textwidth]{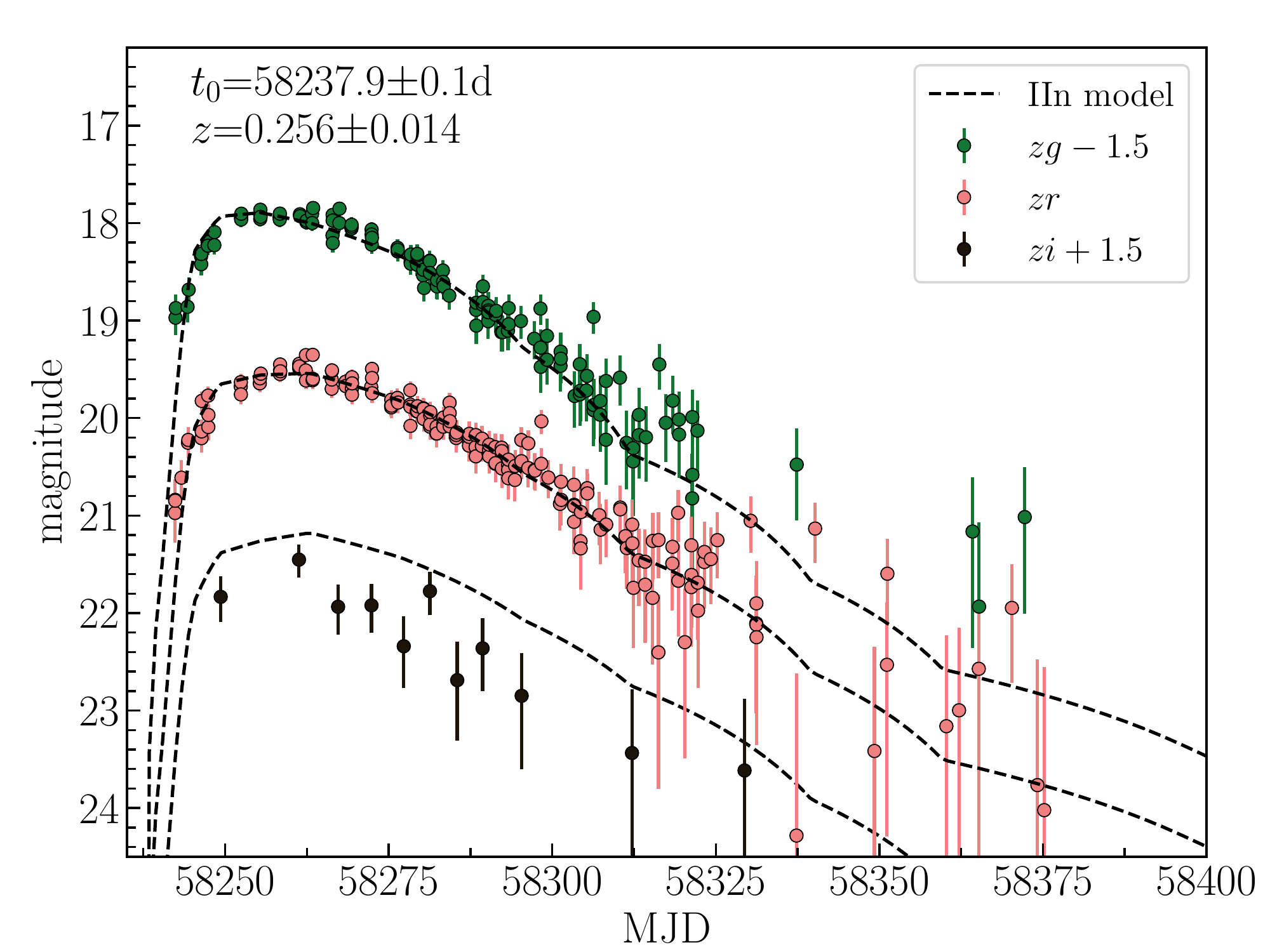}
    \caption{Light curve fit of {\tt SNAD137} by Nugent's Type IIn supernova model. Observational data correspond to OIDs: {\tt 825102200009050} ($zg$), {\tt 825202200039582} ($zr$), and {\tt 825302200018371} ($zi$).}
    \label{snad_137_fit}
\end{figure}

Although this classification should be treated with caution, it follows closely the behaviour of light curves with a sufficient number of observations before and after  maximum light. Using the \textsc{sncosmo} library, we also performed a multi-band light-curve fit for a few objects with the models suggested by the preliminary classification. \maria{The parameters of the fit are $z$, $t_0$, and the amplitude. Then,} {\tt SNAD112}, {\tt SNAD142}, {\tt SNAD165}, {\tt and SNAD137} fitted by Nugent's Type Ia, IIP, Ibc, and IIn models are given in Fig.~\ref{snad_112_fit},~\ref{snad_142_fit},~\ref{snad_165_fit}, and~\ref{snad_137_fit}, respectively. The quality of the fit allows us to conclude that those supernovae belong to the suggested types.

%-----------------------------------------------------------------

\section{Discussion}
\label{sec:discussion}

\subsection{Superluminous supernovae candidates}
\label{subsec:slsn}

Four \maria{supernova candidates} from our list possess significantly  broader light curves in comparison with Nugent's models and other candidates: {\tt SNAD120}, {\tt SNAD121}, {\tt SNAD160},  and {\tt SNAD187} (see Appendix~\ref{ap:fit_results}). In this section we explore the possibility of these objects belonging to the \maria{SLSN} class.

{\tt SNAD120} (AT2018lxa) is located at  $\alpha~=~17^h00^m16.296^s$, $\delta=+70^{\circ}30^{'}49.55^{''}$. In the official ZTF alert stream, it is denoted as ZTF18aazydub. According to ~\cite{2021ApJ...907...99S}, the transient has a spectroscopic redshift \varsha{of} $z_{\rm sp}=0.202$ and was classified as SN~IIn. Assuming this redshift, the estimated absolute magnitude at maximum brightness is $M_r\simeq-20.5$~mag, which is slightly dimmer than the threshold of $-21$~mag established for \maria{SLSNe}~\citep{2012Sci...337..927G}.

{\tt SNAD121} (AT2018lxb, ZTF18abklshn) is located at $\alpha~=~16^h33^m19.937^s$, $\delta=+71^{\circ}06^{'}54.50^{''}$. On  archival images provided by the Legacy Surveys Sky Viewer\footnote{\url{https://www.legacysurvey.org/viewer}}, a possible host is detected with \varsha{an estimated photometric redshift of} $z_{\rm ph} = 0.240\pm0.166$~\citep{2021MNRAS.501.3309Z}. Taking into account redshift uncertainty, the absolute magnitude of this source is estimated to be brighter than $-21$~mag, thus, it is compatible with \maria{SLSNe}. 

{\tt SNAD160} (AT2018lzi, ZTF18aautopz) is located at $\alpha~=~13^h43^m53.357^s$, $\delta=+61^{\circ}33^{'}17.24^{''}$. The ALeRCE ZTF Explorer\footnote{\url{https://alerce.online/object/ZTF18aautopz}} automatically classified ZTF18aautopz as a \maria{SLSN}. The spectroscopic redshift is $z_{\rm sp}=0.295$~\citep{2021ApJ...907...99S}, which gives $M_r\simeq-21.6$~mag at maximum light. {\tt SNAD160} is reported by SNAD team in \citet{2022RNAAS...6..122P} as a possible pair-instability supernova -- a theoretical class of thermonuclear explosions which takes place at the end of life of very massive stars with highly increased production of $^{56}$Ni \cite[e.g.][]{2019ARA&A..57..305G,2014A&A...565A..70K}. 

{\tt SNAD187} (AT2018mcb, ZTF18aaqctvg) is located at $\alpha~=~13^h53^m7.366^s$, $\delta=+40^{\circ}48^{'}7.42^{''}$. There are several photometric redshift estimations of its possible host provided by different surveys: $z_{\rm ph} = 0.204\pm0.084$ by the Legacy Surveys Sky Viewer~\citep{2021MNRAS.501.3309Z}, $z_{\rm ph} = 0.343\pm0.128$ by SDSS DR16~\citep{2020ApJS..249....3A}, and $z_{\rm ph} \simeq 0.201$ by Gaia DR3~\citep{2022yCat.1356....0G}. Also, according to Gaia variability classification results there is an AGN at the transient position~\citep{2022yCat.1356....0G}. It is possible that {\tt SNAD187} is not associated with the host AGN activity and could be a SLSN. Recently, the ANTARES broker \maria{AD} filter reported \maria{the} discovery of a SLSN --- SN~2022mnj at the central region of an \maria{AGN}~(\citealt{2022TNSTR1633....1A,2022TNSCR1690....1A}; see also \citealt{2017ApJ...843L..19M}).

\begin{figure}
    \centering
    \includegraphics[width=0.5\textwidth]{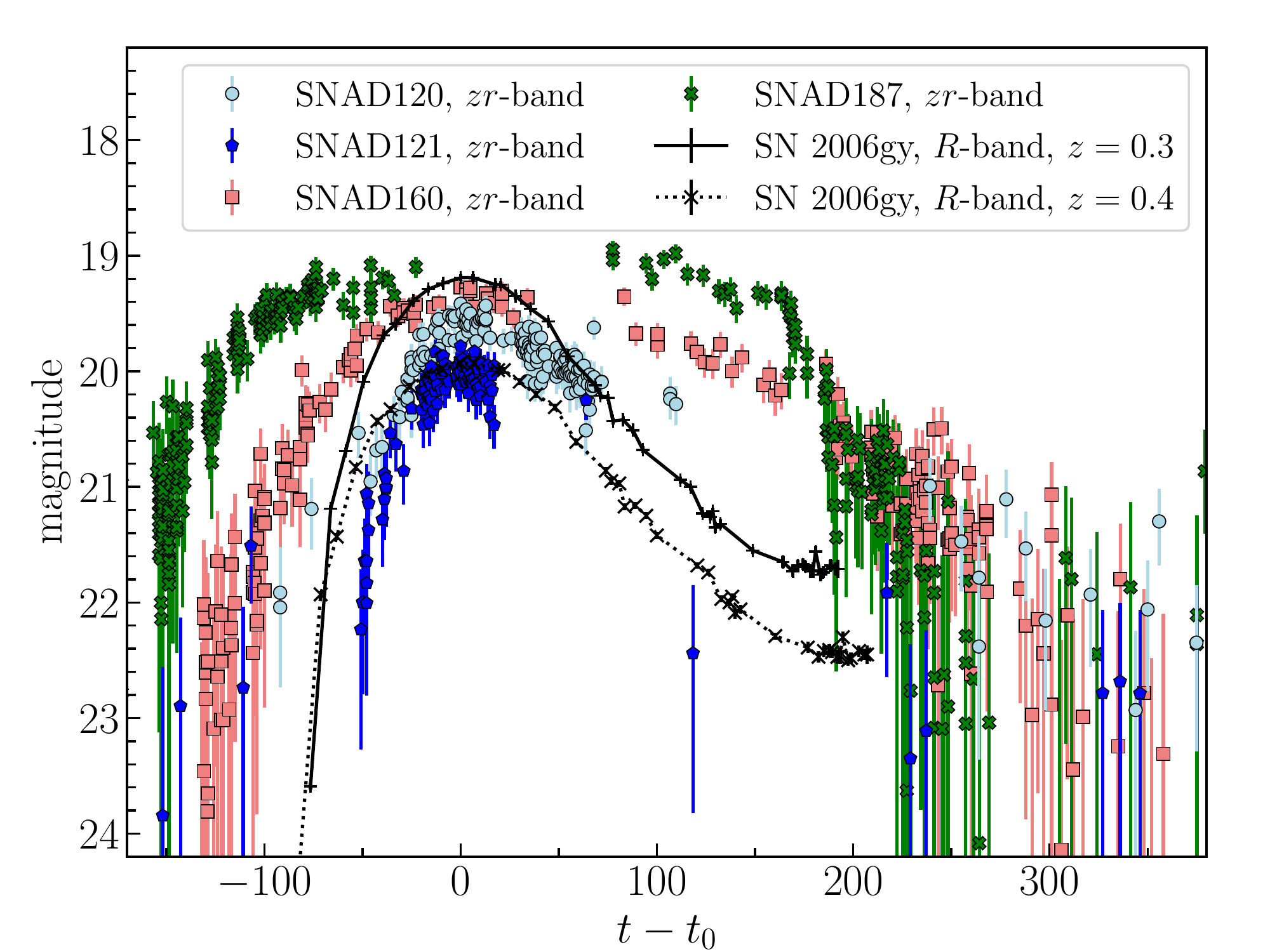}
    \caption{Light curves of SNAD SLSN candidates in $zr$-band in comparison with the \maria{R-band light curve of} well-studied SLSN SN~2006gy shifted to $z = 0.3$ (black pluses) and $z=0.4$ (black crosses). The observed magnitudes of SN 2006gy are taken from \citet{2007ApJ...666.1116S}. All the light curves are shown relative to the maximum light.}
    \label{SN2006gya}
\end{figure}

Figure~\ref{SN2006gya} shows the observed light curves of {\tt SNAD120}, {\tt SNAD121}, {\tt SNAD160}, and {\tt SNAD187} in the $zr$-band in comparison with SN~2006gy~\citep{2007ApJ...666.1116S} --- one of the brightest among the well-studied SLSNe, shifted to $z = 0.3$ and 0.4. SN~2006gy has a very broad light curve, but it is clear that the SNAD candidates 
have even broader light curves, making them 
really peculiar objects among known SNe. The discovery of four slow-evolving  
transients among the SNAD objects, non-reported by previous searches provides clear evidence that the AAD is efficient in searching for rare classes of astronomical objects within large and complex data sets.

%----------------------------------------------------------------

\subsection{SNAD knowledge database}
\label{subsec:labels}

Beyond the transient candidates discussed previously, this work also produced a valuable knowledge database incorporated within the SNAD viewer~\maria{\citep{2022arXiv221107605M}}. 
The viewer is a specially designed web-interface, which allows the expert to visualise ZTF DR light curves, provides access to the individual exposure images, and performs cross-matches with different databases and catalogues. For  authorised users, there is a possibility to assign the labels (tags) to ZTF objects (see Fig.~\ref{viewer}).

We defined a system of tags that includes some general classes:\ variable star of unspecified type (\textsc{VAR}), transient (\textsc{transient}), active galactic nucleus (\textsc{AGN}), quasar (\textsc{QSO}), normal star without strong variability (\textsc{STAR}), and galaxy (\textsc{Galaxy}), as well as the most popular types and subtypes of variable stars and transients,\footnote{Variable star types follow the convention used \maria{by the International Variable Star Index}, \url{https://www.aavso.org/vsx/index.php?view=about.vartypes}} such as:

\begin{itemize}
    \item Supernova (\textsc{SN}): Type Ia supernova (\textsc{SNIa}), core-collapse supernova (\textsc{CCSN}), and super-luminous supernova (\textsc{SLSN});
    \item Eclipsing variable (\textsc{Eclipsing}): $\beta$ Persei-type (Algol) eclipsing system (\textsc{EA}), $\beta$ Lyrae-type eclipsing system (\textsc{EB}), and W Ursae Majoris-type eclipsing variable (\textsc{EW});
    \item Pulsating variable (\textsc{Pulsating}): cepheid (\textsc{CEP}), classical cepheid or $\delta$ Cephei-type variable (\textsc{DCEP}), slow irregular variable (\textsc{L}), long period variable (\textsc{LPV}),  $o$~Ceti-type (Mira) variable (\textsc{M}), variable of the RR Lyrae type (\textsc{RR}), RR Lyrae variable with asymmetric light curve (\textsc{RRAB}), red supergiant (\textsc{RSG}), semi-regular variable (\textsc{SR}), and variable of the $\delta$ Scuti type (\textsc{DSCT});
    \item Cataclysmic variable (\textsc{Cataclysmic}): AM Herculis-type variable (\textsc{AM}), nova (\textsc{N}), U Geminorum-type variable or dwarf nova (\textsc{UG}), SS Cygni-type variable (\textsc{UGSS}), and Z Camelopardalis-type star (\textsc{UGZ});
    \item Eruptive variable (\textsc{Eruptive}): Orion variable with rapid light variations (\textsc{INS}), variable of the S Doradus type (\textsc{SDOR}), T Tauri star (\textsc{TTS}), young stellar object of unspecified variable type (\textsc{YSO}), M dwarf flare (\textsc{M\_dwarf\_flare});
    \item Rotating variable (\textsc{Rotating}): BY Draconis-type variable (\textsc{BY}), RS Canum Venaticorum-type binary system (\textsc{RSCVn}).
\end{itemize}

There are \varsha{a} few custom tags for internal purposes, such as transients with one outlier point (\textsc{1-point}) or candidates to be send to TNS (\textsc{TNS\_candidate}). Also, tags of non-astrophysical origin such as artefacts and their subtypes are present. Several tags can be assigned to one object, the history of tag changes is also stored in the database (Fig.~\ref{viewer}, on the right).

The choice of tags is determined by the experts, based on the most frequent types of objects appearing in the output of the \maria{AD} algorithms and also determined by the project needs. Therefore, we do not claim to be complete in covering all possible types of variables and transients. 

\begin{figure*}
\includegraphics[width=\textwidth]{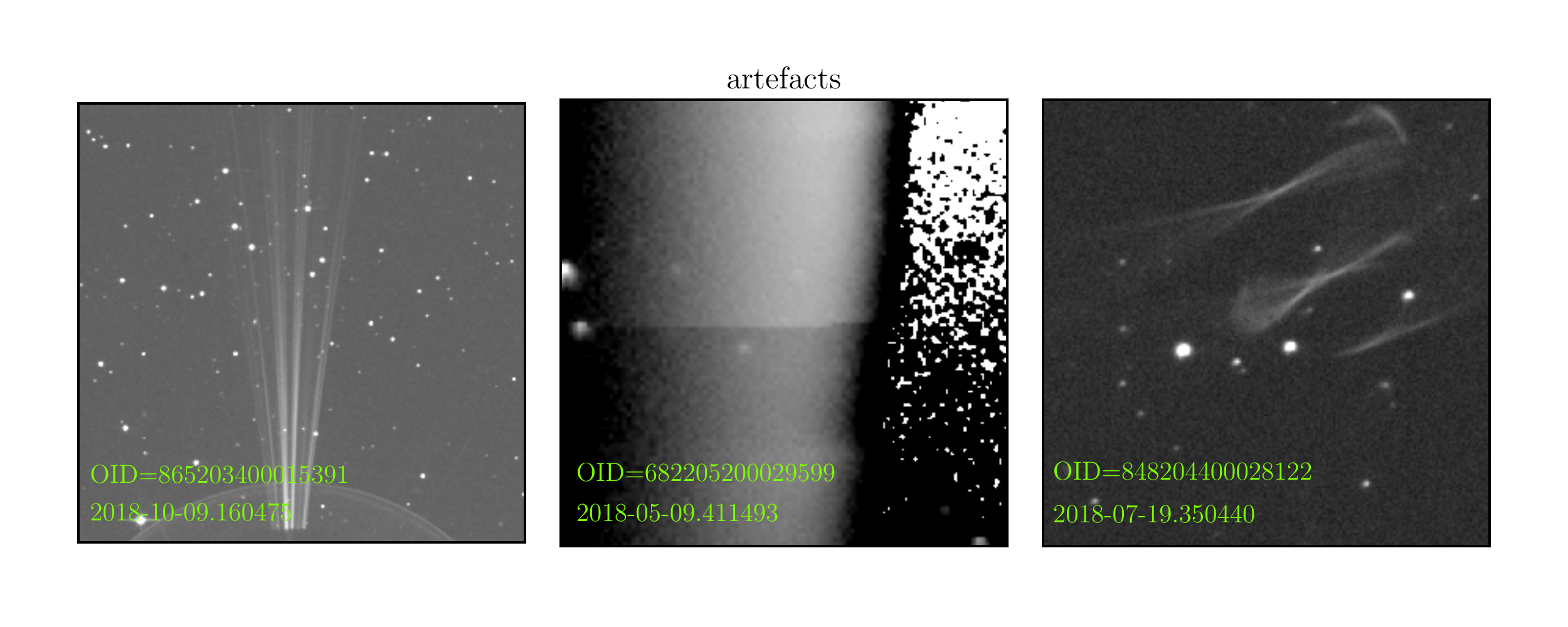}
\caption{Examples of artefacts found during the supernova search with AAD. The outlier is located in the image centre. The image sizes are 600$\times$600, 100$\times$100, and 200$\times$200 CCD pixels, respectively.}
\label{artefacts}
\end{figure*}

During the supernova search a total of 1482 objects were labelled. Despite the fact that ZTF data processing pipeline includes a procedure to separate the astrophysical events from bogus ones, namely, false positive detections \citep{2019PASP..131a8003M}, fields with SNe consists of $\sim$45\%\  of artefacts. Examples of found artefacts are given in Fig.~\ref{artefacts}\footnote{The SNAD catalogue of selected artefacts found in ZTF data is available at  \url{https://snad.space/art/}}.

For real variables, among the most common types in fields containing  SNe, are eclipsing ($N=51$, $\sim$5\%) and pulsating ($N=53$, $\sim$5\%) variables, as well as AGNs ($N=176$, $\sim$17\%). The assigned labels can be used to further improve the ZTF pre-processing pipeline (in case of artefacts) as well as for %the 
machine-learning classification tasks (in case of astrophysical labels).

%-----------------------------------------------------------------

\subsection{Other non-catalogued objects}
\label{subsec:other}

\begin{figure}
    \centering
    \includegraphics[width=0.5\textwidth]{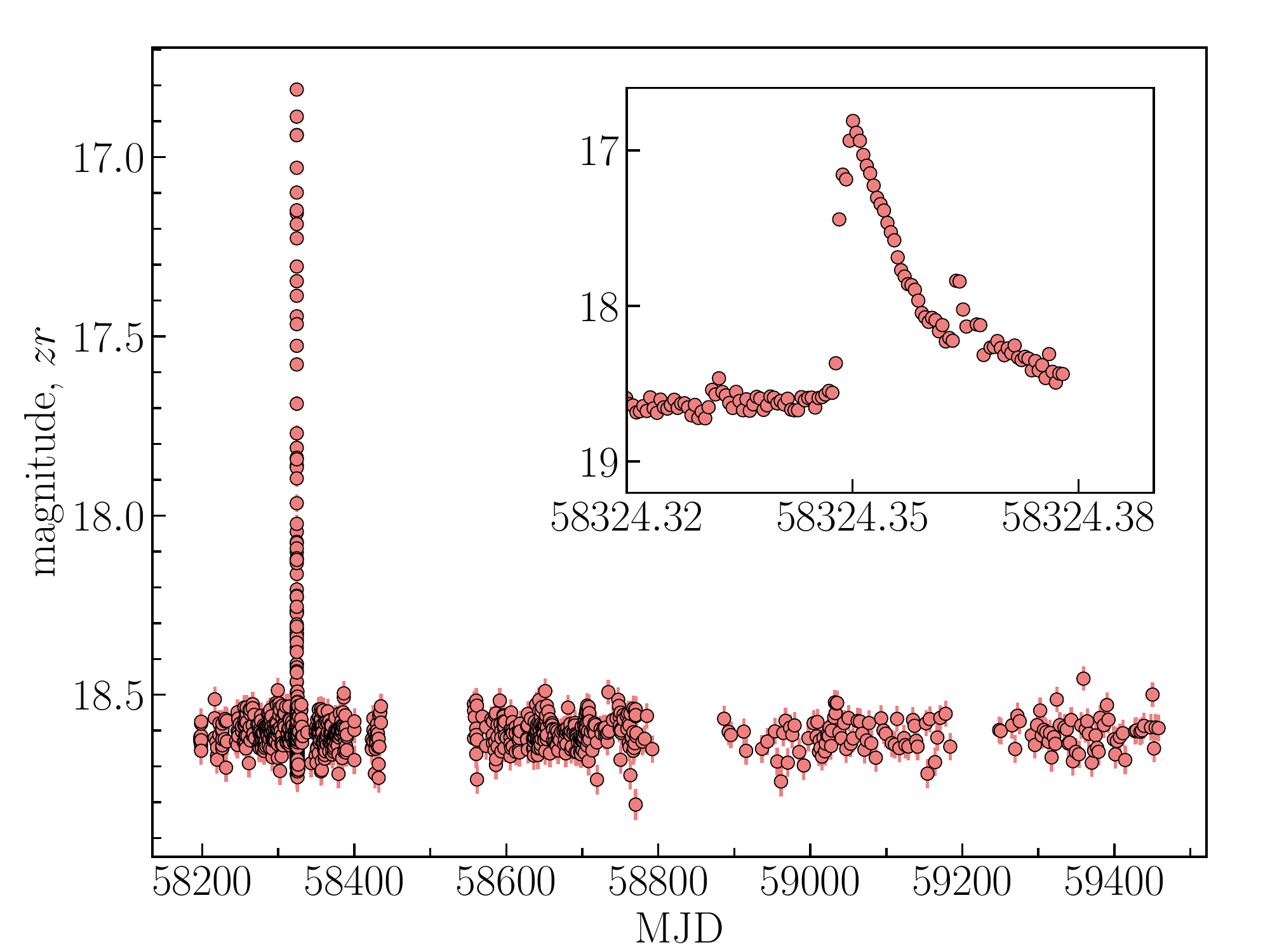}
    \caption{Light curve of a complex red dwarf flare, OID: {\tt 726209400028833} ($zr$). Inset plot shows a zoomed high-cadence light curve with two-peaks flare.}
    \label{rdf}
\end{figure}

During the supernova search, a number of interesting non-catalogued objects of other types have been found. Among those there are red dwarf flares, namely, transients caused by \varsha{the} sudden release of stored magnetic energy from surface magnetic loops into the outer stellar atmosphere~\citep{1989SoPh..121..299P,1991ARA&A..29..275H}, and AGNs. For example, a two-peak flare of a red dwarf, OID~=~{\tt 726209400028833}, located at a distance of $\sim$162~pc~\citep{2018AJ....156...58B} is shown in Fig.~\ref{rdf}. The amplitude of the flare  is $\sim$1.8~mag, the minimum duration is $\sim$46 minutes. There are many unsolved questions related to flare physics, red dwarf distribution in the Galaxy, and habitability of host planets, which can benefit from a systematic study of a large sample of such events (e.g.~\citealt{2010AsBio..10..751S,2011ASPC..451..285E,2013ApJ...763..149F,2021MNRAS.506.2089W}). Moreover, good observational cadence of the flare ($\sim$70 points in 46~min) also opens up a possibility to search for fast transients in ZTF data.

\begin{figure}
    \centering
    \includegraphics[width=0.5\textwidth]{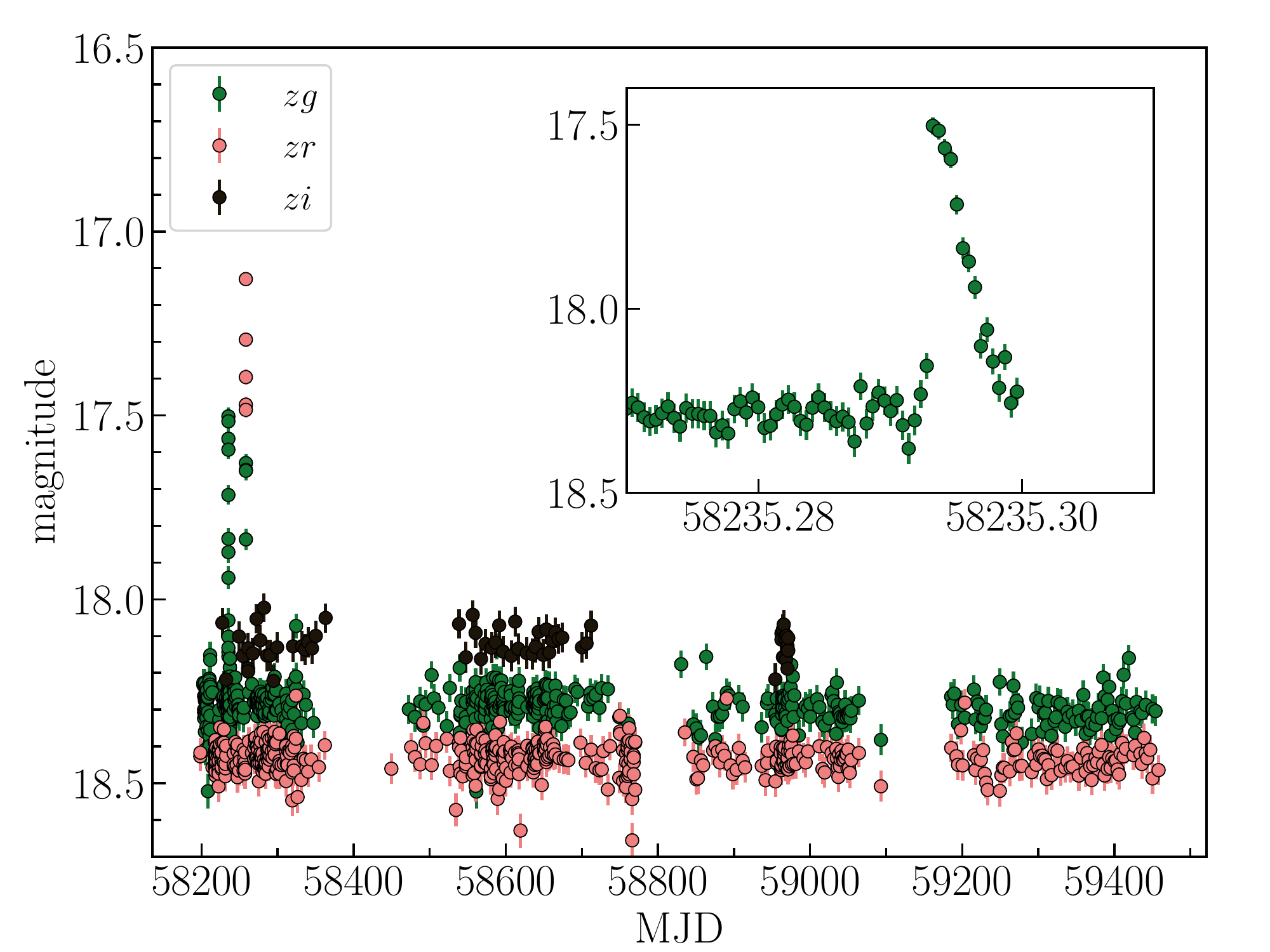}
    \caption{Light curves of a white dwarf-M-dwarf binary system. Observational data correspond to OIDs: {\tt 676113300003418} ($zg$), {\tt 676213300006792} ($zr$), and {\tt 676313300009030} ($zi$). Inset plot shows a zoomed high-cadence $zg$-band light curve with flare and possible periodicity.}
    \label{wdms}
\end{figure}

Another interesting object, OID~=~{\tt 676213300006792}, located at a distance of $\sim$234~pc~\citep{2018AJ....156...58B}, shows two outbursts, one of which is observed at a high frequency (see Fig.~\ref{wdms}). Based on its SDSS spectrum, {\tt 676213300006792} was previously identified as a white dwarf-main sequence binary with a secondary M-dwarf companion~\citep{2012MNRAS.424.1841L}.  
{\tt 676213300006792} is a weak UV source and does not appear in any X-ray database. Its SDSS spectrum does not show a significant $H_\alpha$ emission. The high cadence $zg$-band light curve shows a periodicity with $P\simeq4.25$~minutes, just before the flare. We assume that there is no stable mass transfer in the system, and the M-dwarf has not overflowed its Roche lobe.
We attribute the outbursts to the low accretion rate of the unstable stellar wind on the white dwarf during the increase in  magnetic activity from the M-dwarf. The periodic variation before the flare may be related to a hot spot in the temporary accretion disc.

\begin{figure}
    \centering
    \includegraphics[width=0.5\textwidth]{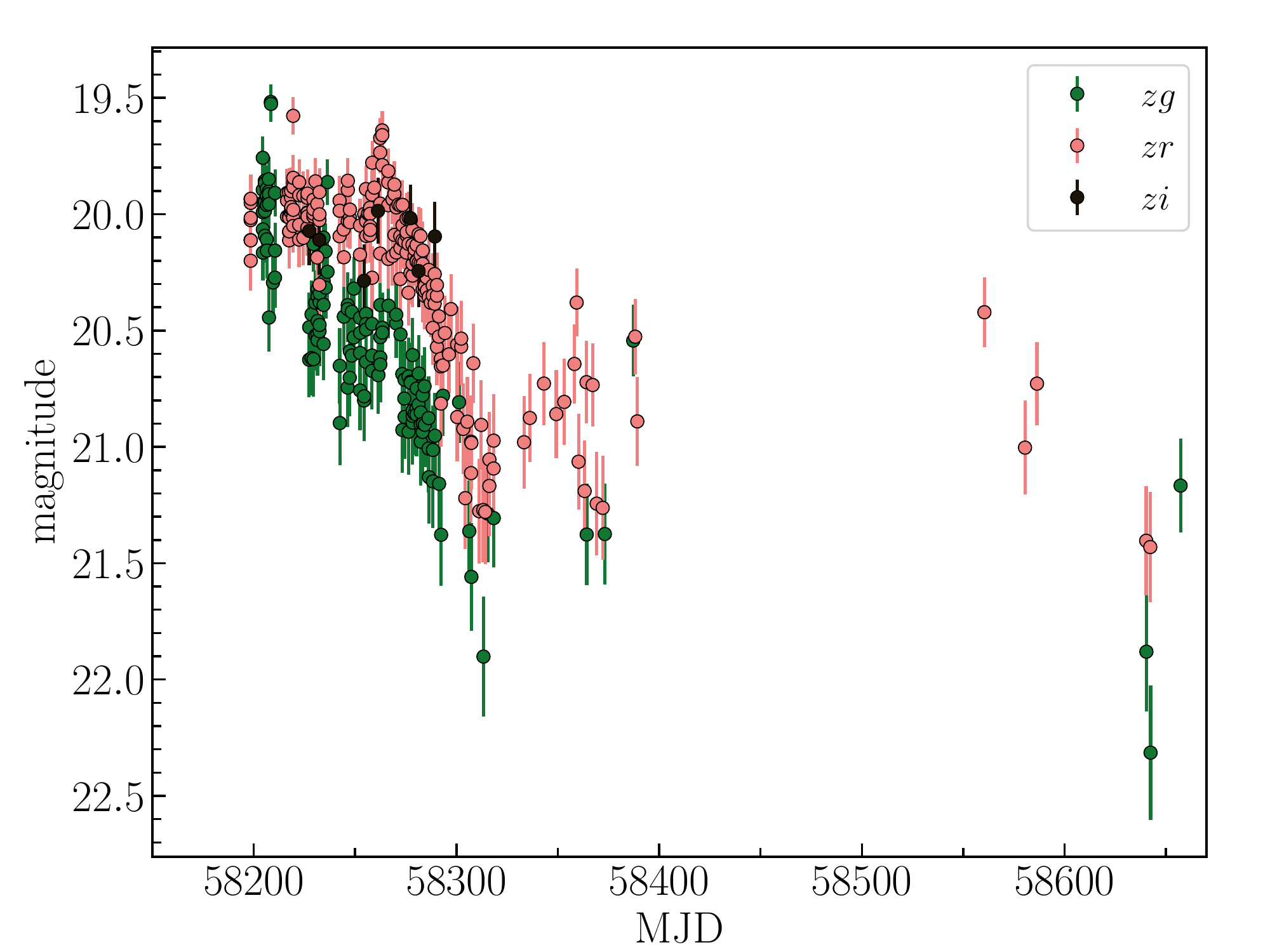}
    \caption{Light curves of an AGN candidate. Observational data correspond to OIDs: {\tt 763114100009685} ($zg$), {\tt 763214100020120} ($zr$), and {\tt 763314100028589} ($zi$).}
    \label{agn}
\end{figure}

\begin{figure}
    \centering
    \includegraphics[width=0.5\textwidth]{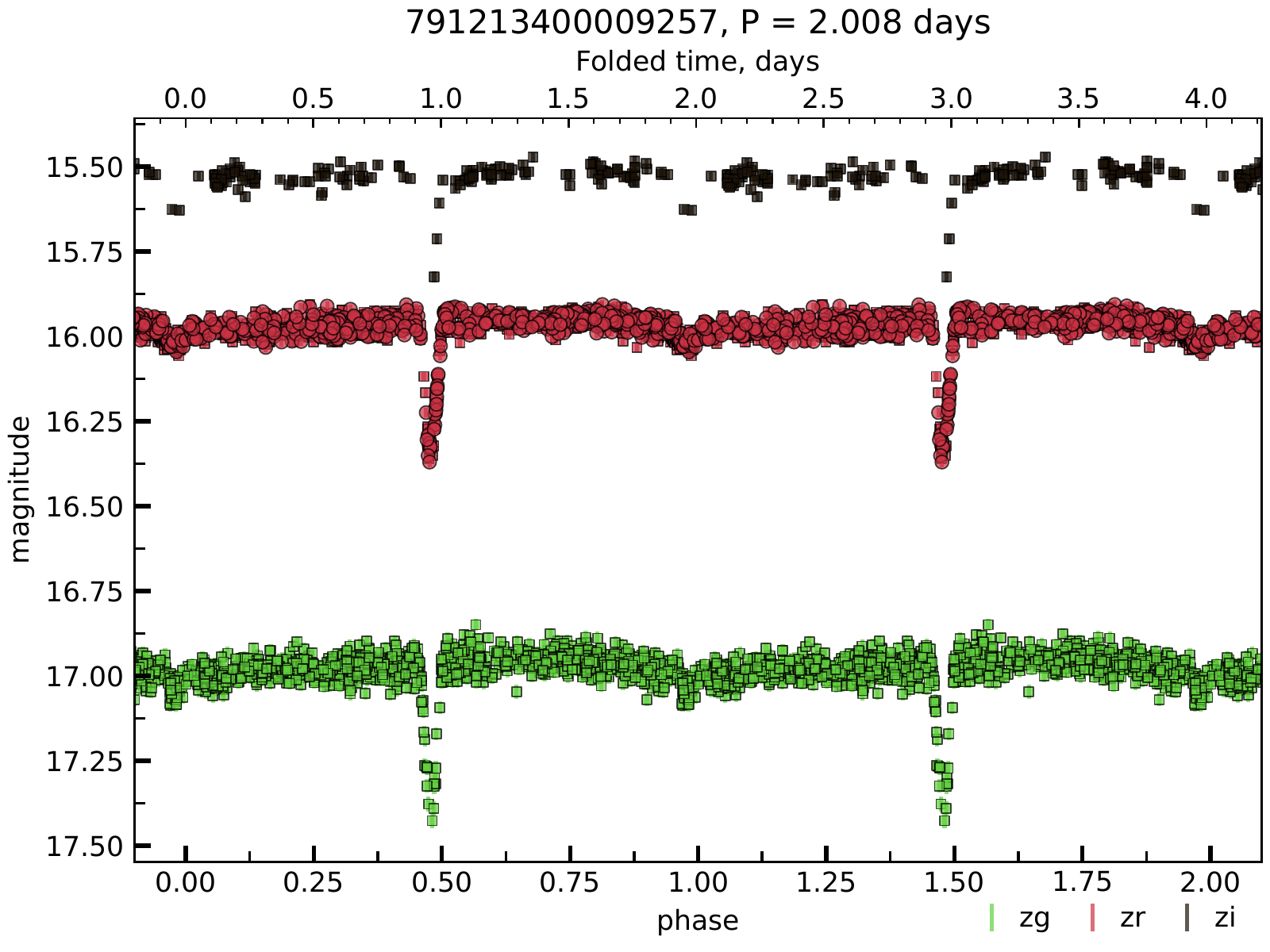}
    \caption{Folded light curves of a non-catalogued eclipsing binary candidate. Observational data correspond to OIDs: {\tt 791113400002334}, {\tt 792116300002277 } ($zg$); {\tt 791213400009257}, {\tt 792216300003876 } ($zr$); {\tt 791313400005610}, and {\tt 792316300013740} ($zi$).}
    \label{EB}
\end{figure}

Other non-catalogued objects include candidates \maria{for} AGNs (e.g.~Fig.~\ref{agn}) and variable stars of different nature (e.g. eclipsing binary candidate in Fig.~\ref{EB}). All these objects can be studied separately in the future by the domain experts.

%-----------------------------------------------------------------

\section{Conclusions}
\label{sec:conclusions}

In this work, we provide the first results from the complete SNAD adaptive learning pipeline  
in the presence of big data from large-scale astronomical surveys. The SNAD team became aware of the existence of non-reported supernova candidates within the ZTF DRs once they appeared in a non-targeted anomaly detection search \citep{Malanchev2021}. A new experiment was then designed to develop a tailored machine learning model which would explore this possibility by taking advantage of the SNAD adaptive learning pipeline \citep{Ishida2021} and our experts\varsha{'} long-term experience studying supernovae. 

We selected 70 ZTF fields in high galactic latitude, employed a series of quality cuts followed by designed feature extraction (Section \ref{subsec:ZTF}). The resulting homogeneous feature sets (one per field) were submitted independently of 30 iterations of the active anomaly discovery algorithm, where at each iteration, the domain expert would input a positive feedback to any outlier whose light curve resembles a \maria{SN} and a negative one otherwise. During this process, human-assigned labels were added to the SNAD knowledge database, opening the way for future deeper analysis of the same data  (Section \ref{subsec:labels}). From the 2100 objects  visually inspected, we found 104 SN-like events, 57 of which \varsha{were} reported for the first time. These transients received an internal name, were reported to TNS and added to the SNAD catalogue\footnote{\url{https://snad.space/}} (see Section \ref{subsec:results}). 

In order to evaluate probable classification types for the newly found transients, we performed light curve fits using different supernova models (Section \ref{sec:models}). Among the newly found transients, we reported three objects (SNAD121, SNAD160 and SNAD187) with broad, slowly evolving light curves that stand as promising superluminous supernova candidates 
\citep[see Fig. \ref{SN2006gya} and ][]{2022RNAAS...6..122P}. 

Despite the fact that the AAD was aimed at supernova search, other potentially interesting objects have been found, including non-catalogued AGNs and red dwarf flares. The high cadence data of discovered flares opens the possibility of searching for fast transients in ZTF. Moreover, the visual inspection of AAD outliers during the SN search led to the creation of the SNAD knowledge database that can be used for different machine learning tasks in \varsha{the} future\footnote{Results from the Active Anomaly Discovery algorithm and light-curve feature set are available in Zenodo, at \url{https://zenodo.org/record/6998913}.}.

The overall efficiency of the pipeline is highly dependent on the total number of objects being analysed, feature choices, and maximum iterations budget, among other parameters. Nevertheless, the results presented here confirm the effectiveness of adaptive learning approaches in filtering large astronomical data sets for expert analysis. They reveal important characteristics of ZTF data releases that ought to be further scrutinised to avoid similar losses in the future \citep{aleo2022}.  

\begin{acknowledgements}
We thank Anastasia Voloshina and Alexandra Zubareva for the assistance in variable star classification and analysis. We also thank Stephane Blodin and Alexandra Kozyreva for discussion involving PISN modelling.  The reported study was funded by RFBR and CNRS according to the research project №21-52-15024. We used the equipment funded by the Lomonosov Moscow State University Program of Development. The authors acknowledge the support by the Interdisciplinary Scientific and Educational School of Moscow University “Fundamental and Applied Space Research”. P.D.A. is supported by the Center for Astrophysical Surveys (CAPS) at the National Center for Supercomputing Applications (NCSA) as an Illinois Survey Science Graduate Fellow. V.V.K. is supported by the Ministry of science and higher education of Russian Federation, topic № FEUZ-2020-0038.  E.E.O.I.  received financial support from CNRS International Emerging Actions under the project \textit{Real-time analysis of astronomical data for the Legacy Survey of Space and Time} during 2021-2022.
\end{acknowledgements}


\bibliographystyle{aa}
\bibliography{ref}

\begin{appendix}
\onecolumn

\section{AAD results}
\label{ap:catalogue}

We report the complete set of SN-like transients shown to the expert by the AAD pipeline below.

    \begin{longtable}{p{0.3cm}p{1.3cm}p{1.3cm}p{1.3cm}p{2.5cm}p{1.2cm}p{2.2cm}p{2.6cm}p{2.5cm}}
        \caption{Complete list of supernovae and supernova candidates found by active anomaly discovery algorithm in ZTF DR3.}
      \label{tab:sn_list}
      \\

\hline
\hline
№     &       Name    &       R.A.    &       Dec.    &       OID     &       Type$^*$        &       \maria{Target alert name}     &       Other names     &       Remarks \\
\hline
\endfirsthead
\caption{continued}\\
\hline
№     &       Name    &       R.A.    &       Dec.    &       OID     &       Type$^*$        &       \maria{Target alert name}     &       Other names     &       Remarks \\
\hline
\endhead
\hline
\endfoot
\hline
\endlastfoot
1       &       SNAD101 &       247.45543       &       24.77282        &       633207400004730 &       IIn     &       ZTF18abqkqdm    &       AT 2018lwh &       \\
2       &       SNAD102 &       245.05375       &       28.38220        &       633216300024691 &       IIn     &       ZTF18abdgwos    &       AT 2018lwi &       \\
3       &       SNAD103 &       219.83787       &       47.40378        &       758205100001118 &       IIP     &               &       AT 2018lwj &       \\
4       &       SNAD104 &       218.91620       &       46.38441        &       758205400019523 &       IIP     &       ZTF18aawqbuc    &       AT 2018lwk &       \\
5       &       SNAD105 &       218.34626       &       49.22553        &       758209200010983 &       Ibc     &               &       AT 2018lwl &       \\
6       &       SNAD106 &       219.41935       &       50.16706        &       758213400002862 &       Ia      &               &       AT 2018lwm &       \\
7       &       SNAD107 &       254.41703       &       44.29804        &       762202300014913 &       Ibc     &               &       AT 2018lwn &   \\
8       &       SNAD108 &       257.84004       &       48.21127        &       762209400037712 &       Ia      &       ZTF18abauopo    &       AT 2018lwo &       \\
9       &       SNAD109 &       191.17808       &       54.31252        &       790207100016149 &       IIn     &               &       AT 2018lwp &       \\
10      &       SNAD111 &       263.42634       &       52.62203        &       796201100002136 &       Ibc     &       ZTF18abdldos    &       AT 2018lwr &       \\
11      &       SNAD112 &       263.37725       &       51.25848        &       796201400007564 &       Ia      &       ZTF18aaubejv    &       AT 2018lws &       \\
12      &       SNAD113 &       254.81890       &       58.37547        &       796215200010831 &       Ia      &       ZTF18abmmdnx    &       AT 2018lwt &       825204400024382 \\
13      &       SNAD113 &       254.81983       &       58.37536        &       825204400024382 &       Ia      &       ZTF18abmmdnx    &       AT 2018lwt &       796215200010831 \\
14      &       SNAD114 &       257.27308       &       57.55380        &       796215400002705 &       IIn     &       ZTF18aawkaod    &       AT 2018lwu &       \\
15      &       SNAD115 &       252.82337       &       62.83286        &       824209400038197 &       IIn     &       ZTF18absufwv    &       AT 2018lwv &       \\
16      &       SNAD116 &       242.29542       &       62.48738        &       824212400005043 &       Ia      &       ZTF18abklrmo    &       AT 2018lww &       \\
17      &       SNAD117 &       254.04870       &       65.01217        &       824213100016761 &       IIP     &       ZTF18aanbiig    &       AT 2018lwx &       \\
18      &       SNAD119 &       245.80060       &       68.69152        &       847207200014161 &       IIn     &       ZTF18abgujpm    &       AT 2018lwz &       \\
19      &       SNAD120 &       255.06791       &       70.51376        &       847209200012956 &       IIP     &       ZTF18aazydub    &       AT 2018lxa &       \\
20      &       SNAD121 &       248.33307       &       71.11514        &       847211100017171 &       IIP     &       ZTF18abklshn    &       AT 2018lxb &       \\
21      &       SNAD122 &       217.38363       &       33.54185        &       676212400013135 &       IIP     &       ZTF18aawqcqz    &       AT 2018lxg &       \\
22      &       SNAD123 &       230.71268       &       41.05182        &       720209400014960 &       IIn     &       ZTF18aagrczj    &       AT 2018lxh &       721212300003198 \\
23      &       SNAD123 &       230.71268       &       41.05184        &       721212300003198 &       IIn     &       ZTF18aagrczj    &       AT 2018lxh &       720209400014960 \\
24      &       SNAD124 &       212.14042       &       58.49805        &       792215100001492 &       IIL     &       ZTF18aaigpcr    &       AT 2018lxi &       821201400006340 \\
25      &       SNAD124 &       212.14043       &       58.49805        &       821201400006340 &       IIL     &       ZTF18aaigpcr    &       AT 2018lxi &       792215100001492 \\
26      &       SNAD125 &       222.74691       &       55.19418        &       793211300013937 &       Ia      &       ZTF18aaykuzt    &       AT 2018lxj &       \\
27      &       SNAD126 &       226.65118       &       54.70207        &       793206100016155 &       IIn     &       ZTF18aaziehk    &       AT 2018lxk &       \\
28      &       SNAD127 &       224.96223       &       57.52264        &       793214300002434 &       IIn     &       ZTF18aaqufmy    &       AT 2018lxl &       \\
29      &       SNAD128 &       230.50802       &       53.61611        &       793205400010761 &       IIn     &               &       AT 2018lxm &       \\
30      &       SNAD129 &       237.28114       &       41.91548        &       721210100012349 &       IIL     &       ZTF18aalrsbd         &       AT 2018lxn      &       \\
31      &       SNAD130 &       232.13026       &       51.98997        &       794204400017737 &       Ia      &       ZTF18abedlru    &       AT 2018lxo &       \\
32      &       SNAD131 &       239.69206       &       56.42458        &       794209200012381 &       Ibc     &               &       AT 2018lxp &       \\
33      &       SNAD132 &       239.95822       &       55.34216        &       794209300004376 &       Ibc     &       ZTF18aaumixp    &       AT 2018lxq &       \\
34      &       SNAD133 &       242.93762       &       55.96133        &       795212100007964 &       Ia      &       ZTF18aanbksg    &       AT 2018lxr &       \\
35      &       SNAD134 &       252.30216       &       54.11178        &       795205100007271 &       Ia      &       ZTF18aayatjf    &       AT 2018lxs &       \\
36      &       SNAD135 &       242.30742       &       52.21426        &       795204100013041 &       IIn     &       ZTF18abgvctp         &       AT 2018lxt      &       \\
37      &       SNAD136 &       257.67202       &        63.69693       &       825211200009477 &       Ia      &       ZTF18abrwswl         &       AT 2018lxu      &       \\
38      &       SNAD137 &       260.16229       &        59.49770       &       825202200039582 &       IIn     &       ZTF18aaqzcvy    &       AT 2018lxv &       \\
39      &       SNAD138 &       257.88902       &       62.49717        &       825211300010764 &       Ia      &       ZTF18abedwws         &       AT 2018lxw      &       \\
40      &       SNAD139 &       264.55471       &       44.26862        &       763202300013915 &       IIn     &       ZTF18aajtpsk    &       AT 2018lxx &       \\
41      &       SNAD141 &       268.46498       &       70.79590        &       848210200003752 &       Ia      &               &       AT 2018lxz &       \\
42      &       SNAD142 &       274.19431       &       59.44037        &       826202200030732 &       IIP     &       ZTF18ablvrgt    &       AT 2018lya &       \\
43      &       SNAD143 &       270.03210       &       34.66995        &       682209200018910 &       IIP     &       ZTF18aaqzrpf    &       AT 2018lyb &       \\
44      &       SNAD144 &       269.87561       &       49.88876        &       764216400008221 &       Ia      &       ZTF18abwlrdj    &       AT 2018lyc &       \\
45      &       SNAD145 &       275.79452       &       48.31019        &       764210400028832 &       Ibc     &               &       AT 2018lyd &       \\
46      &       SNAD146 &       197.21596       &       44.96981        &       756202100006670 &       IIn     &       ZTF18aahiqfy    &       AT 2018lyr &       \\
47      &       SNAD147 &       192.55741       &       46.63066        &       756207300010654 &       IIn     &               &       AT 2018lys &       \\
48      &       SNAD160 &       205.97232       &       61.55479        &       821207100004043 &       IIP     &        ZTF18aautopz    &       AT 2018lzi      &       \\
49      &       SNAD161 &       200.90323       &       55.97815        &       791211100013499 &       Ibc     &       ZTF19aaroswc    &       AT 2018lzj &       \\
50      &       SNAD162 &       267.30305       &       62.14472        &       825209400016106 &       Ia      &               &       AT 2018lzk &       \\
51      &       SNAD163 &       265.04514       &       58.79008        &       825201300004134 &       Ibc     &               &       AT 2018lzl &       \\
52      &       SNAD164 &       274.46930       &       62.62716        &       826210300004963 &       Ia      &       ZTF18abnwqje    &       AT 2018lzm & \\
53      &       SNAD165 &       259.39972       &       44.51506        &       763204300004087 &       Ibc     &               &       AT 2018lzn & \\
54      &       SNAD166 &       259.60178       &       48.60846        &       763212400008960 &       Ia      &       ZTF18abrwtqb    &       AT 2018lzo &       \\
55      &       SNAD184 &       223.60115       &       32.77872        &       676205100003608 &       IIL     &       ZTF18abdgucl    &       AT 2018mby &       \\
56      &       SNAD185 &       191.47409       &       42.22174        &       716211100008498 &       IIL     &       ZTF18acnnerq    &       AT 2018mbz &       \\
57      &       SNAD186 &       204.74466       &       43.90742        &       718216200026001 &       IIL     &       ZTF18aagrfaw    &       AT 2018mca &       \\
58      &       SNAD187 &       208.28069       &       40.80206        &       718211400012193 &       IIP     &       ZTF18aaqctvg    &       AT 2018mcb &       \\
59      &       SNAD188 &       210.69041       &       39.65870        &       718206100007526 &       Ia      &               &       AT 2018mcc &       \\
60      &       SNAD189 &       224.46669       &       38.95331        &       720207300005154 &       Ia      &               &       AT 2018mcd &       \\
61      &               &       246.05583       &       25.27213        &       633208100018991 &       Ia-91bg         &       ZTF18aaiajvb    &       SN 2018baz      &               \\
62      &               &       252.75330       &       25.87635        &       634208100013151 &       Ia      &       ZTF18abmxfrc    &       SN 2018fin &               \\
63      &               &       256.52059       &       29.66854        &       634214200012529 &       II      &       ZTF18aainvic    &       SN 2018jo  &               \\
64      &               &       224.06941       &       34.05073        &       676209400030690 &       PSN     &       ZTF18aalsomy    &       AT 2018cff &               \\
65      &               &       266.98595       &       34.93270        &       682210200032862 &       PSN     &       ZTF18aamlqqh    &       AT 2018bcz &               \\
66      &               &       264.50123       &       35.08437        &       682212100012973 &       PSN     &               &       AT 2018bhu &               \\
\multirow{2}{*}{67}     &               &       212.59125       &       39.32474        &       718205300001790 &       PSN     &       ZTF18aacckja    &       \small{MASTER OT}     &               \\
        &               &               &               &               &               &               &       \small{J141021.86+391928.4}     &               \\
\multirow{2}{*}{68}     &               &       213.91916       &       43.53175        &       718213100006185 &       PSN     &       ZTF18aaihqtg    &       \small{MLS180418:}      &       719216200005369 \\
        &               &               &               &               &               &               &       \small{141541+433154}   &               \\
69      &               &       218.33353       &       41.26712        &       719210300033956 &       Ib      &       ZTF18aagrdcs    &       SN 2018alc &               \\
\multirow{2}{*}{70}     &               &       213.91894       &       43.53171        &       719216200005369 &       PSN     &       ZTF18aaihqtg    &       \small{MLS180418:}      &       718213100006185 \\
        &               &               &               &               &               &               &       \small{141541+433154}   &               \\
71      &               &       226.84514       &       38.41658        &       720202200013161 &       Ia      &       ZTF18aabyhlc    &       SN 2018aab &               \\
72      &               &       232.22274       &       37.62043        &       721204400000387 &       PSN     &       ZTF18aaikozr    &       AT 2018doz &               \\
73      &               &       237.51483       &       42.08853        &       721210100010621 &       Ia      &        ZTF18aagstdc    &       SN 2018apn      &               \\
74      &               &       247.27265       &       39.88654        &       722205200035190 &       II      &       ZTF18aalbpll    &       SN 2018cbb &               \\
75      &               &       245.78265       &       39.31497        &       722206400011527 &       II      &       ZTF18aahbfcr    &       SN 2018anx &               \\
76      &               &       247.26026       &       43.62678        &       722213200005805 &       II      &       ZTF18aarpttw    &       SN 2018bqs &               \\
77      &               &       248.04807       &       42.71343        &       722213400026577 &       Ia      &       ZTF18aagtcxj    &       SN 2018aqm &               \\
78      &               &       254.27082       &       39.38169        &       723206400011620 &       IIP     &       ZTF18aajczsi    &       SN 2018aql &               \\
79      &               &       248.97523       &       40.03282        &       723208100004162 &       II      &       ZTF18aaiyqow    &       SN 2018gk  &               \\
80      &               &       247.27265       &       39.88654        &       723211300005112 &       Ic BL      &       ZTF18abukavn    &       SN 2018gep      &               \\
81      &               &       250.36668       &       44.02829        &       723215200033236 &       PSN     &       ZTF18abitwua    &       AT 2018feo &       762204400033332 \\
82      &               &       251.97713       &       42.96844        &       723215400005033 &       Ia      &       ZTF18abpmmpo    &       SN 2018fnd &               \\
83      &               &       240.45421       &       47.39072        &       761208100023231 &       Ia      &       ZTF18aamlhee    &       SN 2018zs  &               \\
84      &               &       254.72370       &       45.28910        &       762202200038305 &       Ic      &       ZTF18abfzhct    &       SN 2018dxt &               \\
85      &               &       252.70920       &       45.39807        &       762203100020850 &       II      &       ZTF18abffyqp    &       SN 2018dfi         &               \\
86      &               &       250.36654       &       44.02827        &       762204400033332 &       PSN     &       ZTF18abitwua    &       AT 2018feo &       723215200033236 \\
87      &               &       254.77103       &       47.23657        &       762206200028741 &       Ia      &       ZTF18abauprj    &       SN 2018cnw &               \\
\multirow{2}{*}{88}             &               &       259.28671       &       48.03253        &       763212400019431 &       PSN     &       ZTF18aarrffk    &       \small{MLS180604:}      &               \\
        &               &               &               &               &               &               &       \small{171709+480157}   &               \\
89      &               &       265.56735       &       50.48871        &       763214400010041 &       II      &       ZTF18aamiman    &       SN 2018arc &               \\
90      &               &       185.39255       &       55.57447        &       789209400002035 &       Ia      &       ZTF18aaiscil         &       SN 2018aae      &       790212300001679 \\
91      &               &       185.39253       &       55.57446        &       790212300001679 &       Ia      &       ZTF18aaiscil         &       SN 2018aae      &       789209400002035 \\
92      &               &       203.77313       &       53.87516        &       791206300017767 &       Ia      &       ZTF18aakecej    &       SN 2018bbj &               \\
93      &               &       200.67064       &       54.35235        &       791207200004054 &       PSN     &       ZTF18aaitbcm    &       AT 2018awa &               \\
\multirow{2}{*}{94}     &               &       208.78603       &       58.49483        &       791213100013510 &       SLSN-I  &       ZTF18aajqcue    &       SN 2018don &       792216100036120, \\
        &               &               &               &       &               &               &               &       792216100000865 \\
95      &               &       209.77291       &       58.26840        &       792216100002676 &       Ia      &               &       SN 2018avz &               \\
\multirow{2}{*}{96}     &               &       208.78615       &       58.49533        &       792216100036120 &       SLSN-I  &       ZTF18aajqcue    &       SN2018don       &       791213100013510, \\
        &               &               &               &       &               &               &               &       792216100000865 \\
\multirow{2}{*}{97}     &               &       208.78601       &       58.49476        &       792216100000865 &       SLSN-I  &       ZTF18aajqcue    &       SN2018don       &       791213100013510, \\
        &               &               &               &       &               &               &               &       792216100036120 \\
98      &               &       230.21754       &       54.21590        &       793205100013372 &       IIP     &       ZTF18abcfdzu    &       SN 2018dfa &       794208200022136 \\
99      &               &       222.69960       &       54.40795        &       793207200003904 &       Ia      &       ZTF18aakglgw    &       SN 2018aoy &               \\
100     &               &       230.21751       &       54.21591        &       794208200022136 &       IIP     &       ZTF18abcfdzu    &       SN 2018dfa &       793205100013372 \\
\multirow{2}{*}{101}    &               &       248.65767       &       52.27841        &       795202100005941 &       PSN     &       ZTF18aanbnjh    &       \small{MLS180307:}      &               \\
        &               &               &               &       &               &               &       \small{163438+521642}   &               \\
102     &               &       244.74136       &       56.71714        &       795211200035931 &       Ia      &       ZTF18aazixbw    &       SN 2018coi &               \\
103     &               &       252.50392       &       55.83982        &       796212200009183 &       PSN     &       ZTF18abrwrch    &       AT 2018giw &               \\
104     &               &       274.99848       &       51.79623        &       798204300012865 &       IIb     &       ZTF18abgrbjb    &       SN 2018efd &               \\
105     &               &       277.67808       &       54.63441        &       798207200002997 &       Ia Pec     &       ZTF18abhpgje    &       SN 2018eul      &               \\
106     &               &       197.86570       &       65.63811        &       821216200012526 &       Ic      &       ZTF18aaisyyp    &       SN 2018avk &               \\
107     &               &       261.41326       &       59.44670        &       825202200015692 &       Ia      &       ZTF18aasdted         &       SN 2018big      &               \\
108     &               &       261.82509       &       58.65272        &       825202400018278 &       PSN     &       ZTF18aalfqdc    &       AT 2018czw &               \\
109     &               &       252.90590       &       61.54532        &       825208200022152 &       Ia      &       ZTF18aaxwjmp    &       SN 2018coe &               \\
110     &               &       271.04939       &       59.81758        &       826203200034373 &       PSN     &       ZTF18aawkdbk    &       AT 2018cjm &               \\
111     &               &       94.51328        &       78.36701        &       858216400012109 &       IIn     &       ZTF18abtgyme    &       SN 2018zd  &               \\
112     &               &       248.00946       &       78.21141        &       864211100001497 &       IIb     &       ZTF18aaxljll    &       SN 2018gj  &               \\
113     &               &       277.17167       &       75.81321        &       865206200014558 &       Ic      &       ZTF18acapyww    &       SN 2018hpq &               \\
\hline
% }
\end{longtable}
$^*$For the SNAD candidates, the type corresponds to the best-fit model according to the fit with Nugent's supernova templates.
 
% --------------------------------------------------------

% \bigskip
\section{Light curves of the SNAD supernova candidates}
\label{ap:fit_results}

We present below a subset of SNAD candidates and their respective light curve fits (Section \ref{subsec:results}).

\begin{figure*}
    \begin{minipage}{0.49\linewidth}
        \centering
        \includegraphics[scale=0.35]{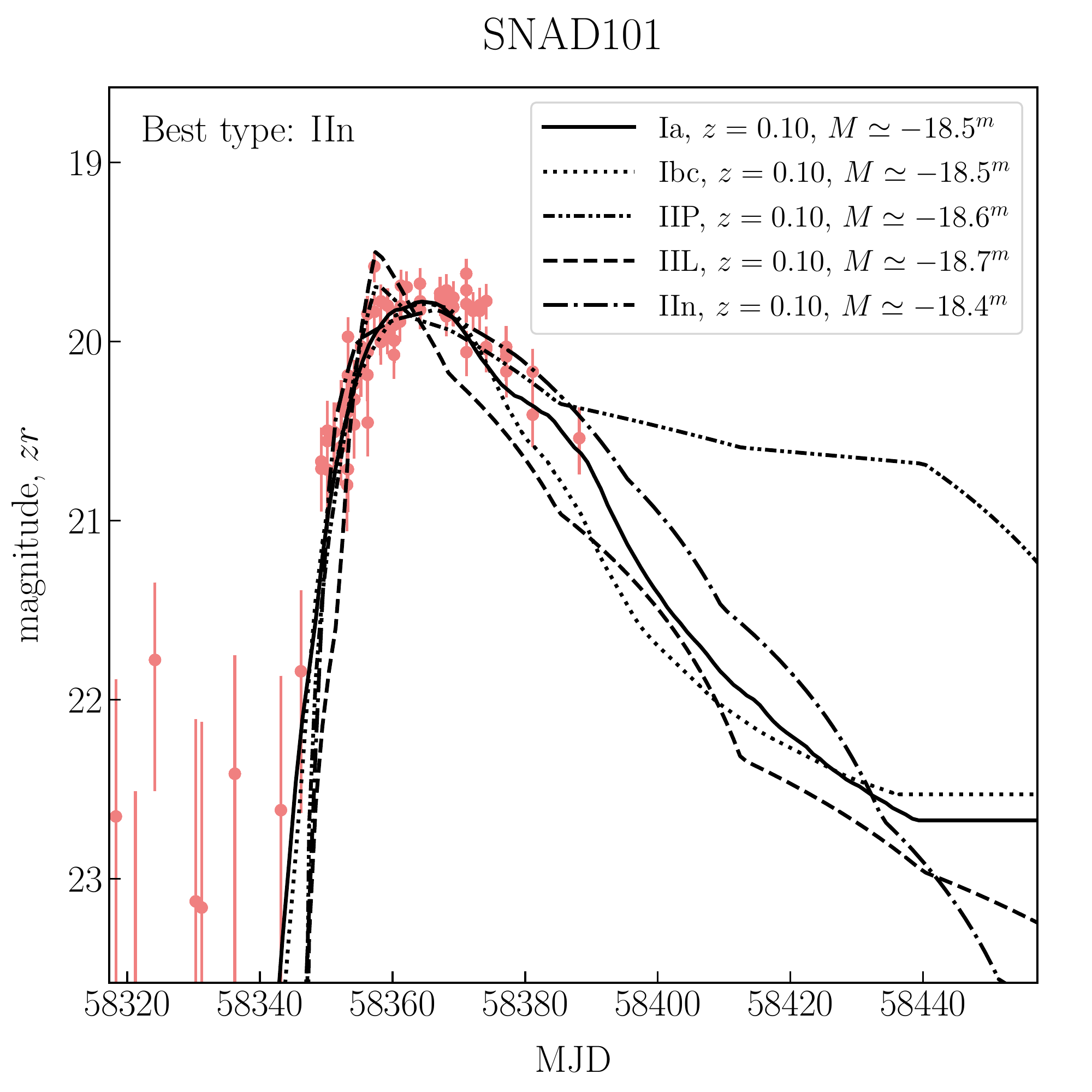}\\a) \\
    \end{minipage}
    \hfill
    \begin{minipage}{0.49\linewidth}
        \centering
        \includegraphics[scale=0.35]{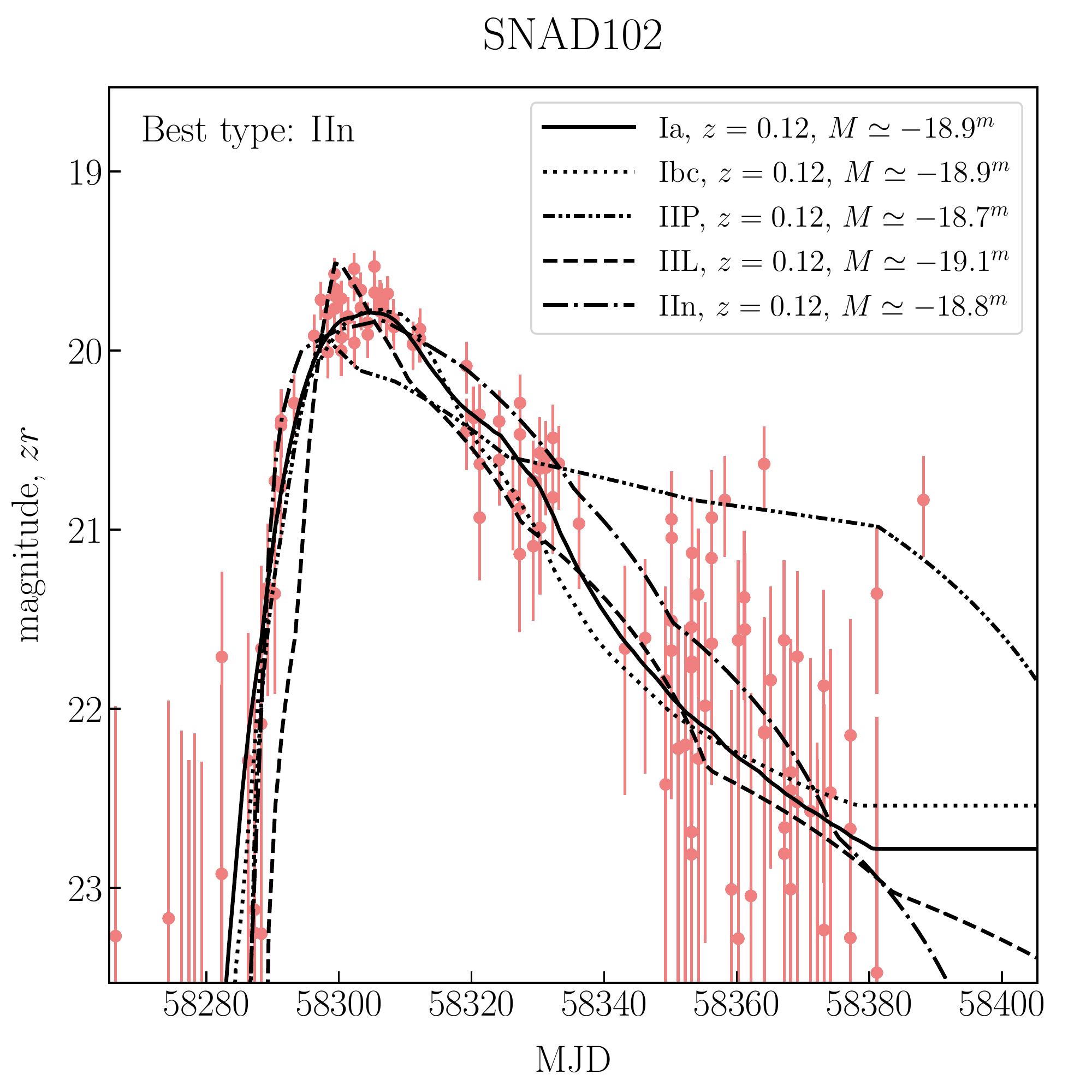}\\b) \\
    \end{minipage}  
    \vfill
    \begin{minipage}{0.49\linewidth}
        \centering
        \includegraphics[scale=0.35]{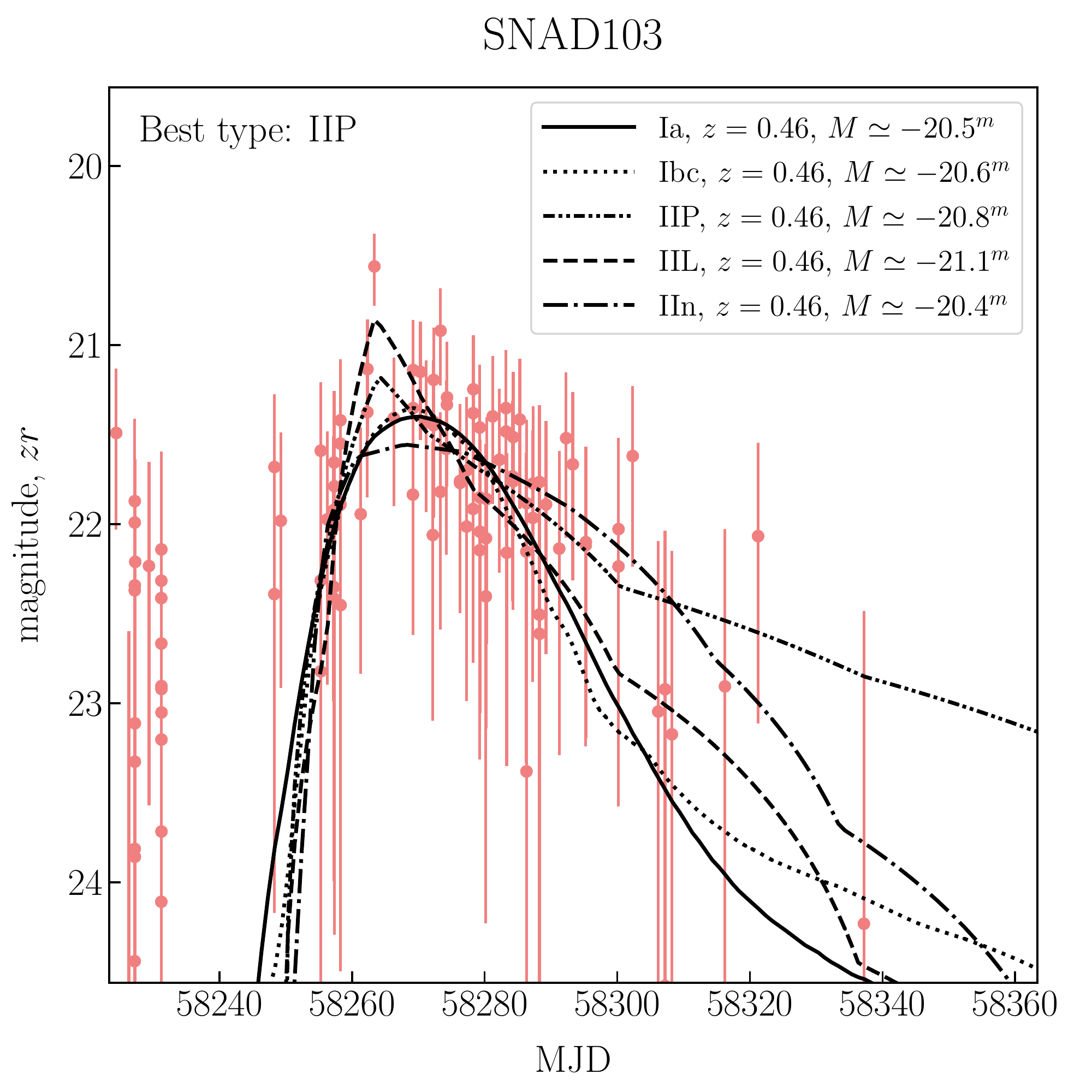}\\c) \\
    \end{minipage}
    \hfill
    \begin{minipage}{0.49\linewidth}
        \centering
        \includegraphics[scale=0.35]{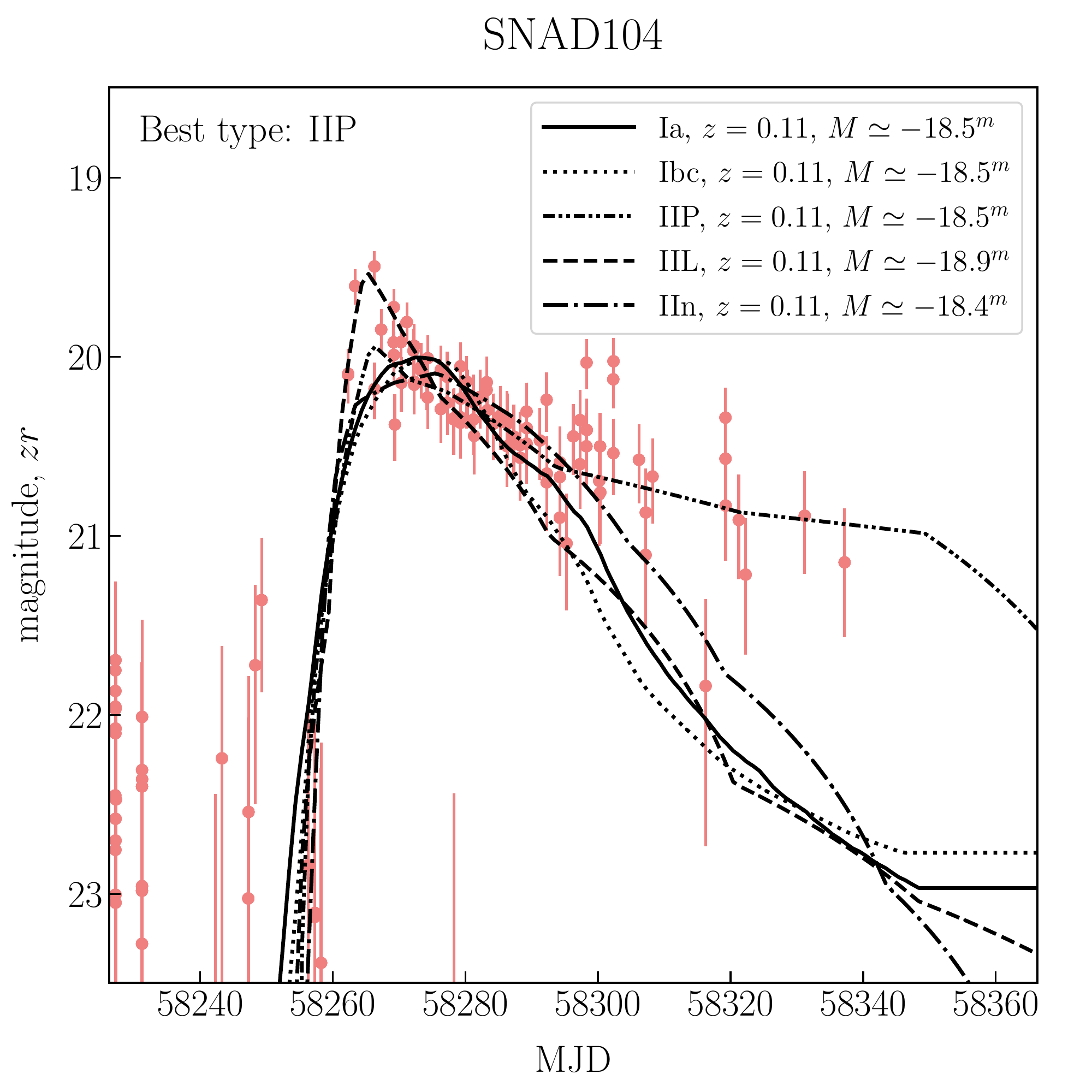}\\d) \\
    \end{minipage}
    \vfill
    \begin{minipage}{0.49\linewidth}
        \centering
        \includegraphics[scale=0.35]{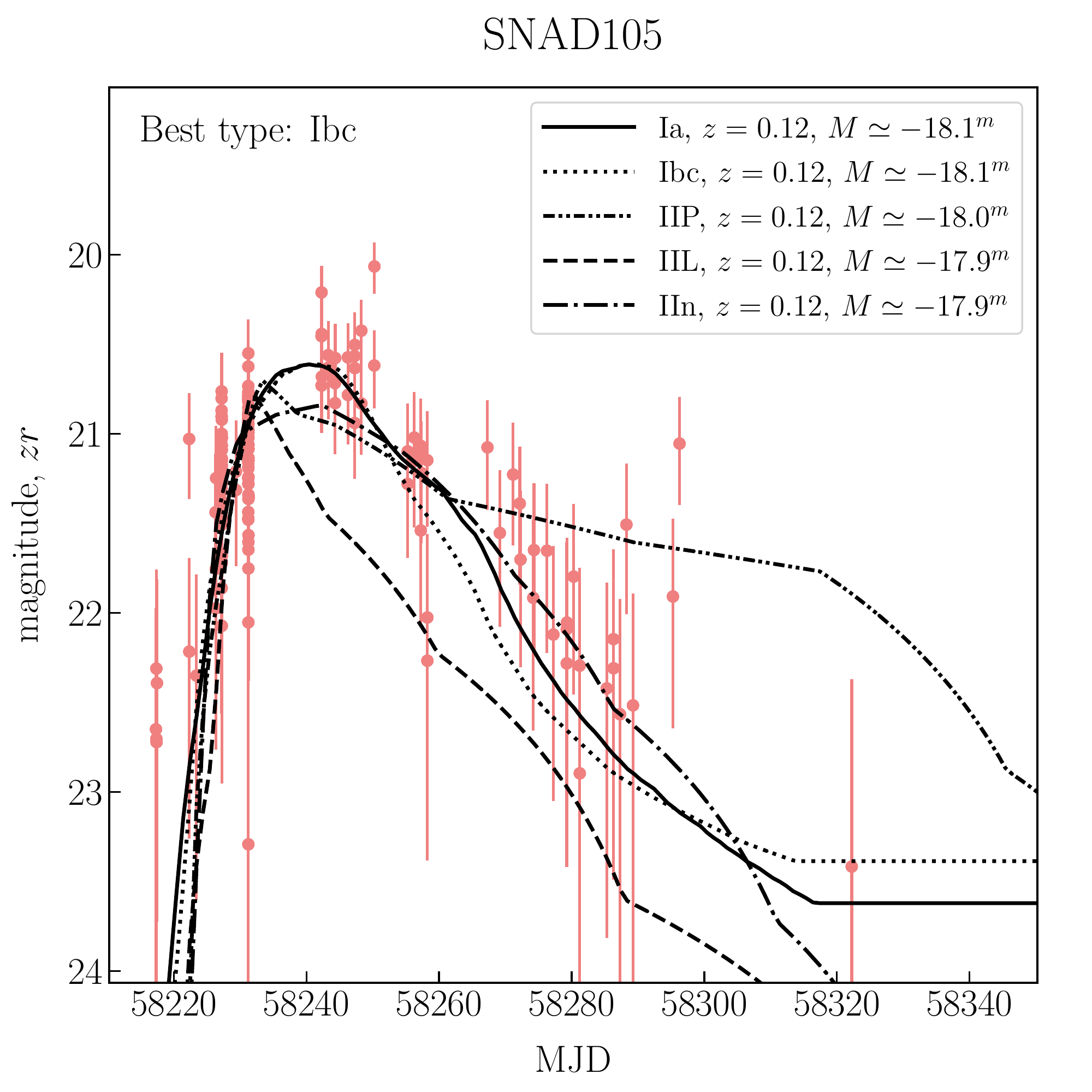}\\e) \\
    \end{minipage}
    \hfill
    \begin{minipage}{0.49\linewidth}
        \centering
        \includegraphics[scale=0.35]{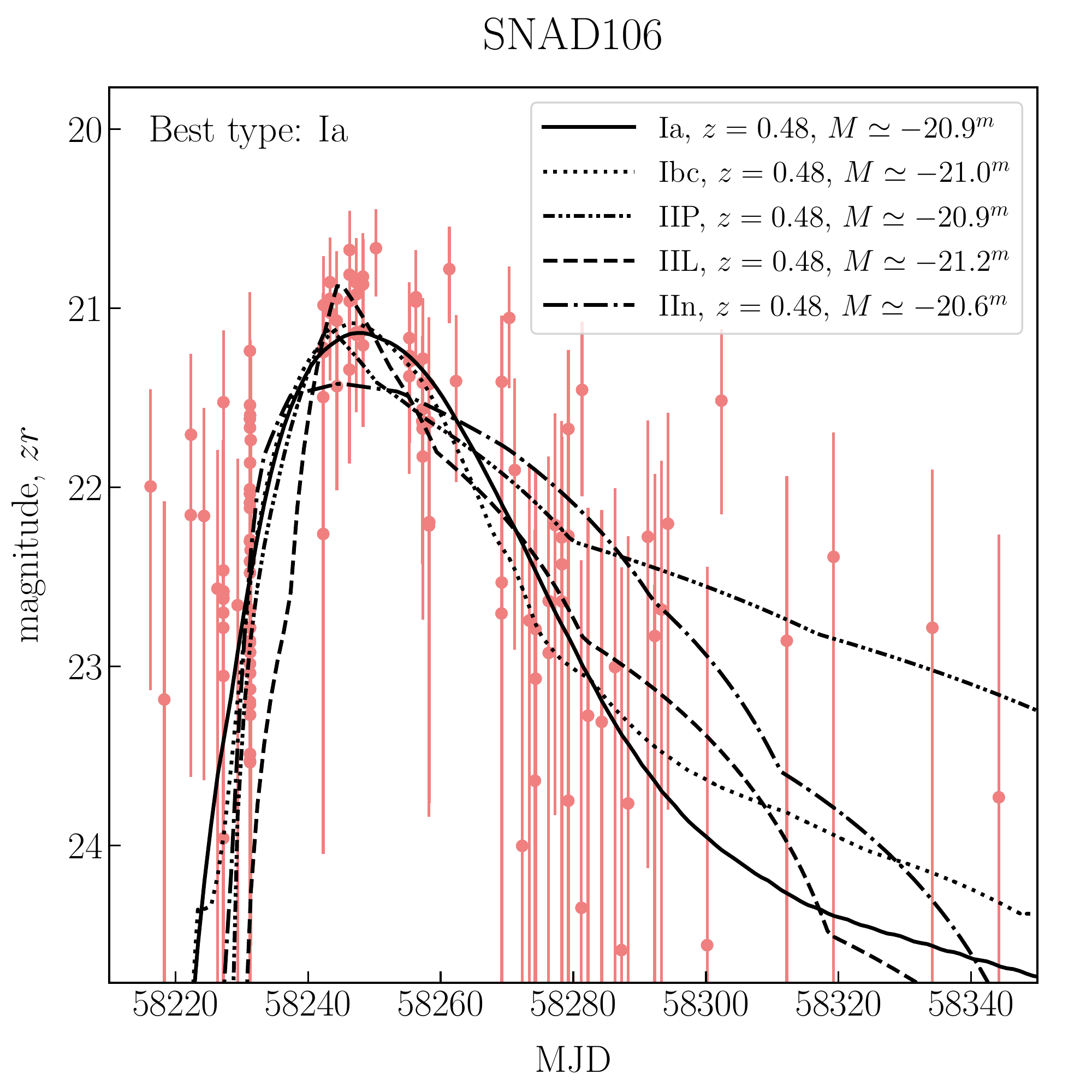}\\f) \\
    \end{minipage} 
    \caption{Light curves of SNAD supernova candidates in $zr$-band and the results of their fit by Nugent's supernova models.}
    \label{fig:snad_LC1}
\end{figure*}

\begin{figure*}
    \begin{minipage}{0.49\linewidth}
        \centering
        \includegraphics[scale=0.35]{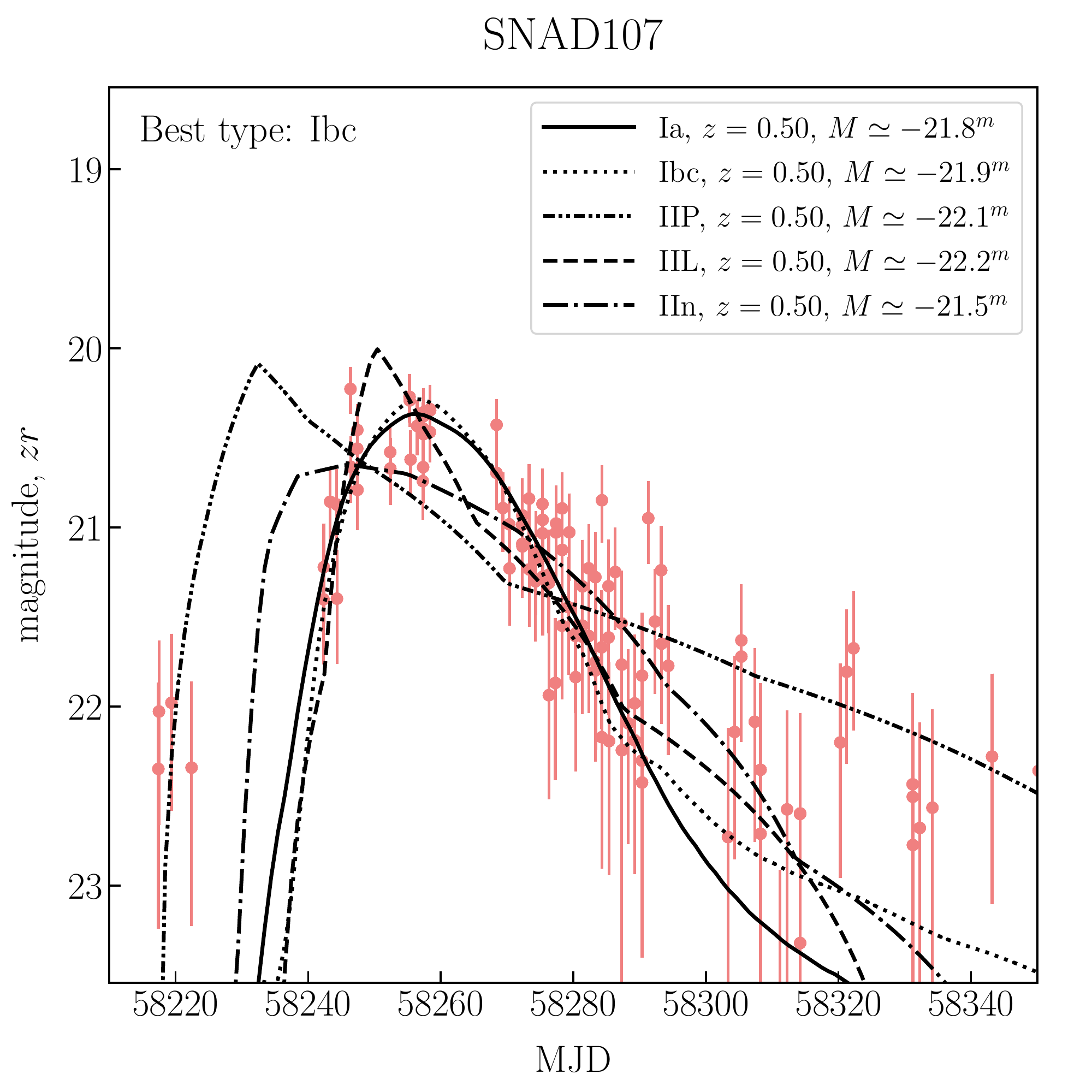}\\a) \\
    \end{minipage}
    \hfill
    \begin{minipage}{0.49\linewidth}
        \centering
        \includegraphics[scale=0.35]{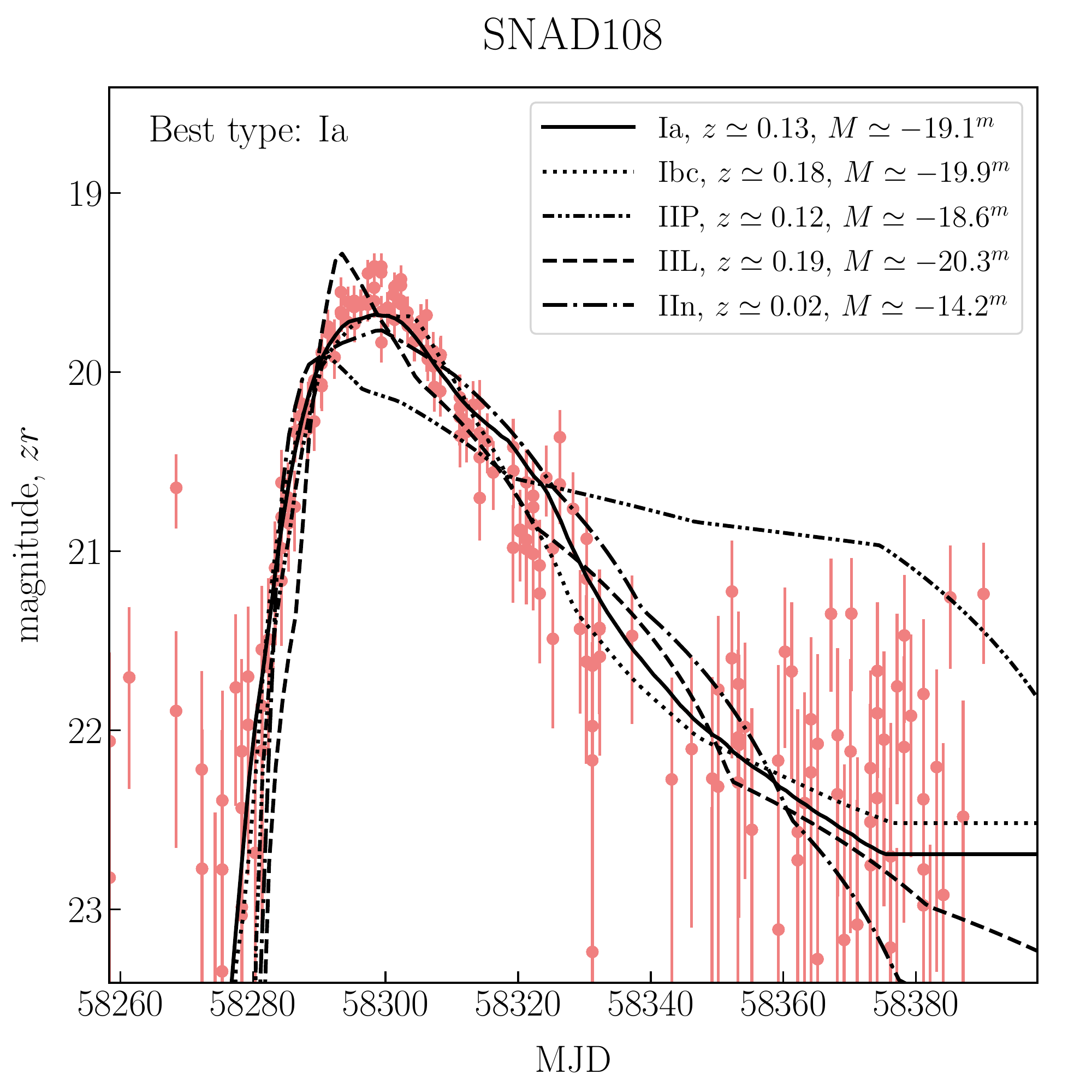}\\b) \\
    \end{minipage}  
    \vfill
    \begin{minipage}{0.49\linewidth}
        \centering
        \includegraphics[scale=0.35]{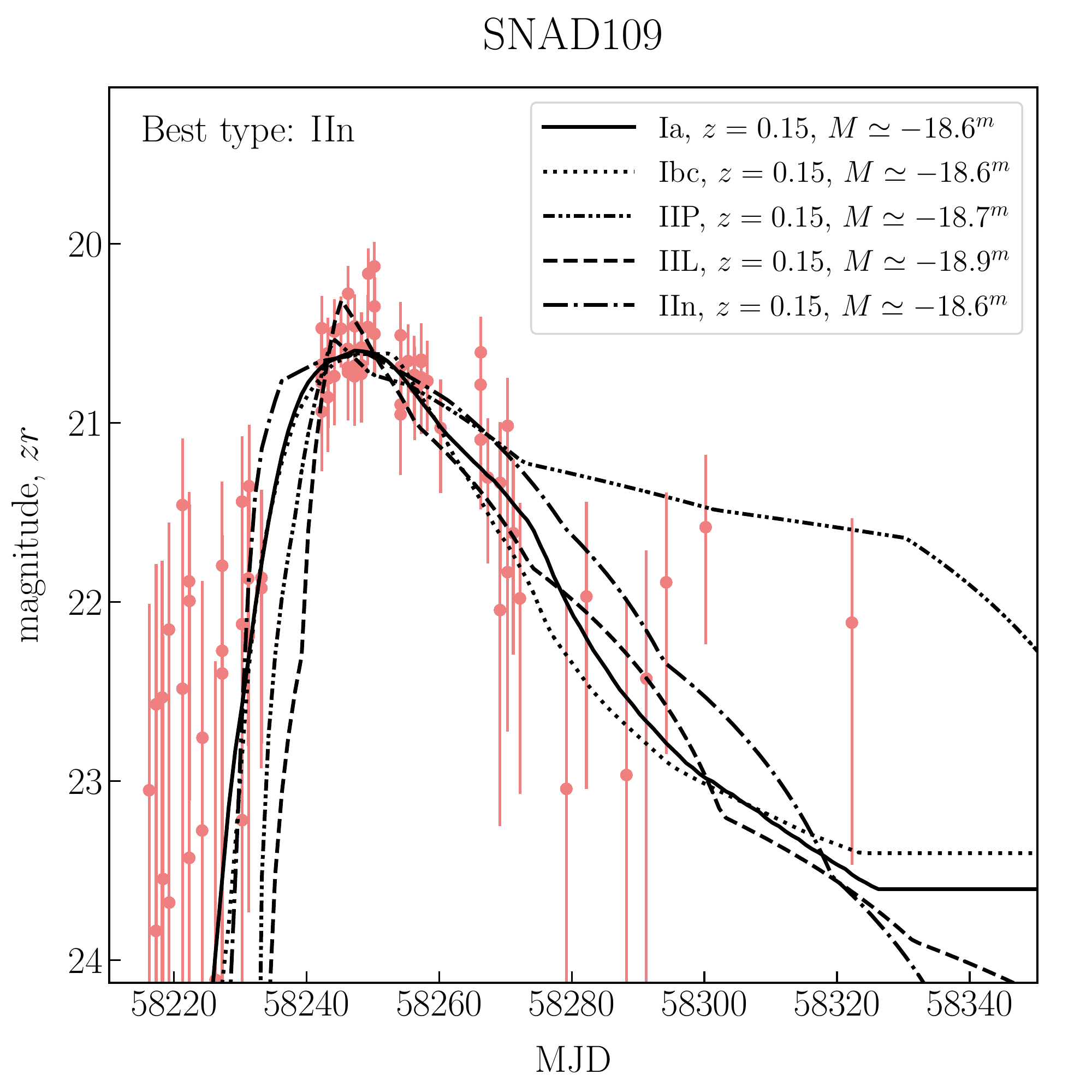}\\c) \\
    \end{minipage}
    \hfill
    \begin{minipage}{0.49\linewidth}
        \centering
        \includegraphics[scale=0.35]{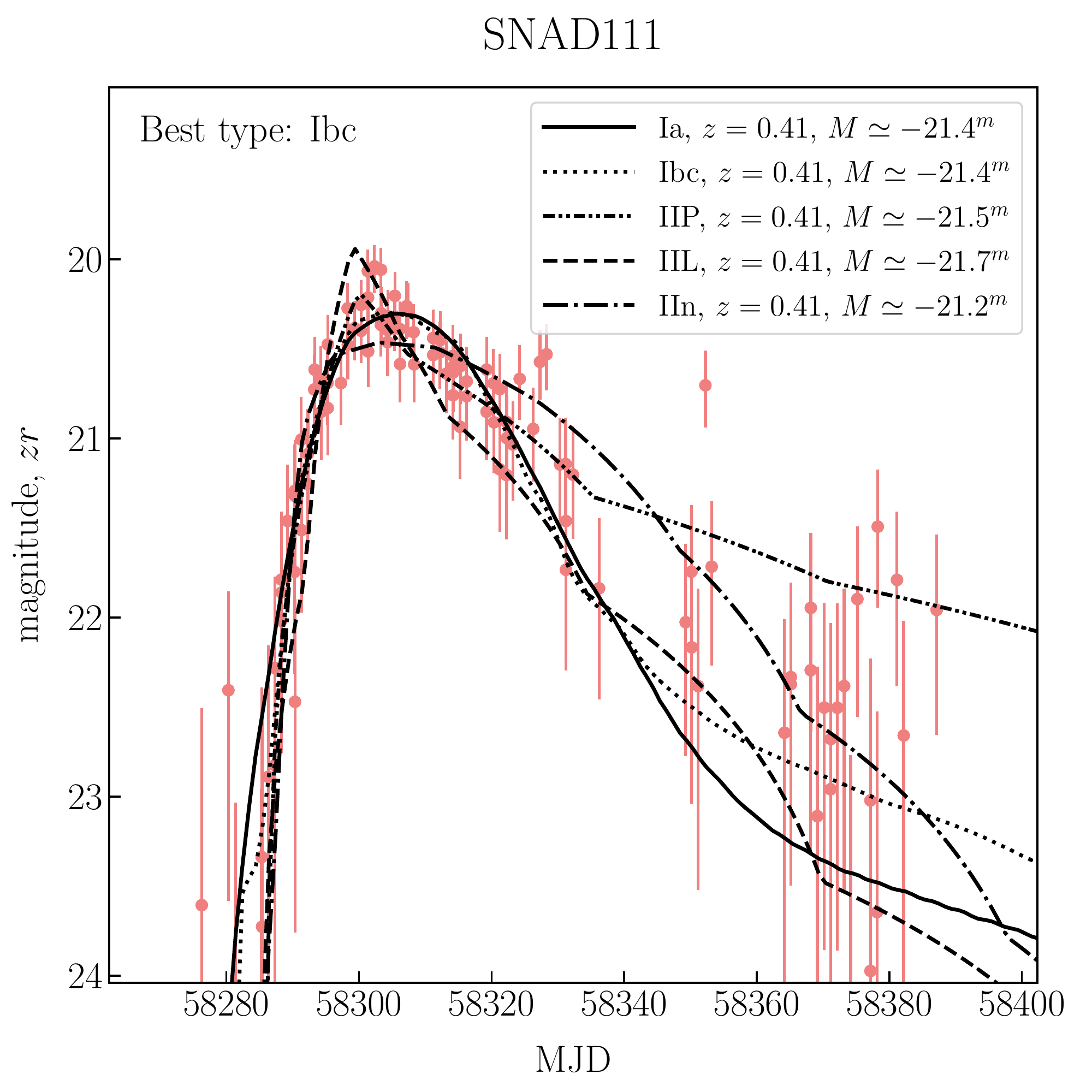}\\d) \\
    \end{minipage}  
    \vfill
    \begin{minipage}{0.49\linewidth}
        \centering
        \includegraphics[scale=0.35]{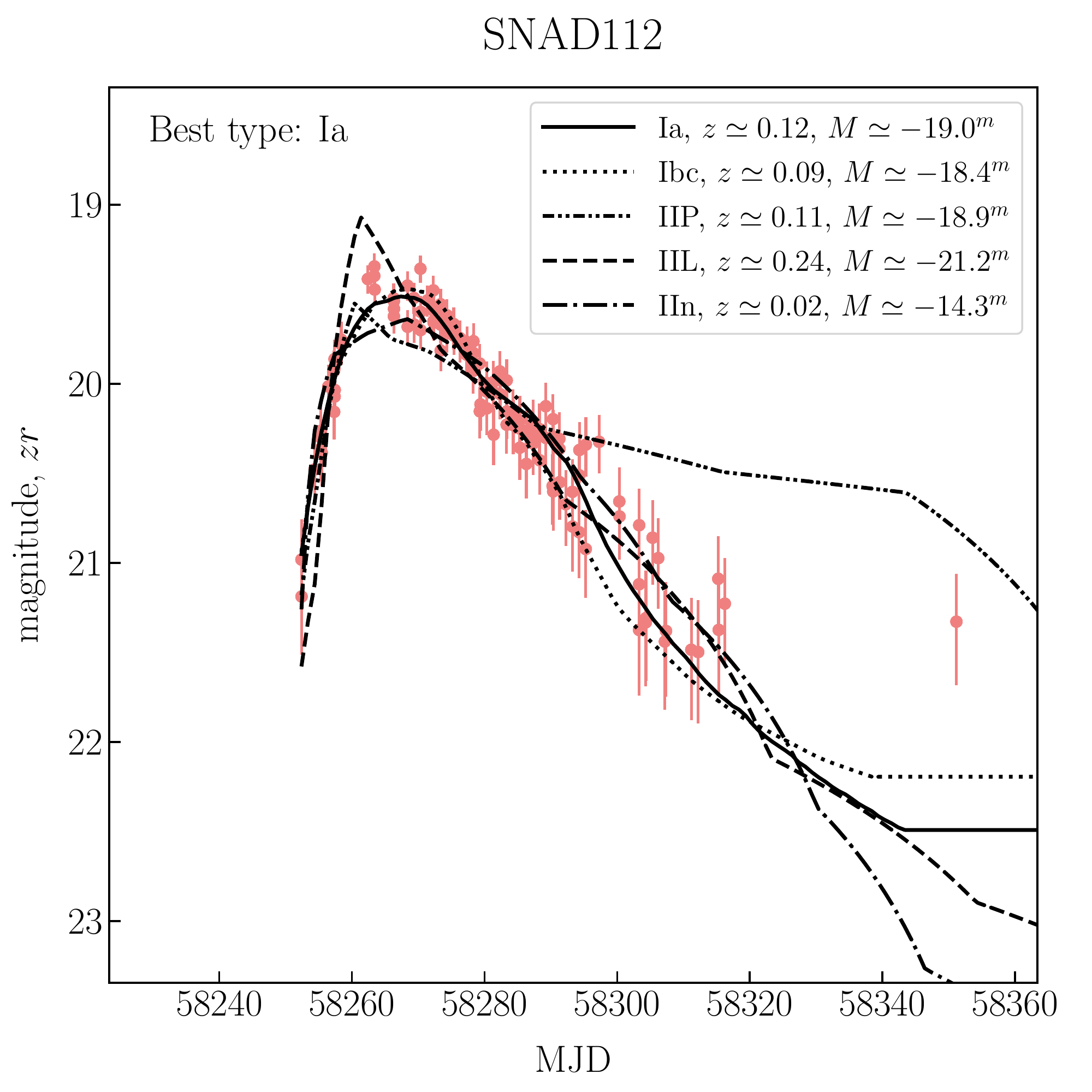}\\e) \\
    \end{minipage}
    \hfill
    \begin{minipage}{0.49\linewidth}
        \centering
        \includegraphics[scale=0.35]{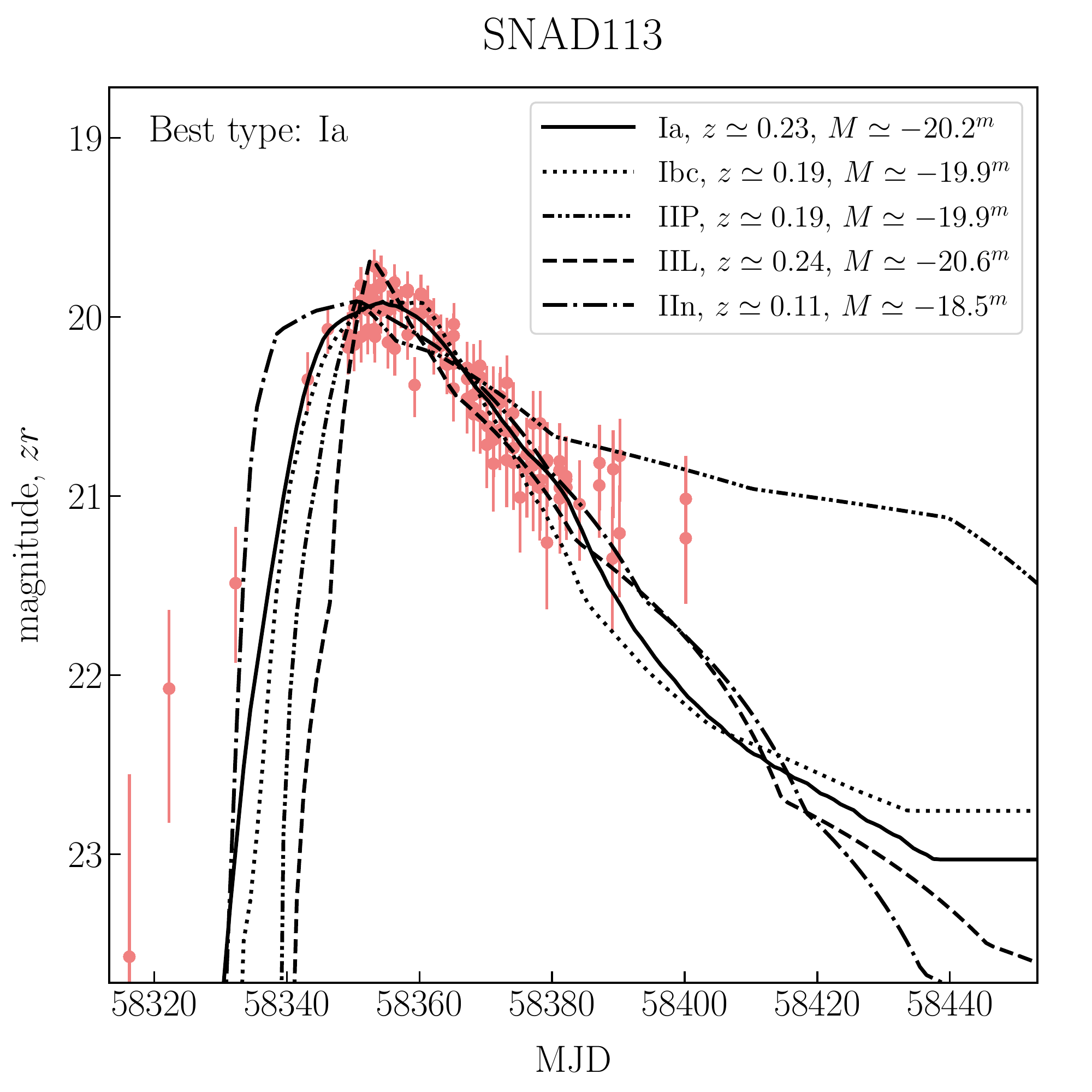}\\f) \\
    \end{minipage}
    \caption{Light curves of SNAD supernova candidates in $zr$-band and the results of their fit by Nugent's supernova models.}
        \label{fig:snad_LC2}
\end{figure*}

\begin{figure*}
    \begin{minipage}{0.49\linewidth}
        \centering
        \includegraphics[scale=0.35]{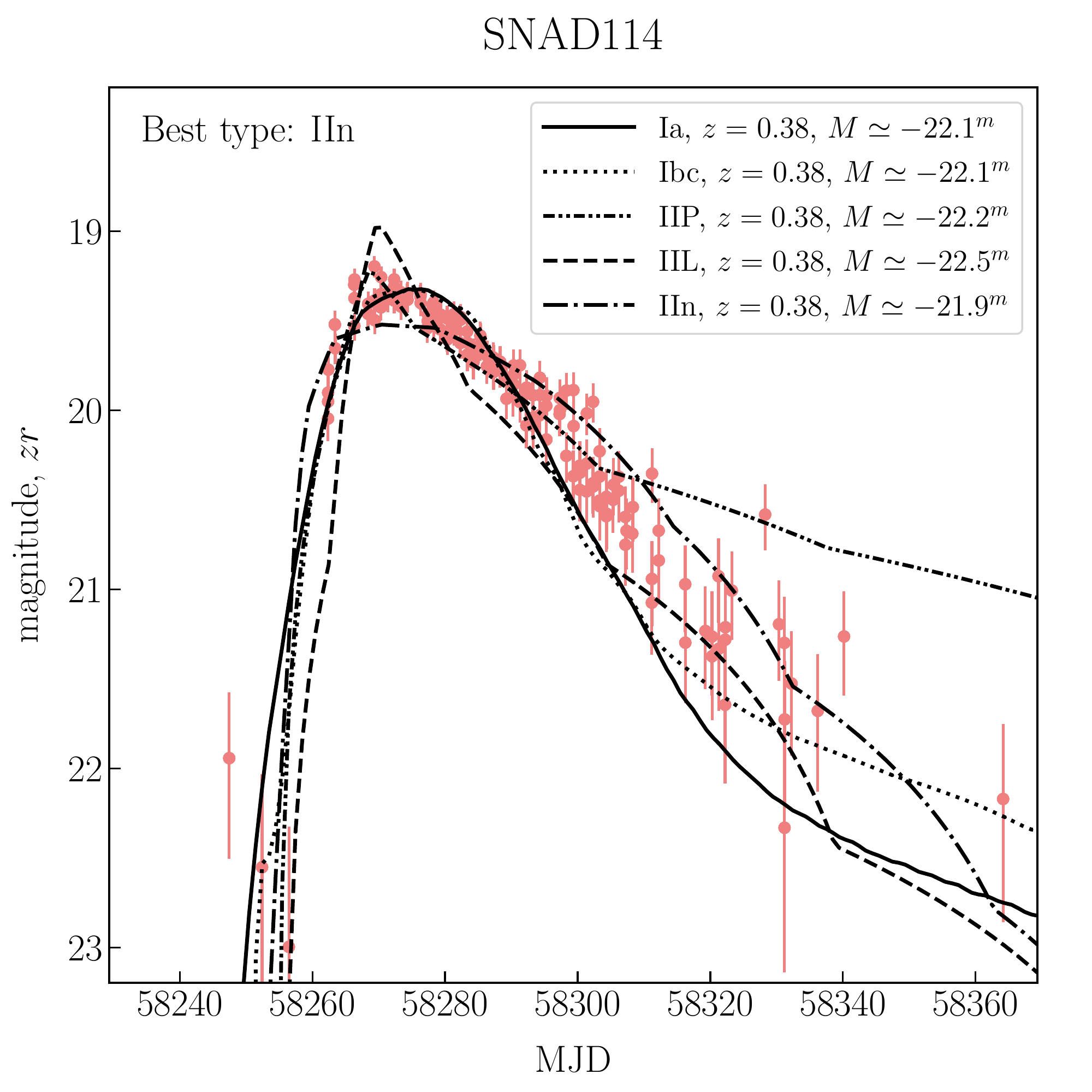}\\a) \\
    \end{minipage}
    \hfill
    \begin{minipage}{0.49\linewidth}
        \centering
        \includegraphics[scale=0.35]{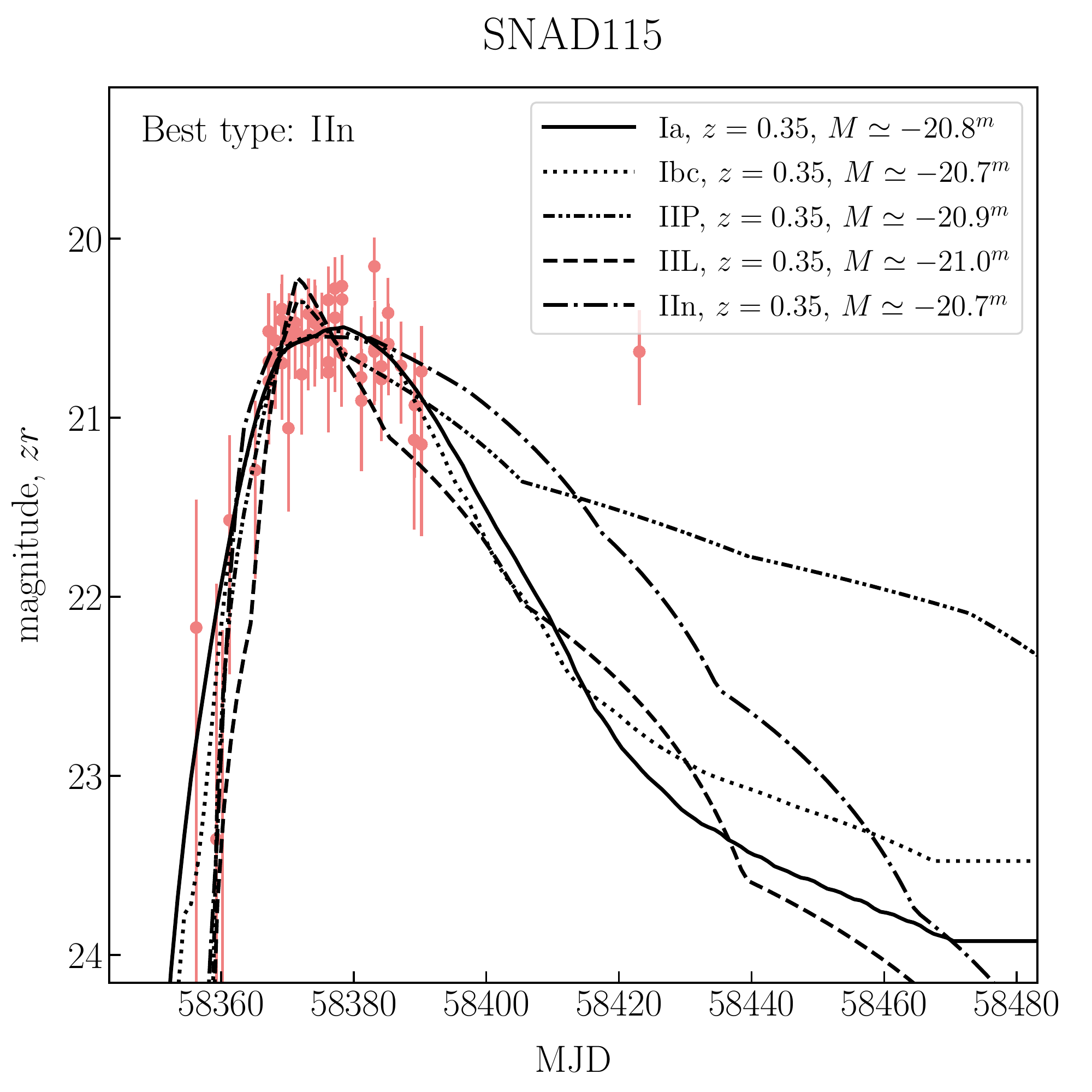}\\b) \\
    \end{minipage} 
    \vfill
    \begin{minipage}{0.49\linewidth}
        \centering
        \includegraphics[scale=0.35]{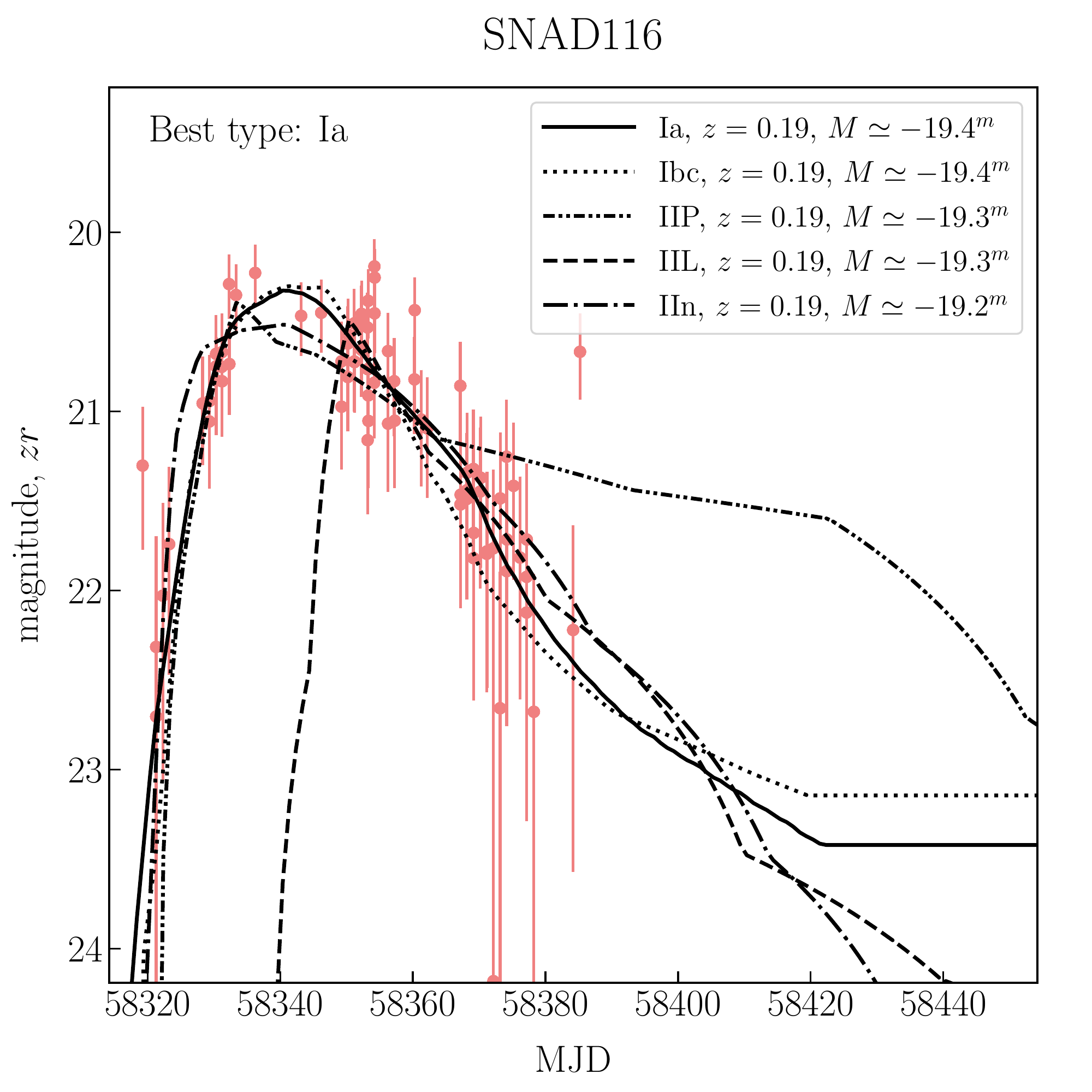}\\c) \\
    \end{minipage}
    \hfill
    \begin{minipage}{0.49\linewidth}
        \centering
        \includegraphics[scale=0.35]{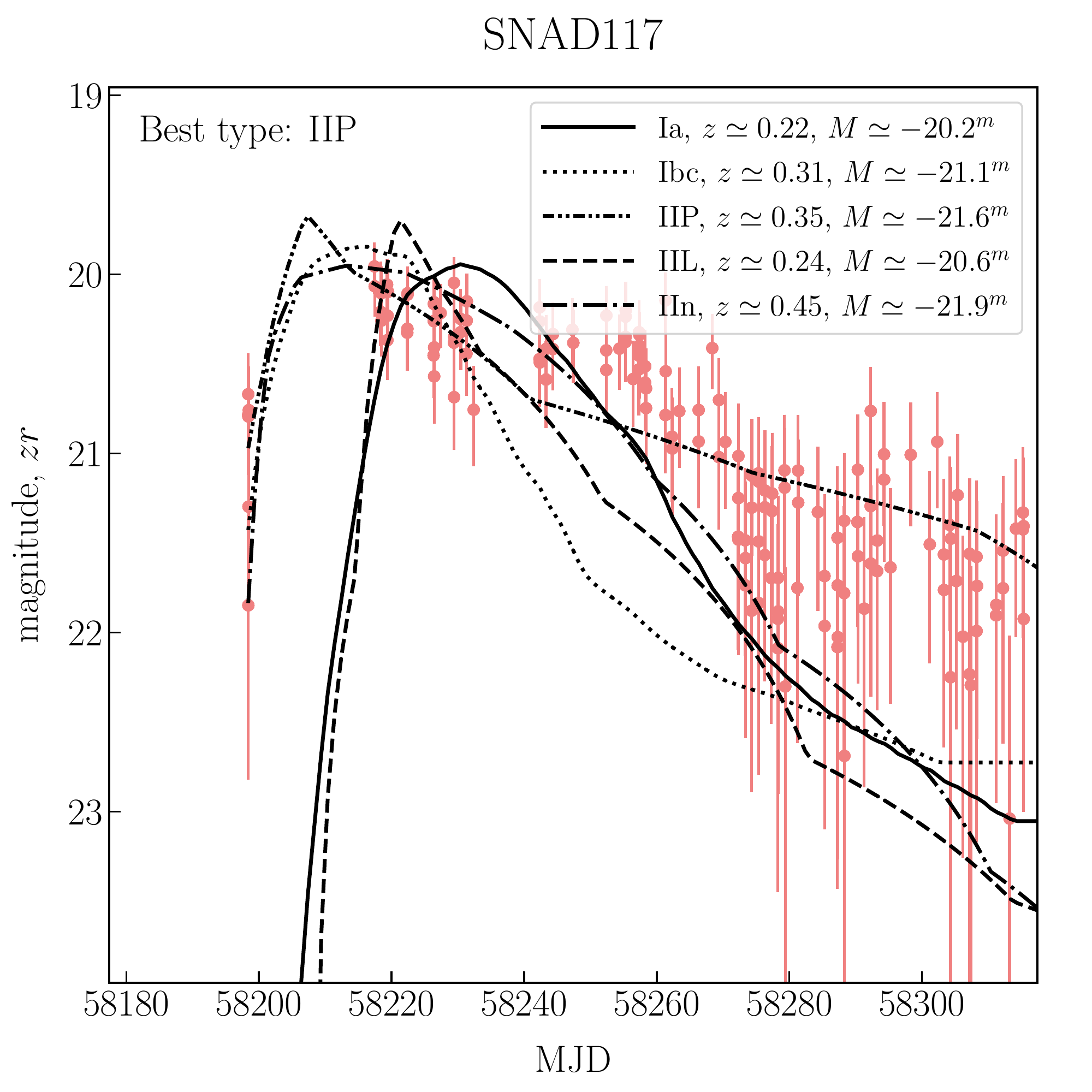}\\d) \\
    \end{minipage}  
    \vfill    
    \begin{minipage}{0.49\linewidth}
        \centering
        \includegraphics[scale=0.35]{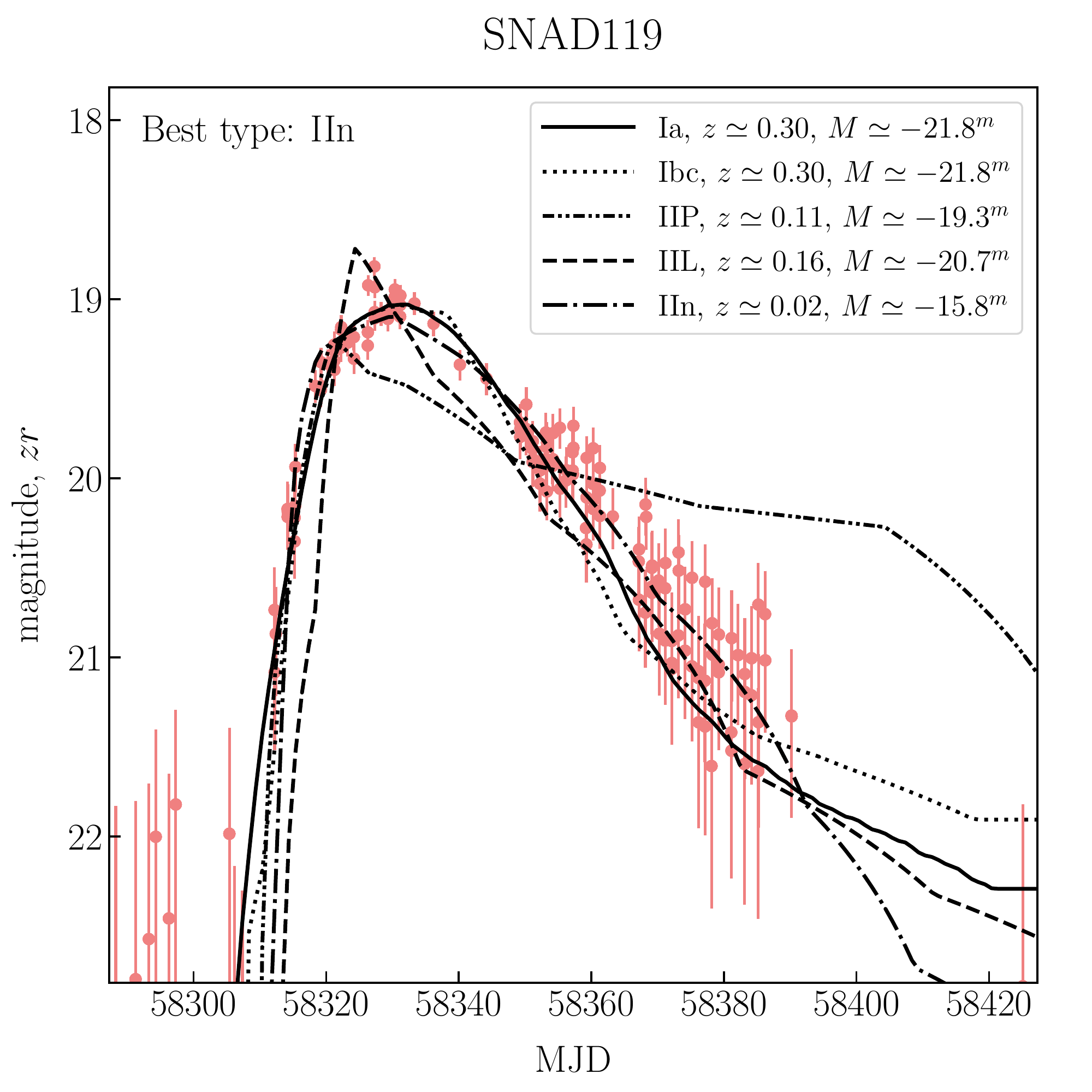}\\e) \\
    \end{minipage}
    \hfill
    \begin{minipage}{0.49\linewidth}
        \centering
        \includegraphics[scale=0.35]{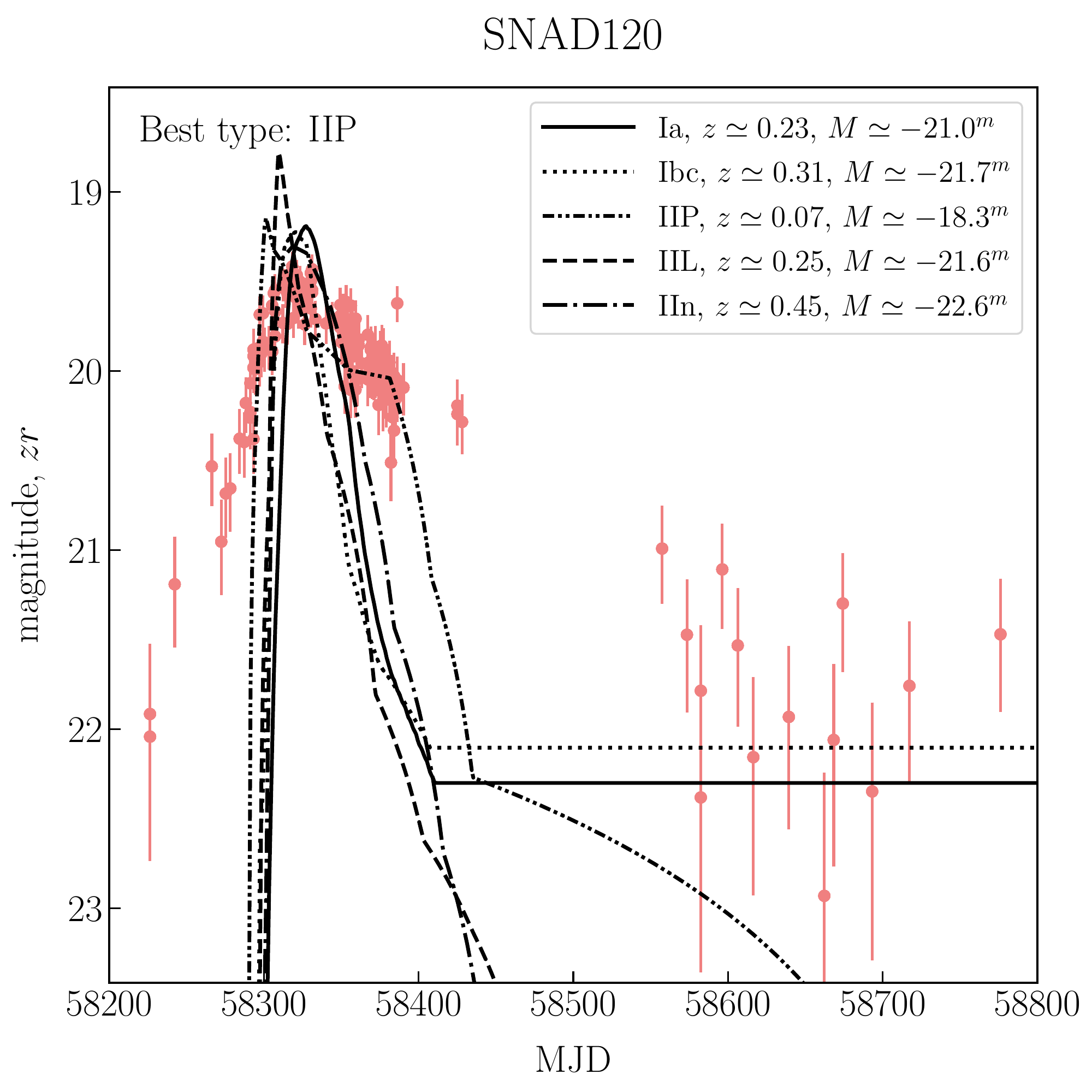}\\f) \\
    \end{minipage}     
    \caption{Light curves of SNAD supernova candidates in $zr$-band and the results of their fit by Nugent's supernova models.}
        \label{fig:snad_LC3}
\end{figure*}

\begin{figure*}
    \begin{minipage}{0.49\linewidth}
        \centering
        \includegraphics[scale=0.35]{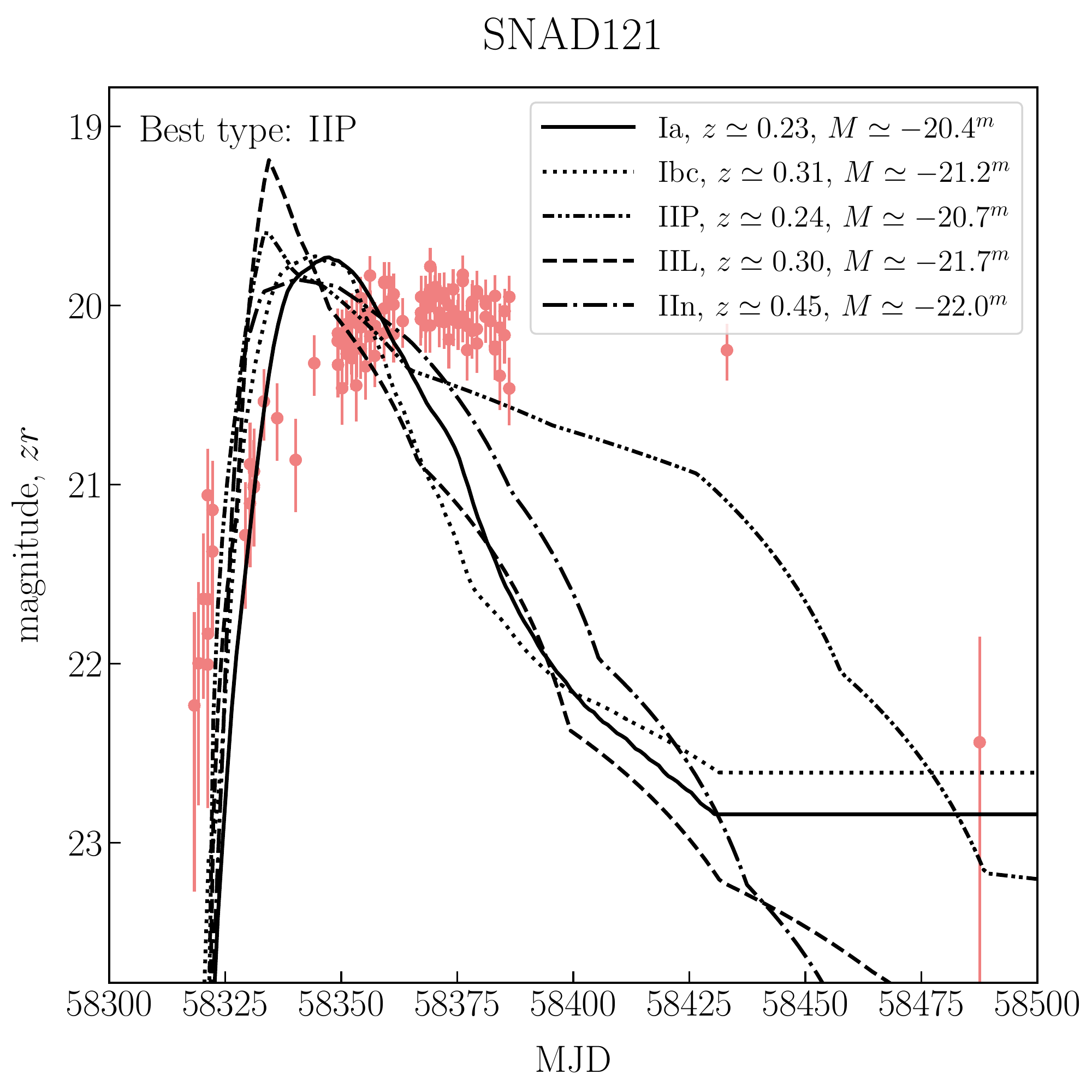}\\a) \\
    \end{minipage}
    \hfill
    \begin{minipage}{0.49\linewidth}
        \centering
        \includegraphics[scale=0.35]{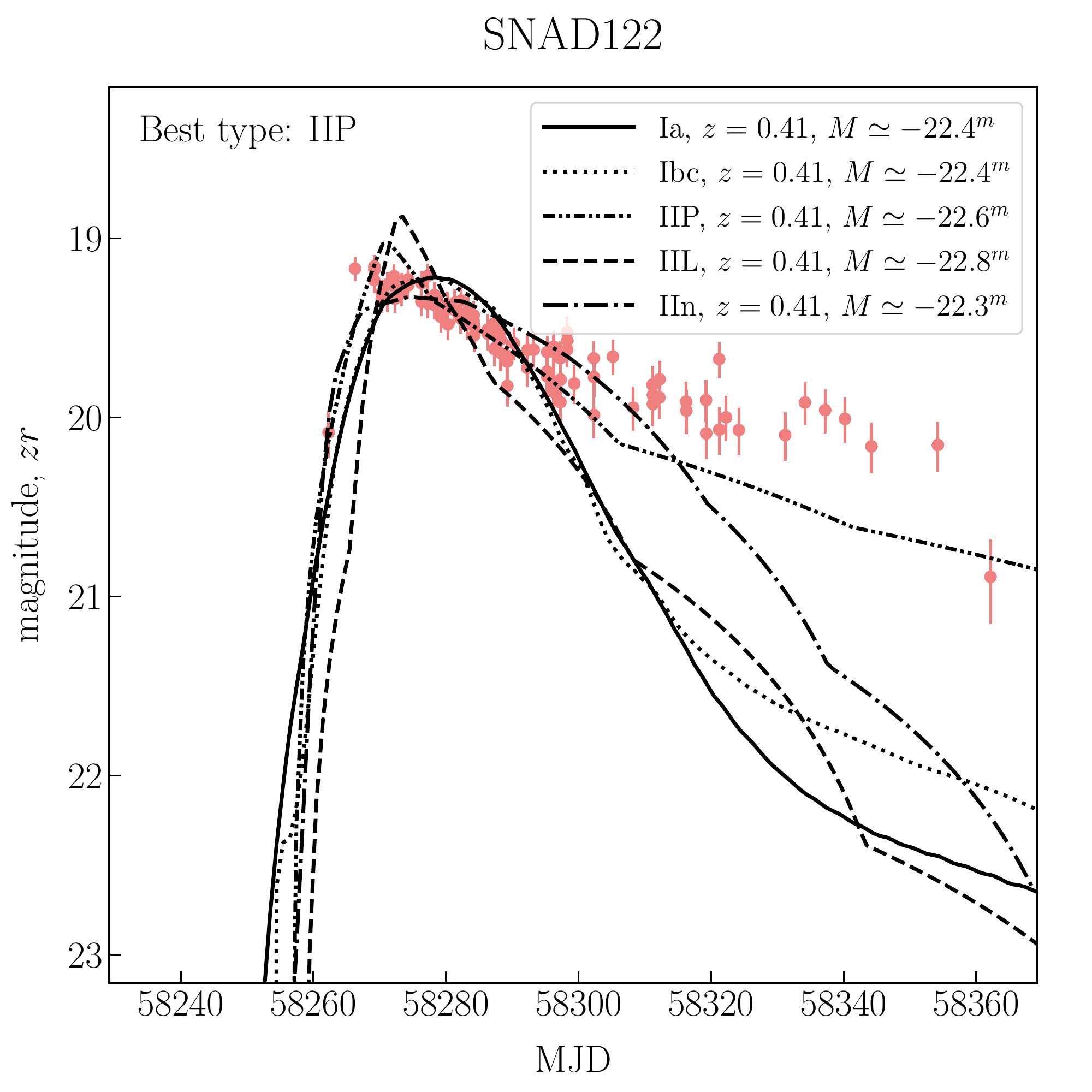}\\b) \\
    \end{minipage}
    \vfill
    \begin{minipage}{0.49\linewidth}
        \centering
        \includegraphics[scale=0.35]{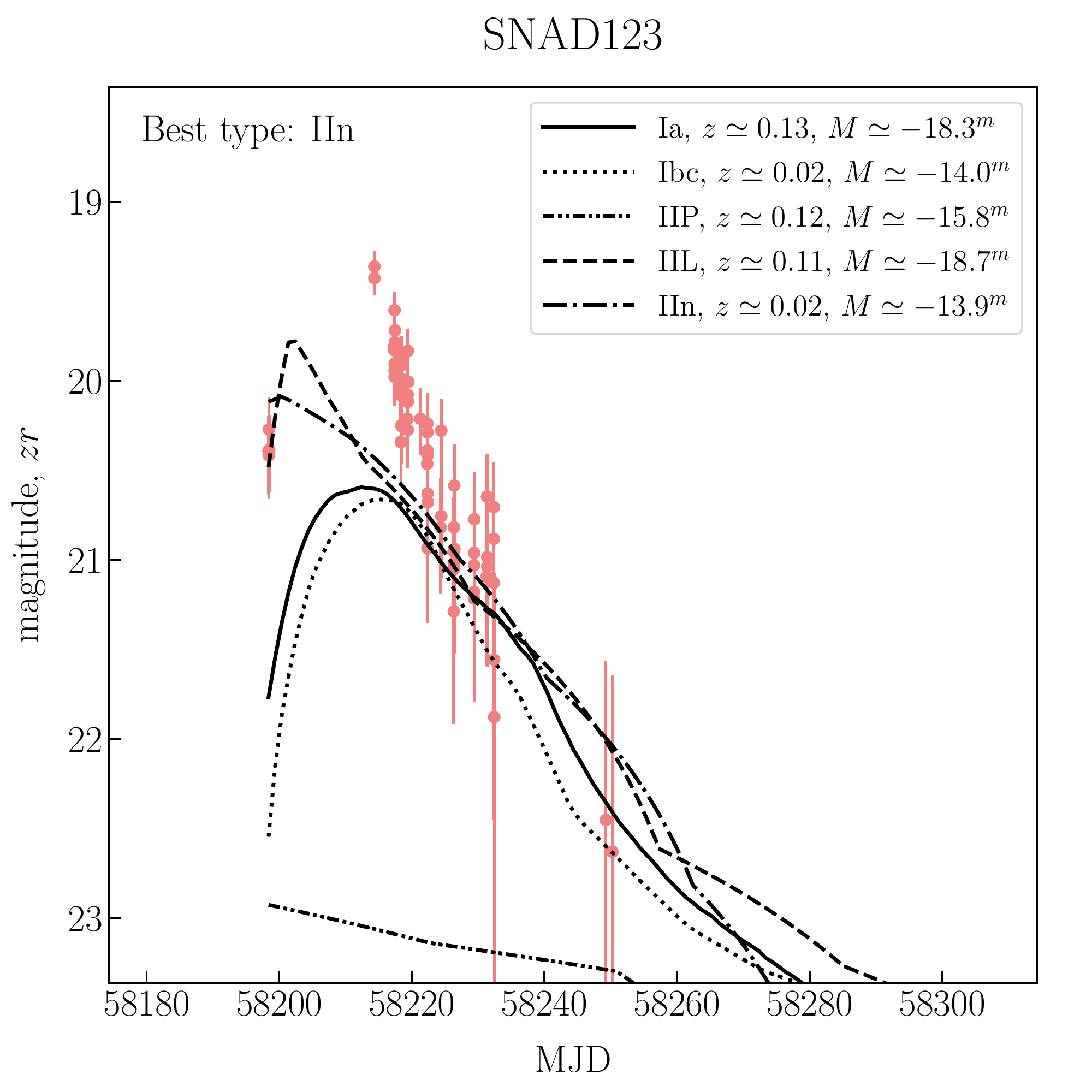}\\c) \\
    \end{minipage}
    \hfill
    \begin{minipage}{0.49\linewidth}
        \centering
        \includegraphics[scale=0.35]{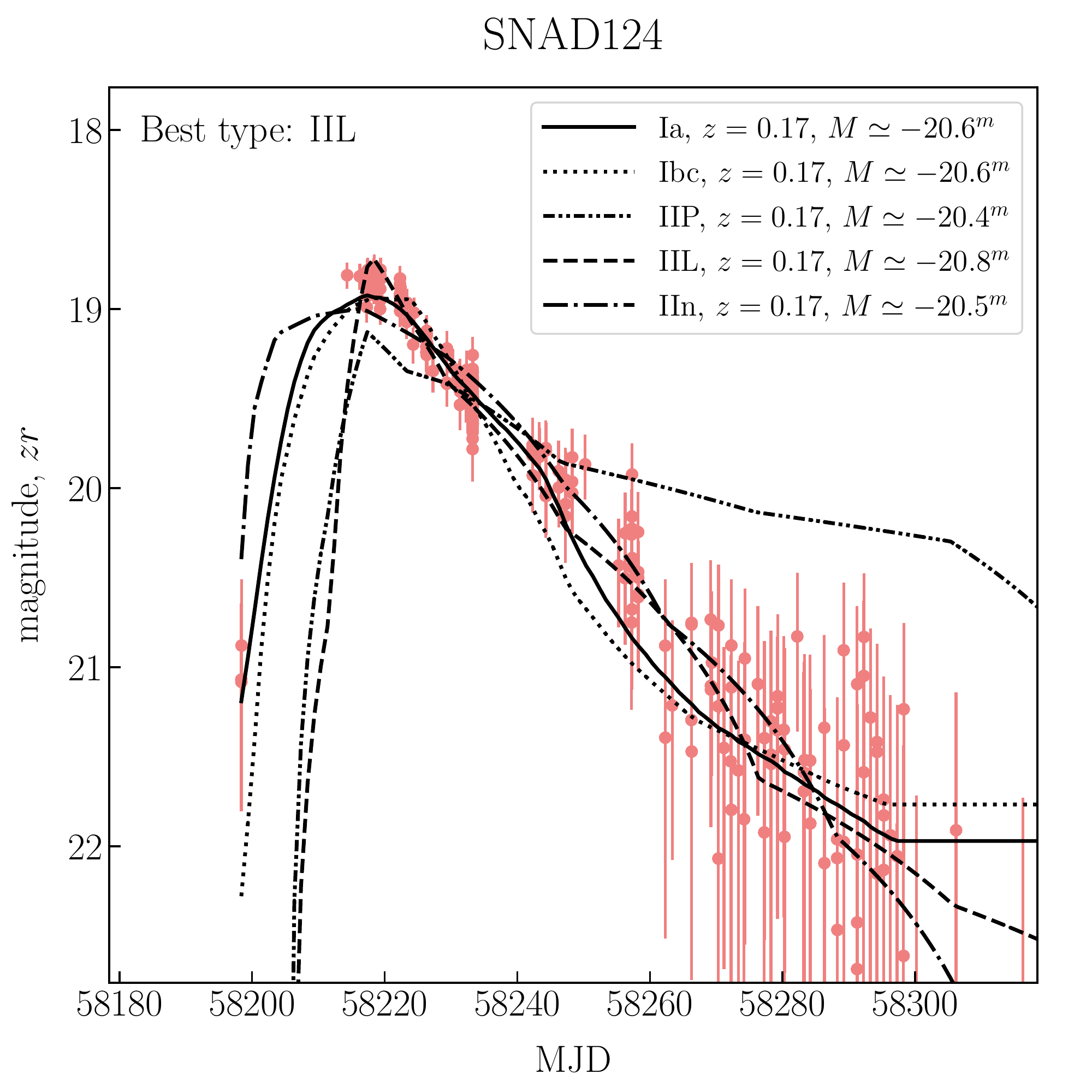}\\d) \\
    \end{minipage} 
    \vfill
    \begin{minipage}{0.49\linewidth}
        \centering
        \includegraphics[scale=0.35]{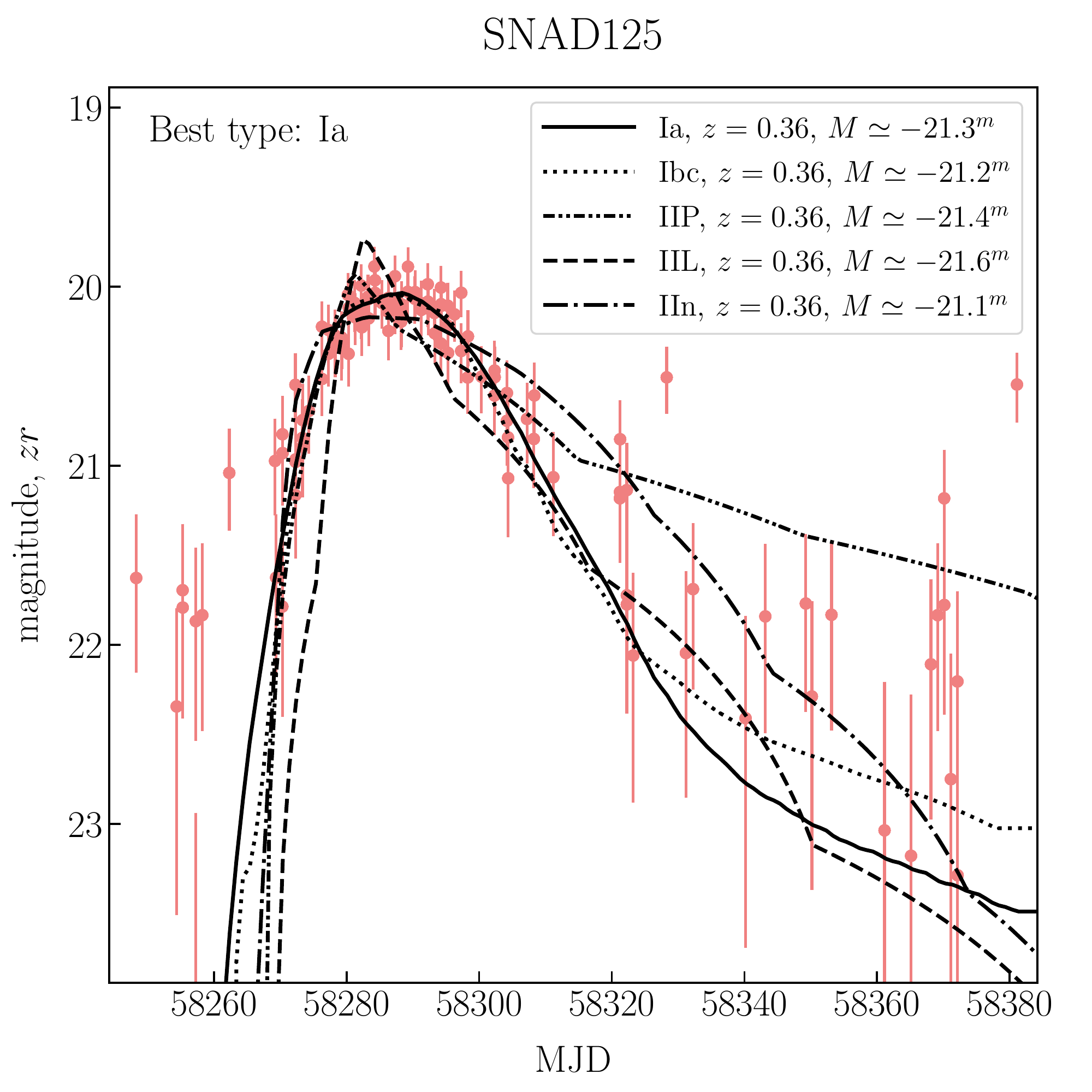}\\e) \\
    \end{minipage}
    \hfill
    \begin{minipage}{0.49\linewidth}
        \centering
        \includegraphics[scale=0.35]{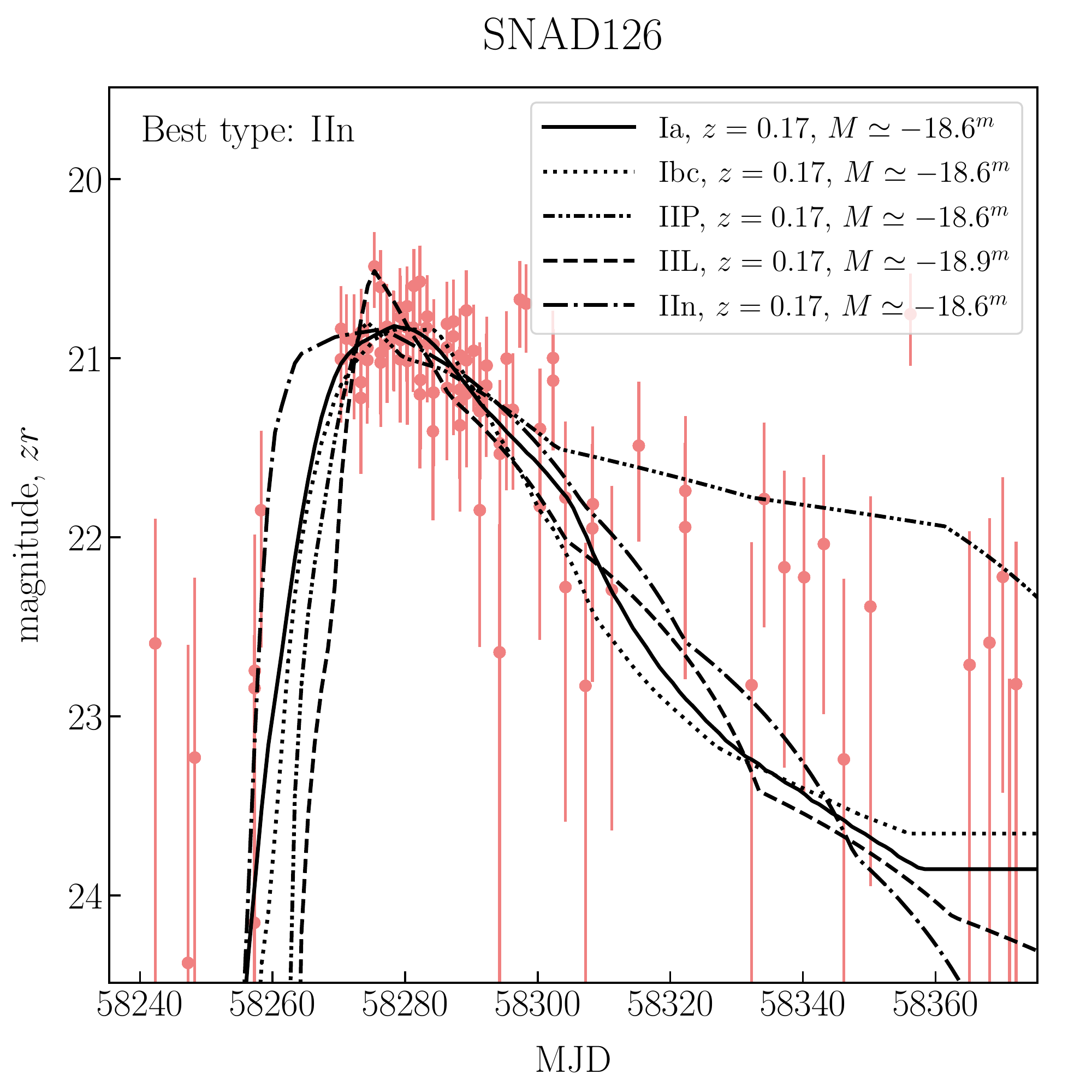}\\f) \\
    \end{minipage}  
    \caption{Light curves of SNAD supernova candidates in $zr$-band and the results of their fit by Nugent's supernova models.}
        \label{fig:snad_LC4}
\end{figure*}
   
\begin{figure*}
    \begin{minipage}{0.49\linewidth}
        \centering
        \includegraphics[scale=0.35]{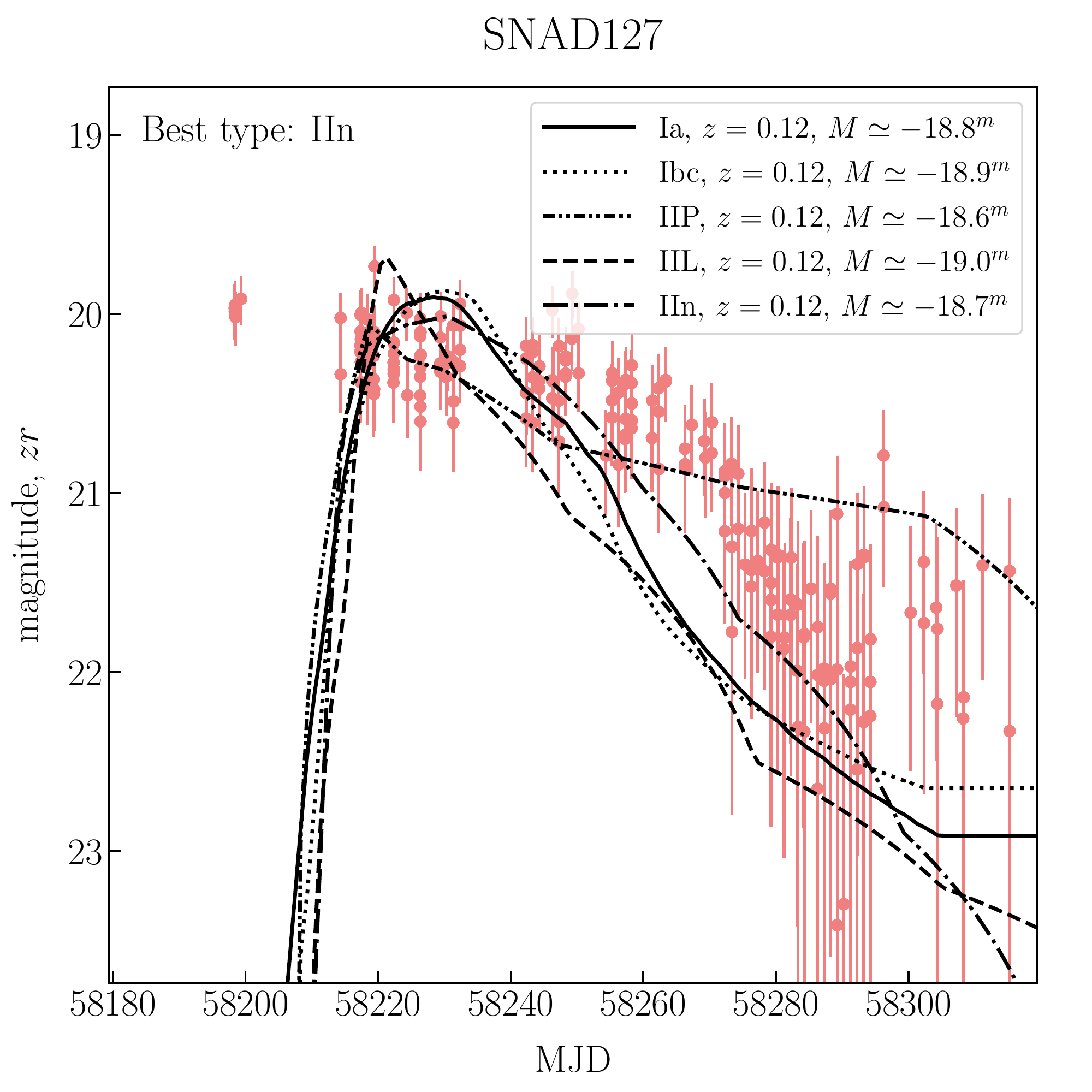}\\a) \\
    \end{minipage}
    \hfill
    \begin{minipage}{0.49\linewidth}
        \centering
        \includegraphics[scale=0.35]{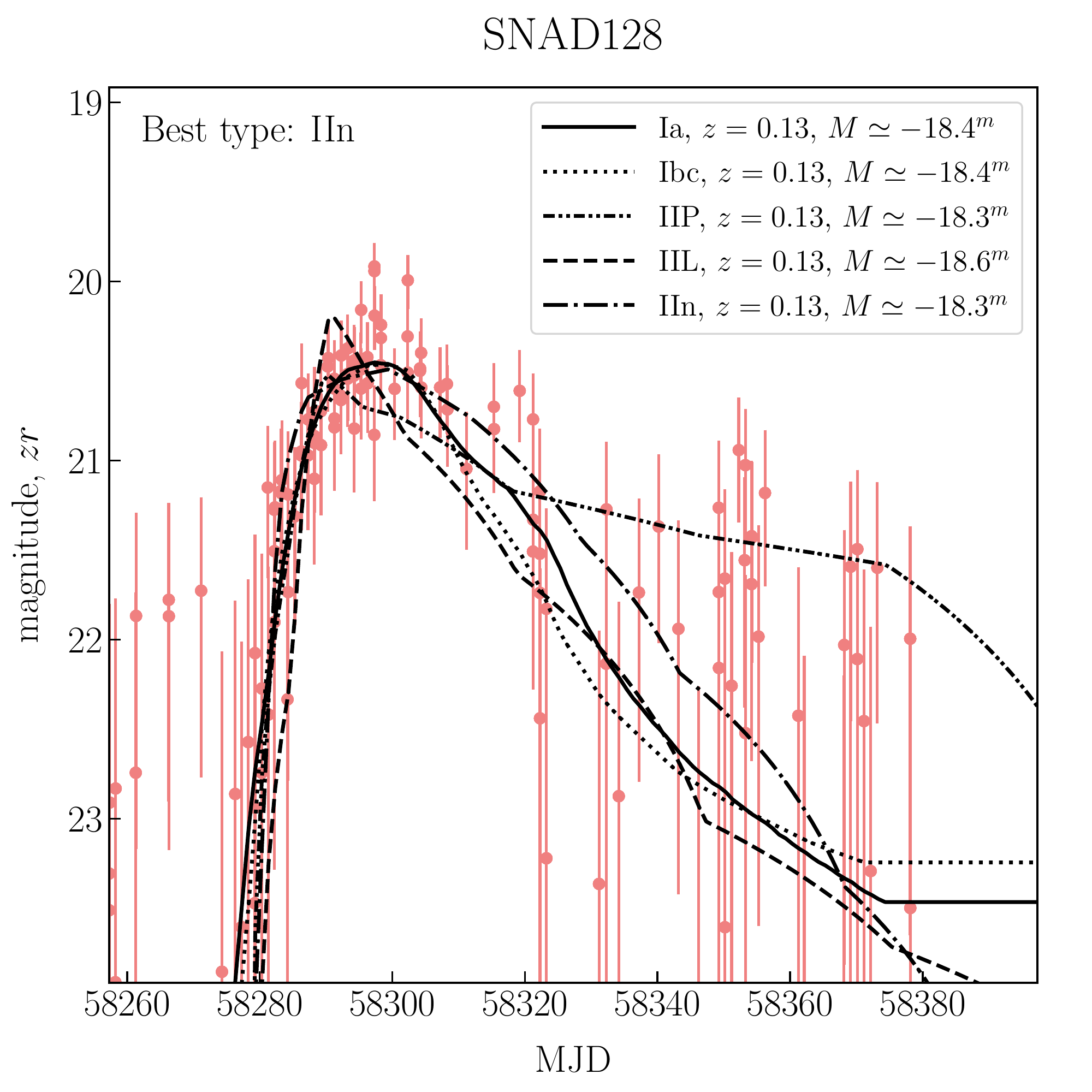}\\b) \\
    \end{minipage}  
    \vfill
    \begin{minipage}{0.49\linewidth}
        \centering
        \includegraphics[scale=0.35]{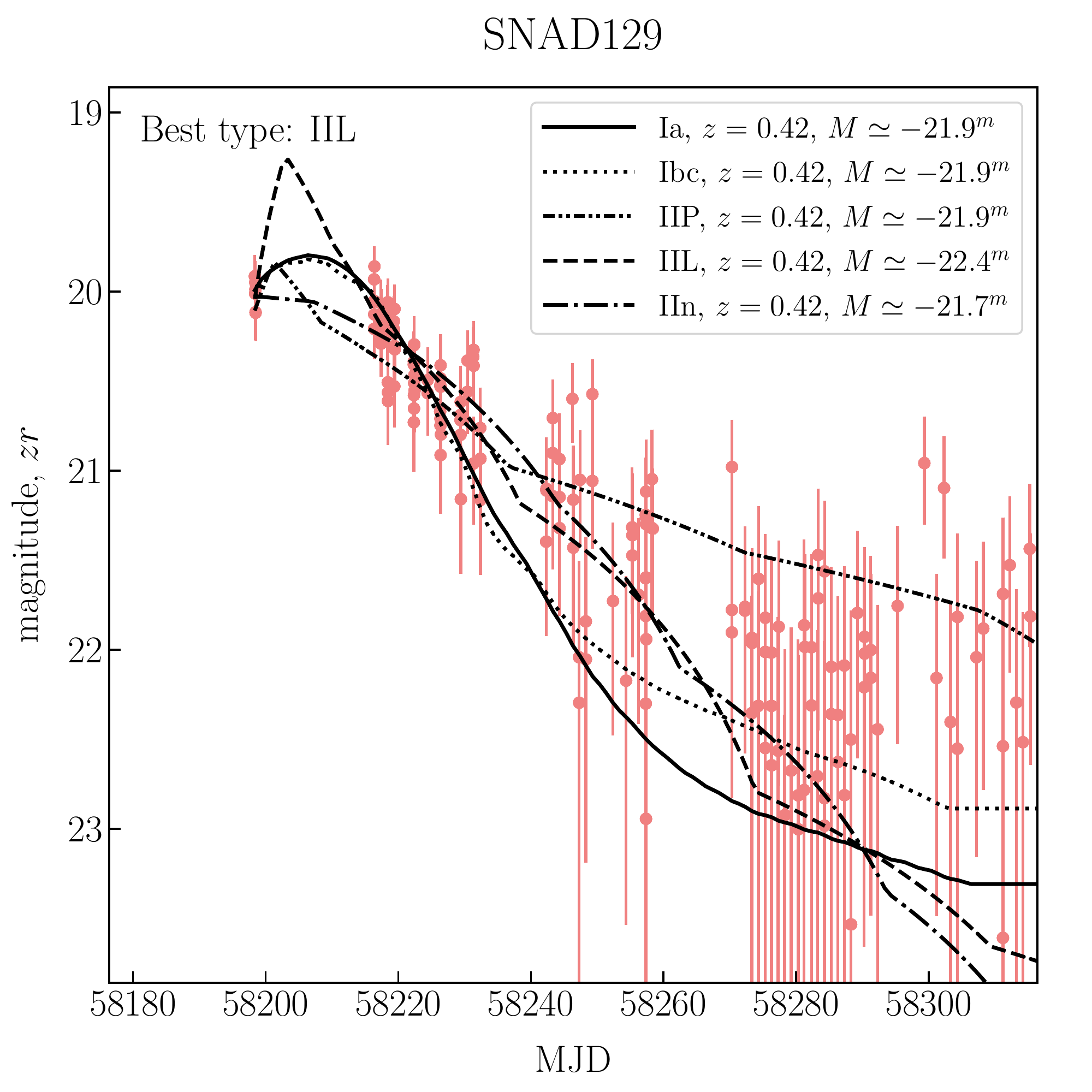}\\c) \\
    \end{minipage}
    \hfill
    \begin{minipage}{0.49\linewidth}
        \centering
        \includegraphics[scale=0.35]{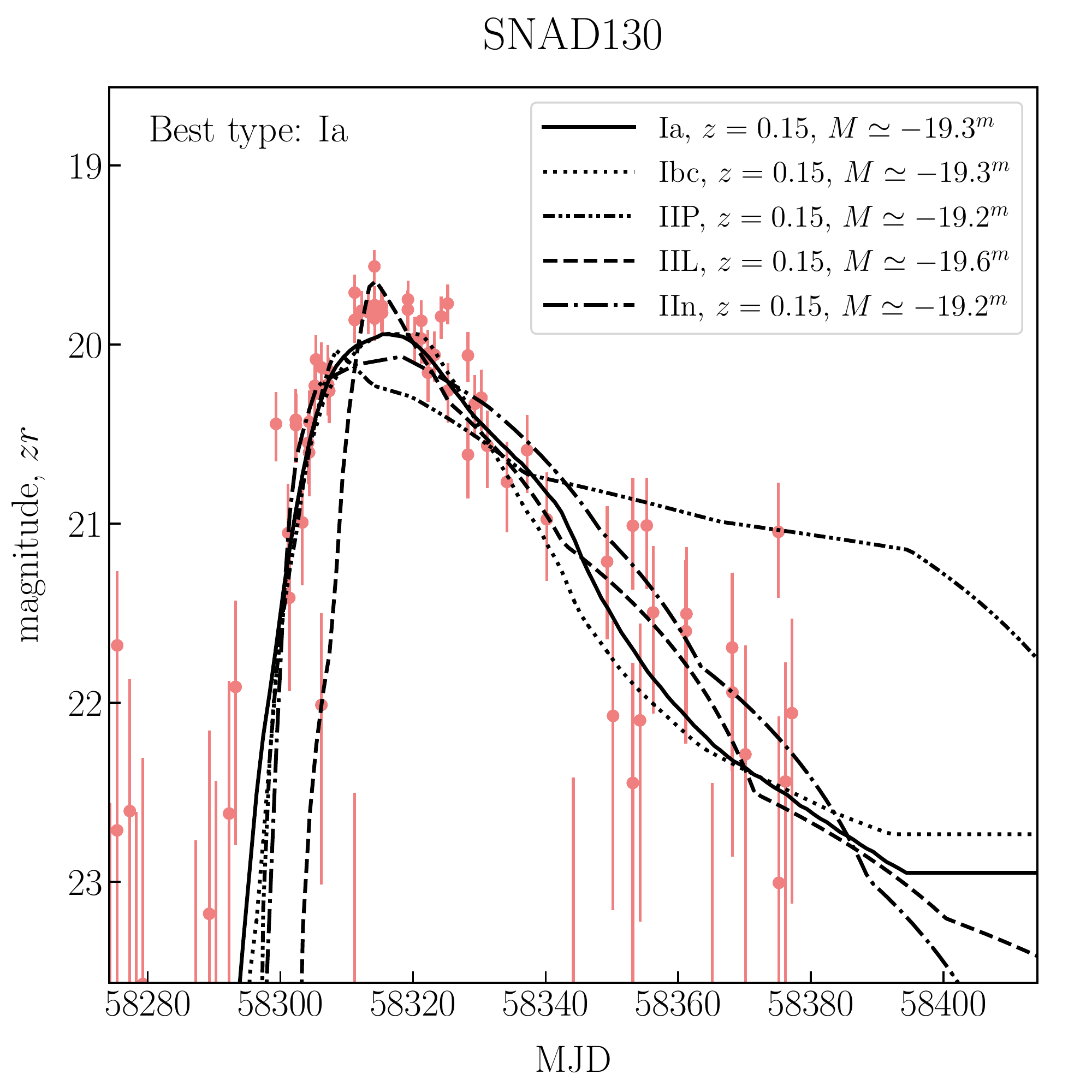}\\d) \\
    \end{minipage}
    \vfill
    \begin{minipage}{0.49\linewidth}
        \centering
        \includegraphics[scale=0.35]{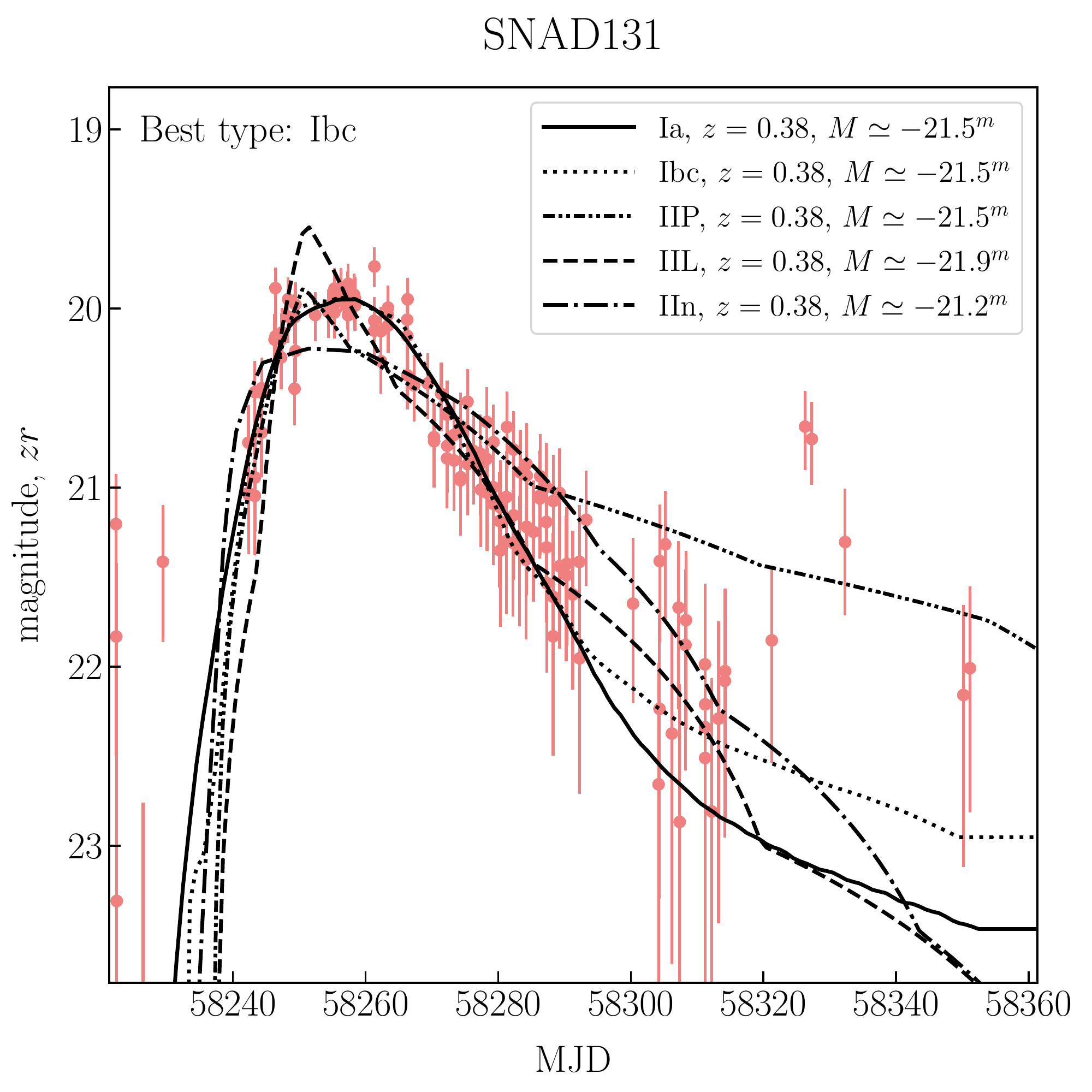}\\e) \\
    \end{minipage}
    \hfill
    \begin{minipage}{0.49\linewidth}
        \centering
        \includegraphics[scale=0.35]{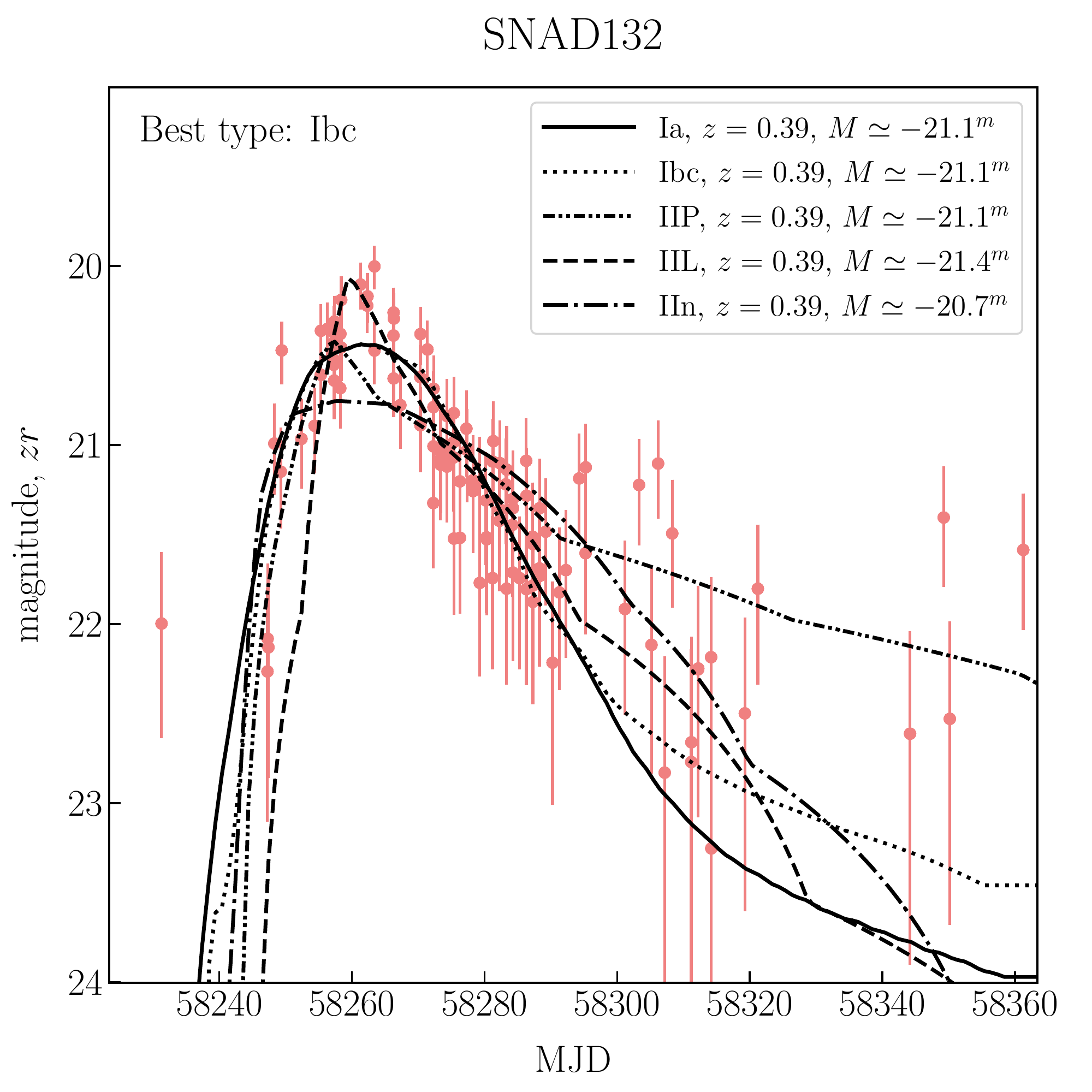}\\f) \\
    \end{minipage} 
    \caption{Light curves of SNAD supernova candidates in $zr$-band and the results of their fit by Nugent's supernova models.}
        \label{fig:snad_LC5}
\end{figure*}

\begin{figure*}
    \begin{minipage}{0.49\linewidth}
        \centering
        \includegraphics[scale=0.35]{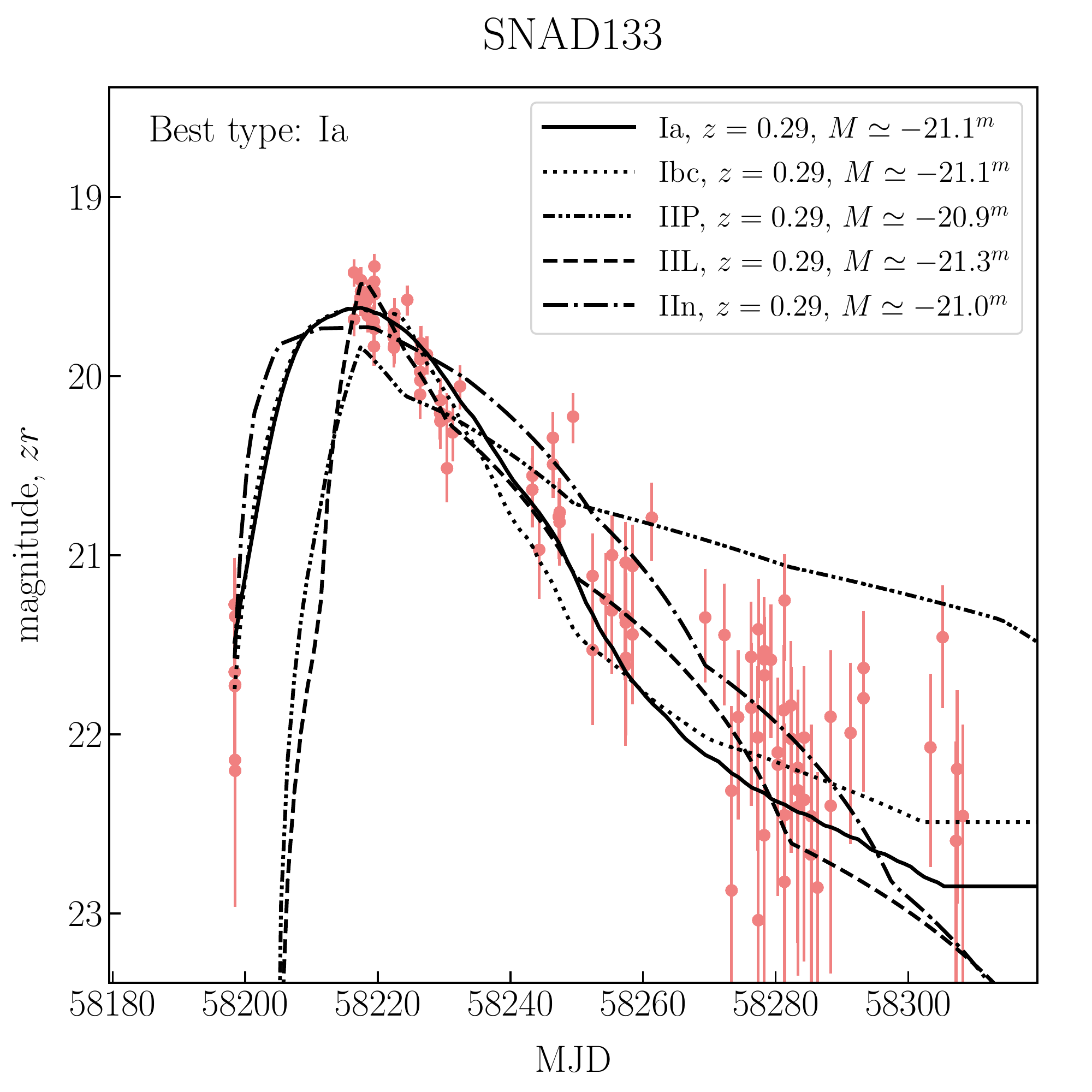}\\a) \\
    \end{minipage}
    \hfill
    \begin{minipage}{0.49\linewidth}
        \centering
        \includegraphics[scale=0.35]{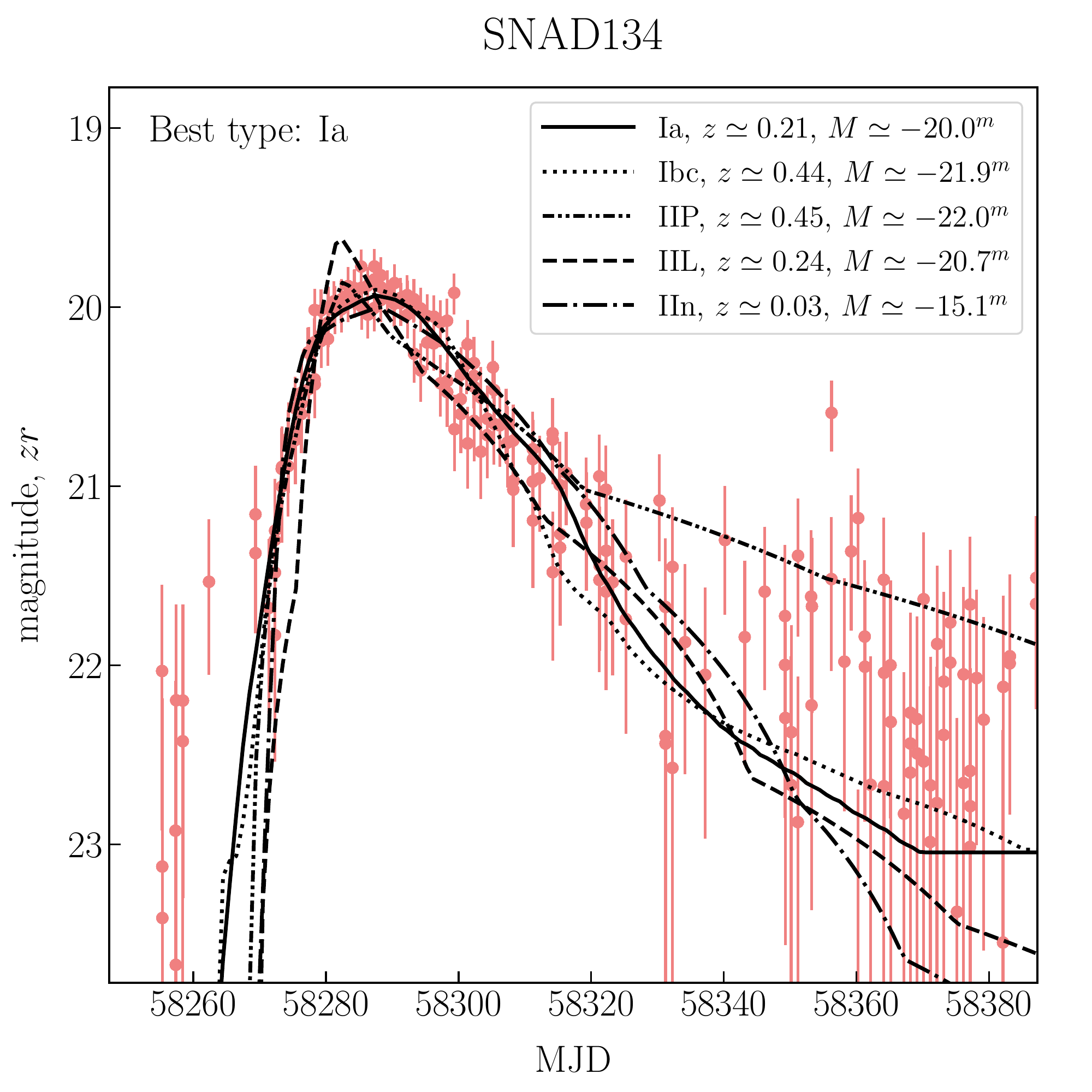}\\b) \\
    \end{minipage}  
    \vfill
    \begin{minipage}{0.49\linewidth}
        \centering
        \includegraphics[scale=0.35]{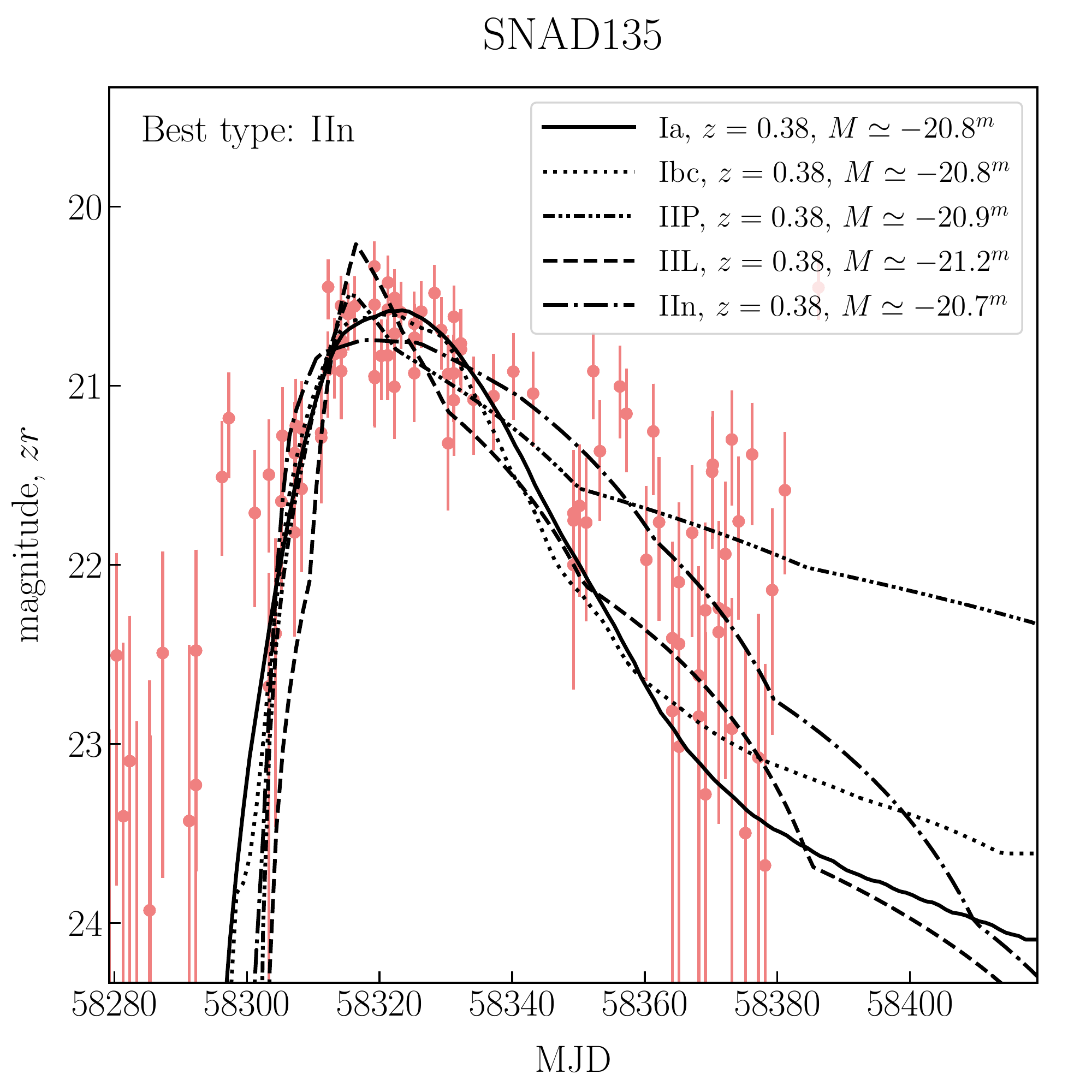}\\c) \\
    \end{minipage}
    \hfill
    \begin{minipage}{0.49\linewidth}
        \centering
        \includegraphics[scale=0.35]{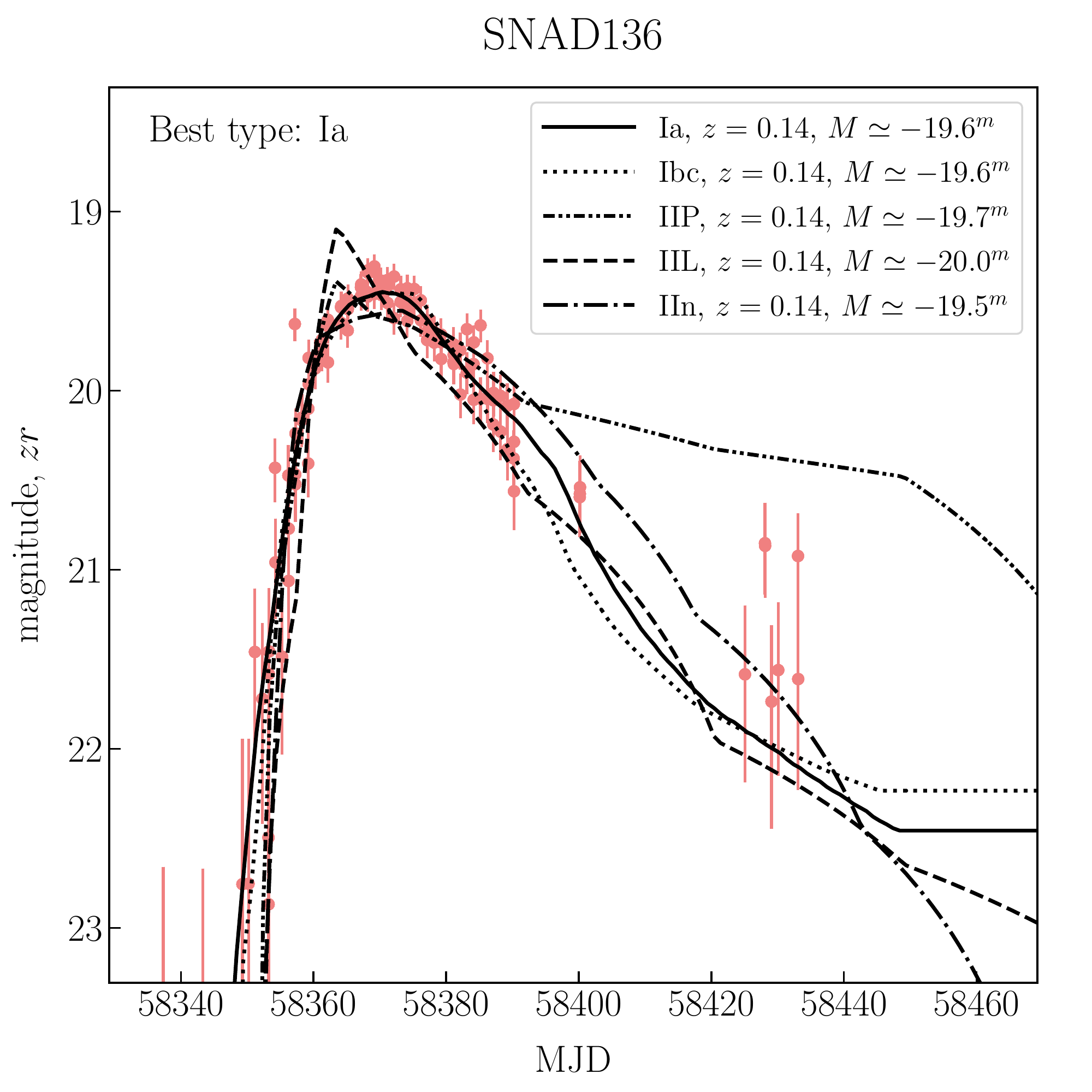}\\d) \\
    \end{minipage}  
    \vfill
    \begin{minipage}{0.49\linewidth}
        \centering
        \includegraphics[scale=0.35]{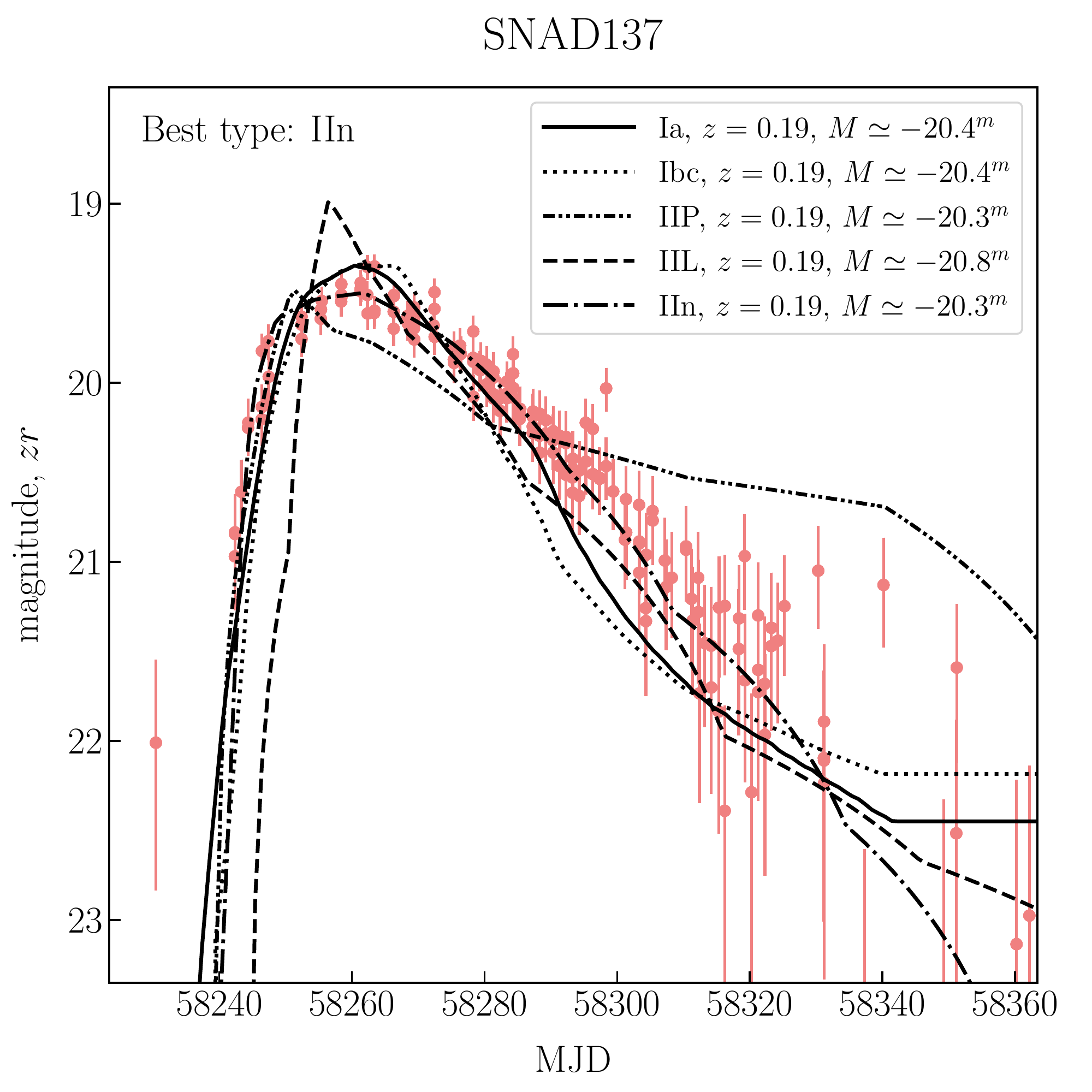}\\e) \\
    \end{minipage}
    \hfill
    \begin{minipage}{0.49\linewidth}
        \centering
        \includegraphics[scale=0.35]{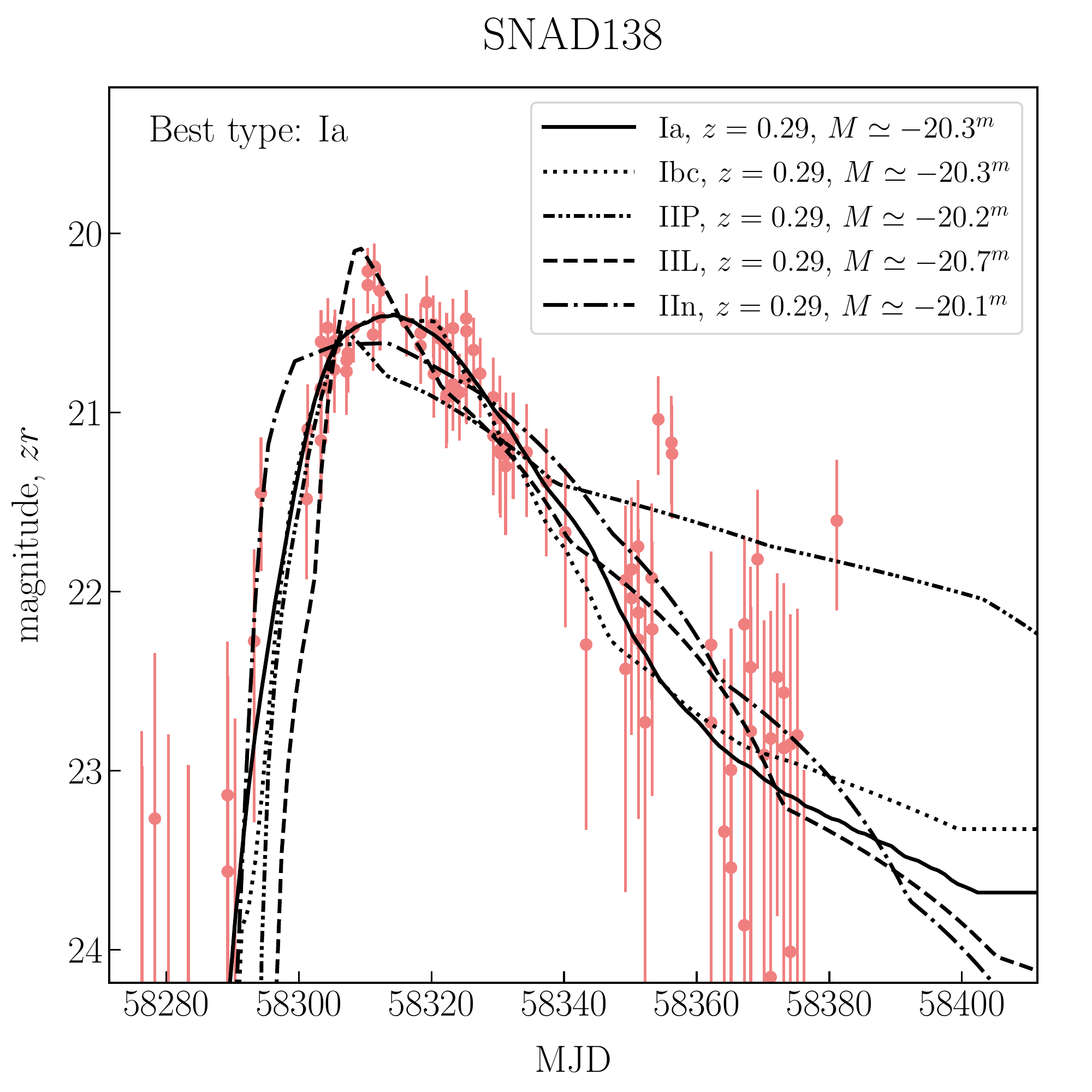}\\f) \\
    \end{minipage}
    \caption{Light curves of SNAD supernova candidates in $zr$-band and the results of their fit by Nugent's supernova models.}
        \label{fig:snad_LC6}
\end{figure*}

\begin{figure*}
    \vfill
    \begin{minipage}{0.49\linewidth}
        \centering
        \includegraphics[scale=0.35]{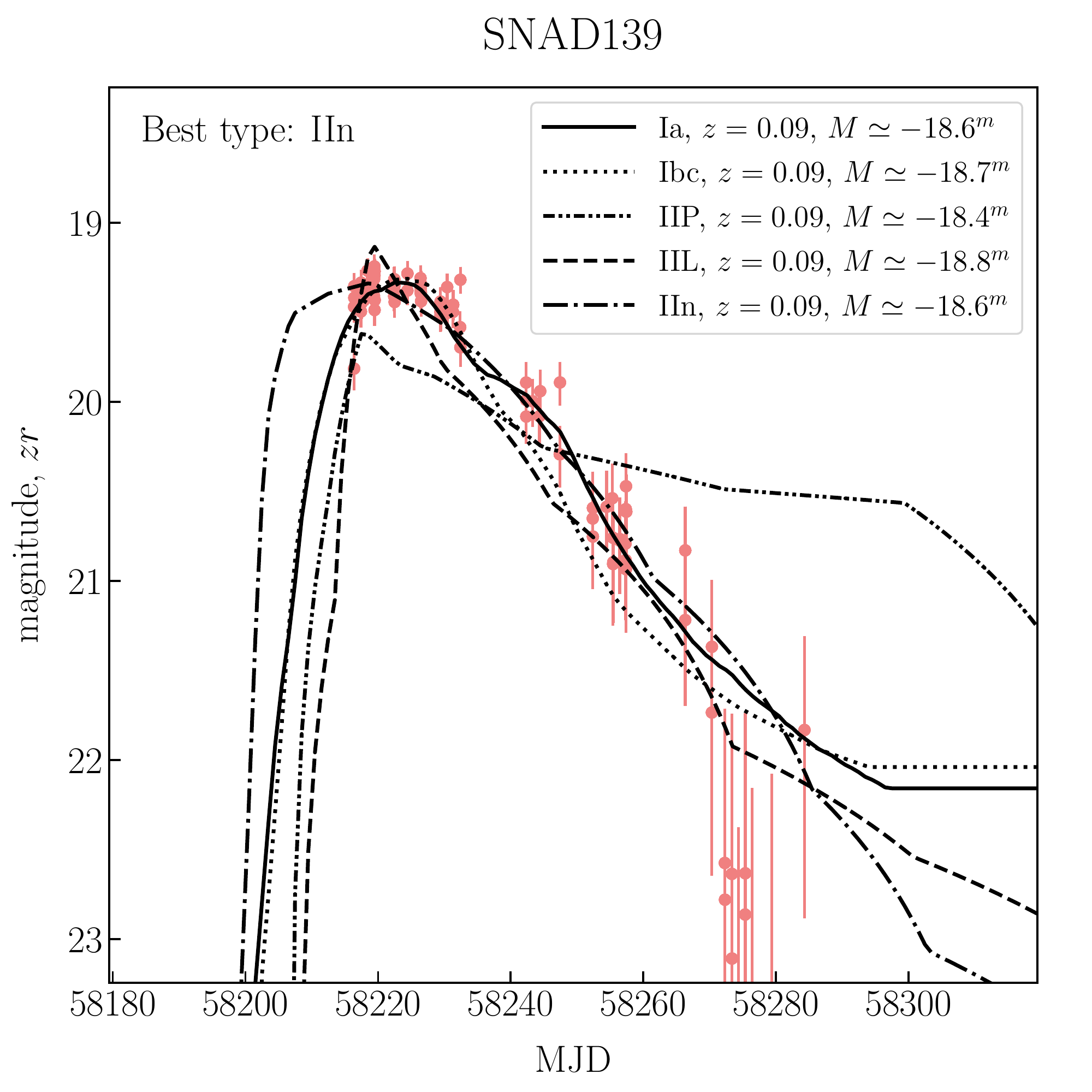}\\a) \\
    \end{minipage}
    \hfill
    \begin{minipage}{0.49\linewidth}
        \centering
        \includegraphics[scale=0.35]{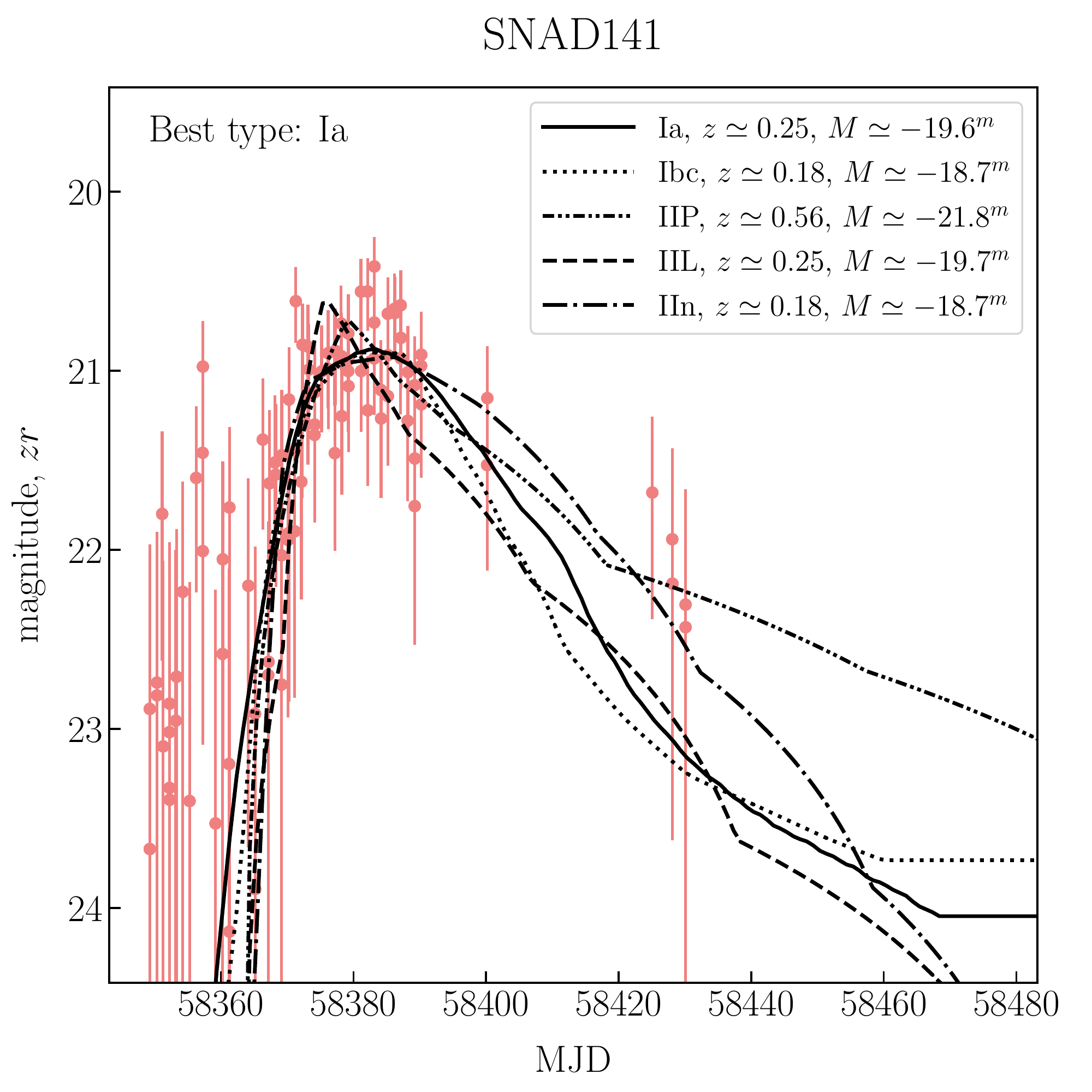}\\b) \\
    \end{minipage} 
    \vfill
    \begin{minipage}{0.49\linewidth}
        \centering
        \includegraphics[scale=0.35]{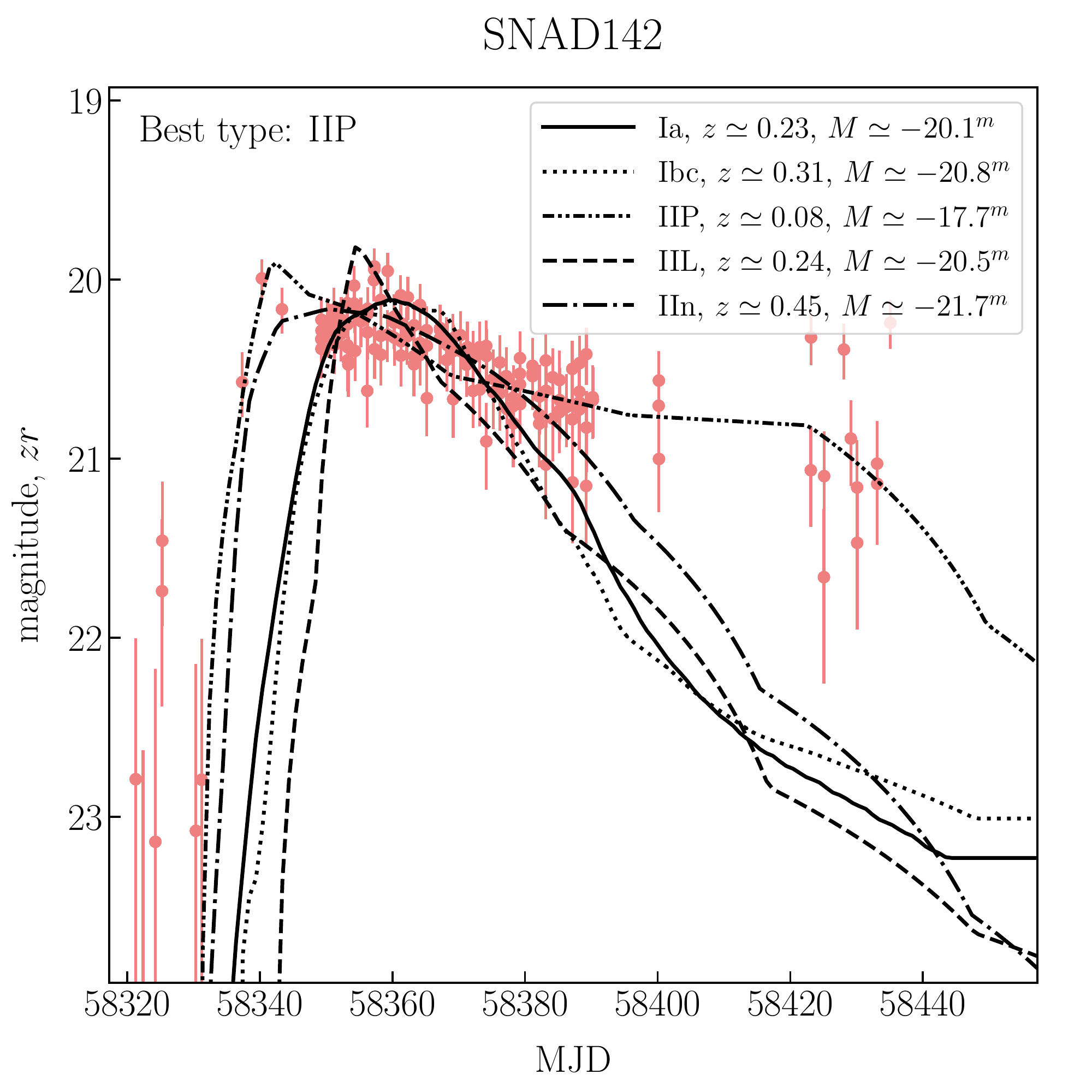}\\c) \\
    \end{minipage}
    \hfill
    \begin{minipage}{0.49\linewidth}
        \centering
        \includegraphics[scale=0.35]{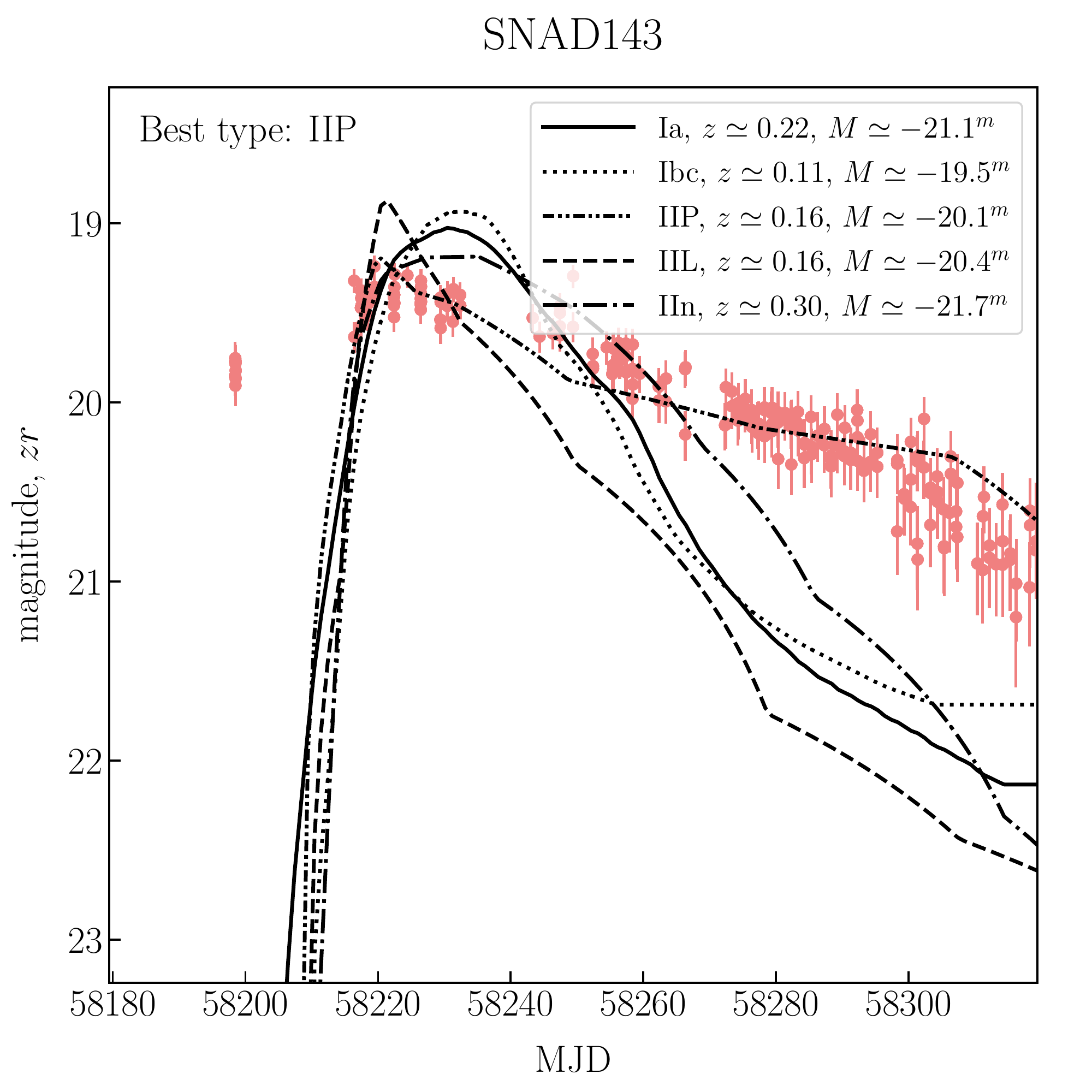}\\d) \\
    \end{minipage}  
    \vfill
    \begin{minipage}{0.49\linewidth}
        \centering
        \includegraphics[scale=0.35]{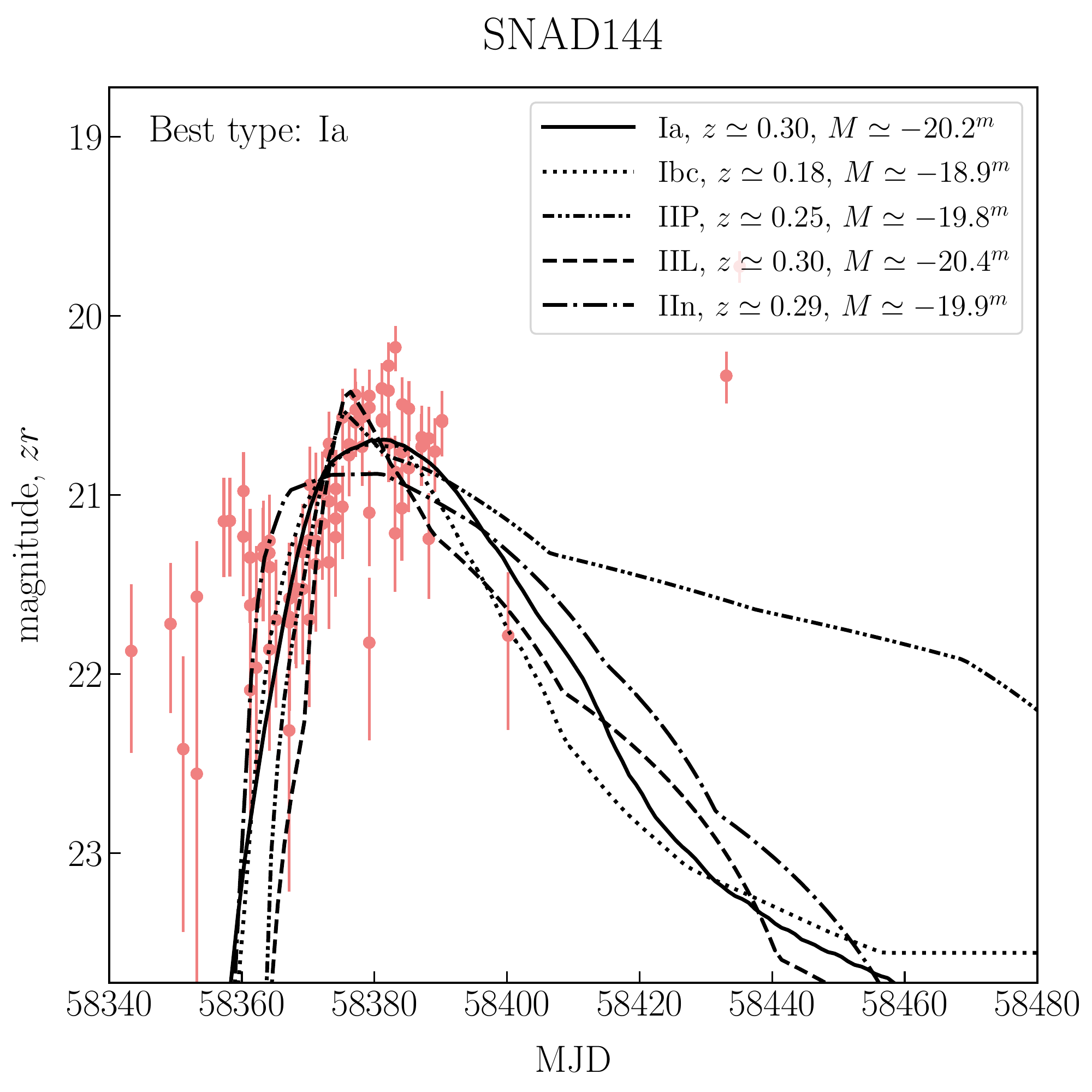}\\e) \\
    \end{minipage}
    \hfill
    \begin{minipage}{0.49\linewidth}
        \centering
        \includegraphics[scale=0.35]{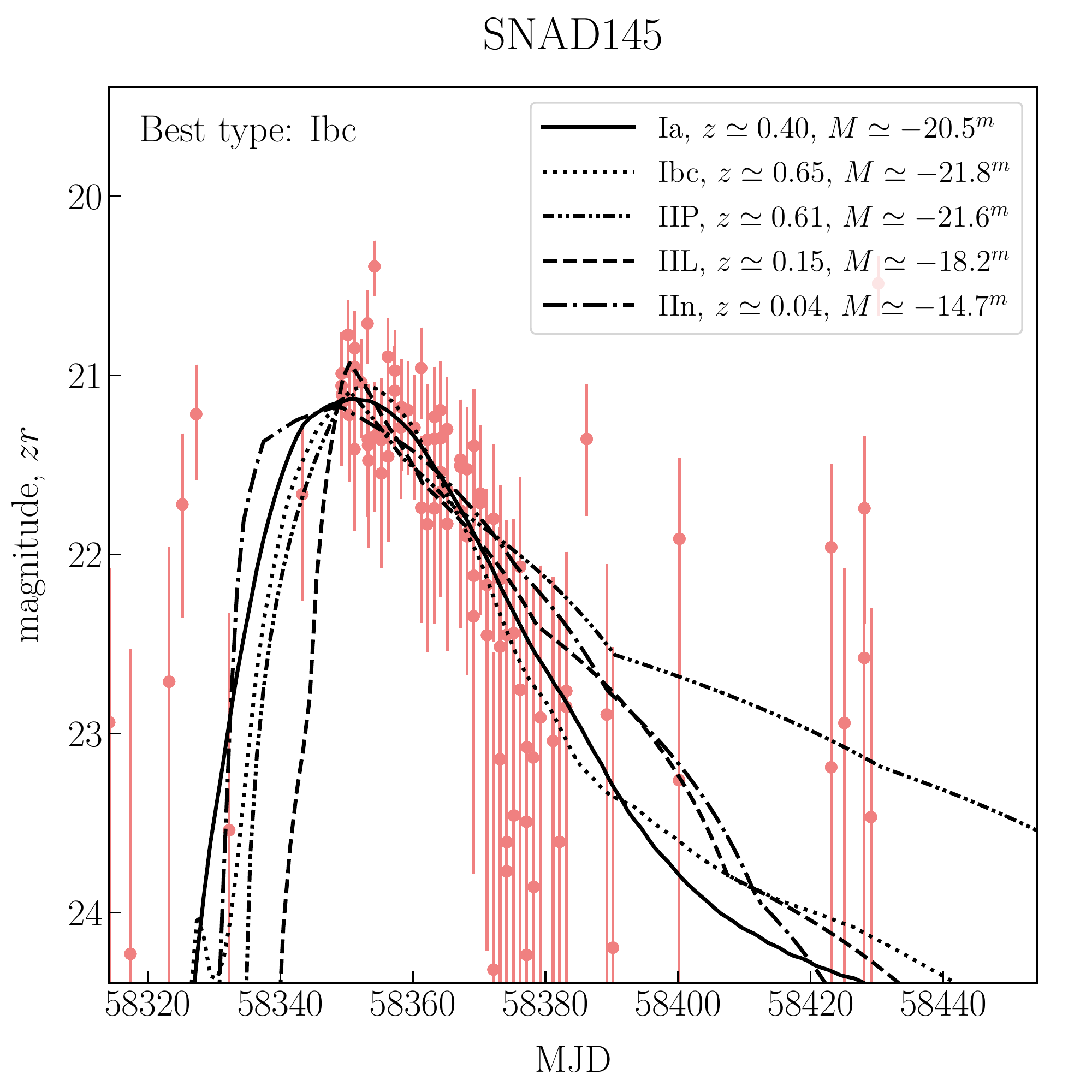}\\f) \\
    \end{minipage}  
    \caption{Light curves of SNAD supernova candidates in $zr$-band and the results of their fit by Nugent's supernova models.}
        \label{fig:snad_LC7}
\end{figure*}

\begin{figure*}
    \begin{minipage}{0.49\linewidth}
        \centering
        \includegraphics[scale=0.35]{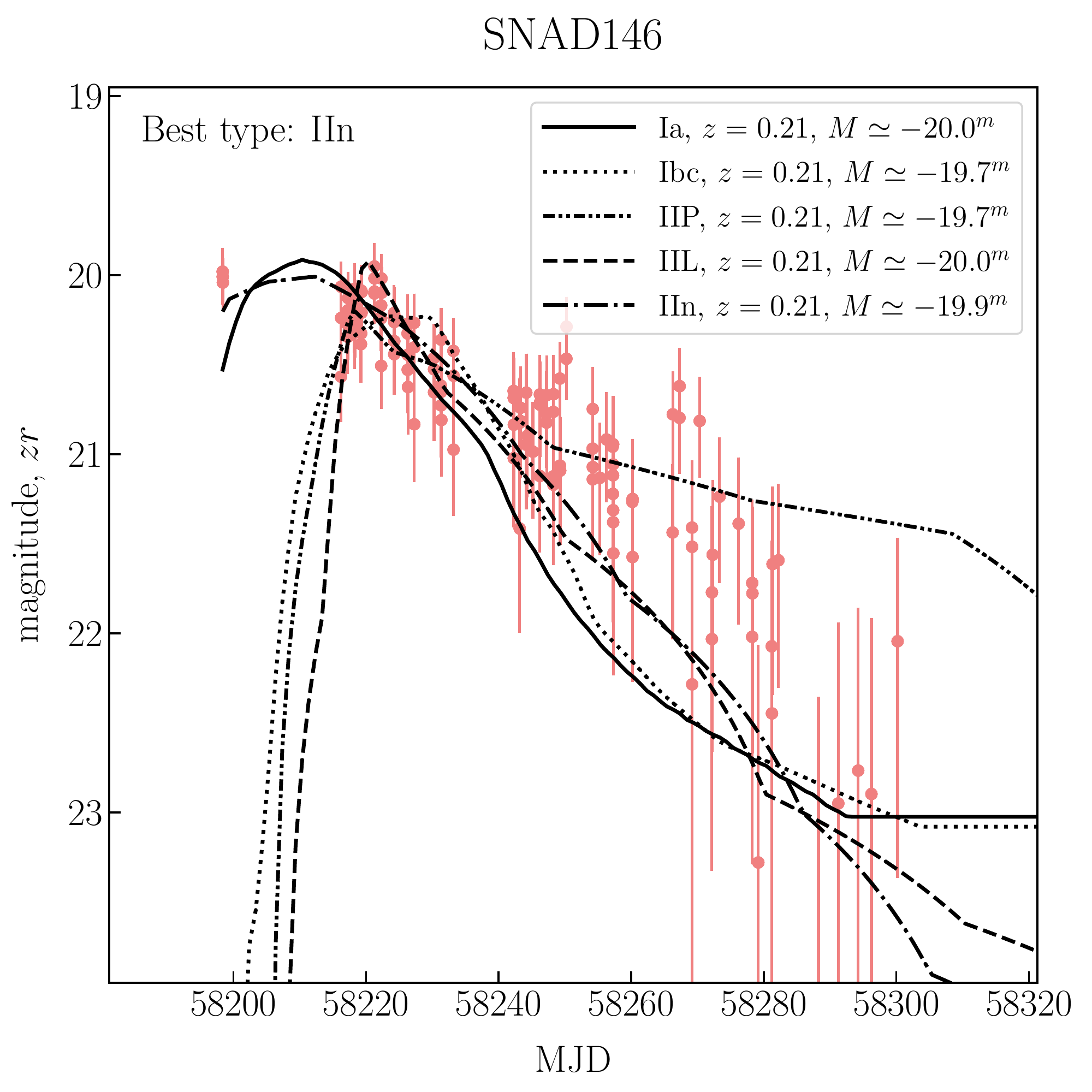}\\a) \\
    \end{minipage}
    \hfill
    \begin{minipage}{0.49\linewidth}
        \centering
        \includegraphics[scale=0.35]{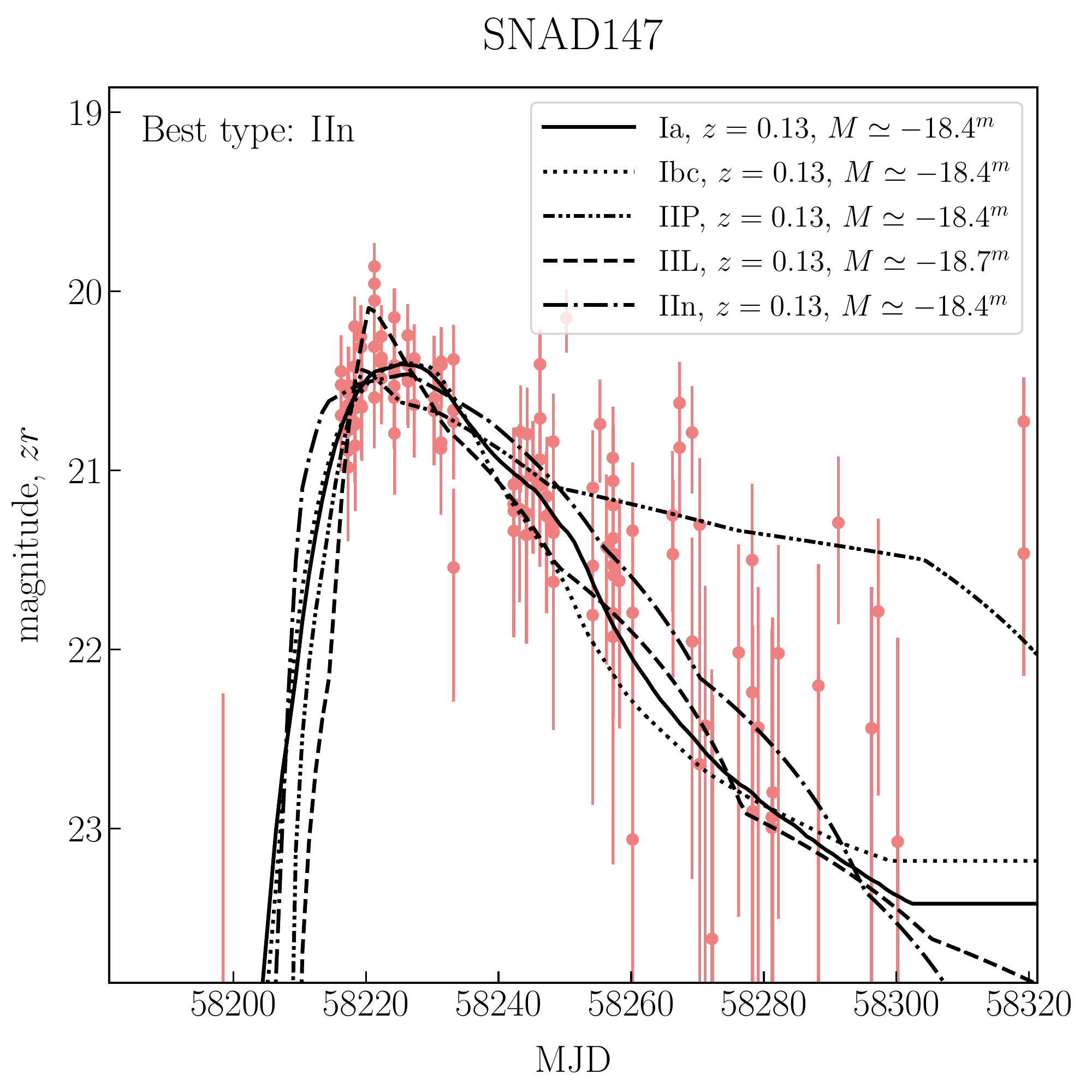}\\b) \\
    \end{minipage}
    \vfill
    \begin{minipage}{0.49\linewidth}
        \centering
        \includegraphics[scale=0.35]{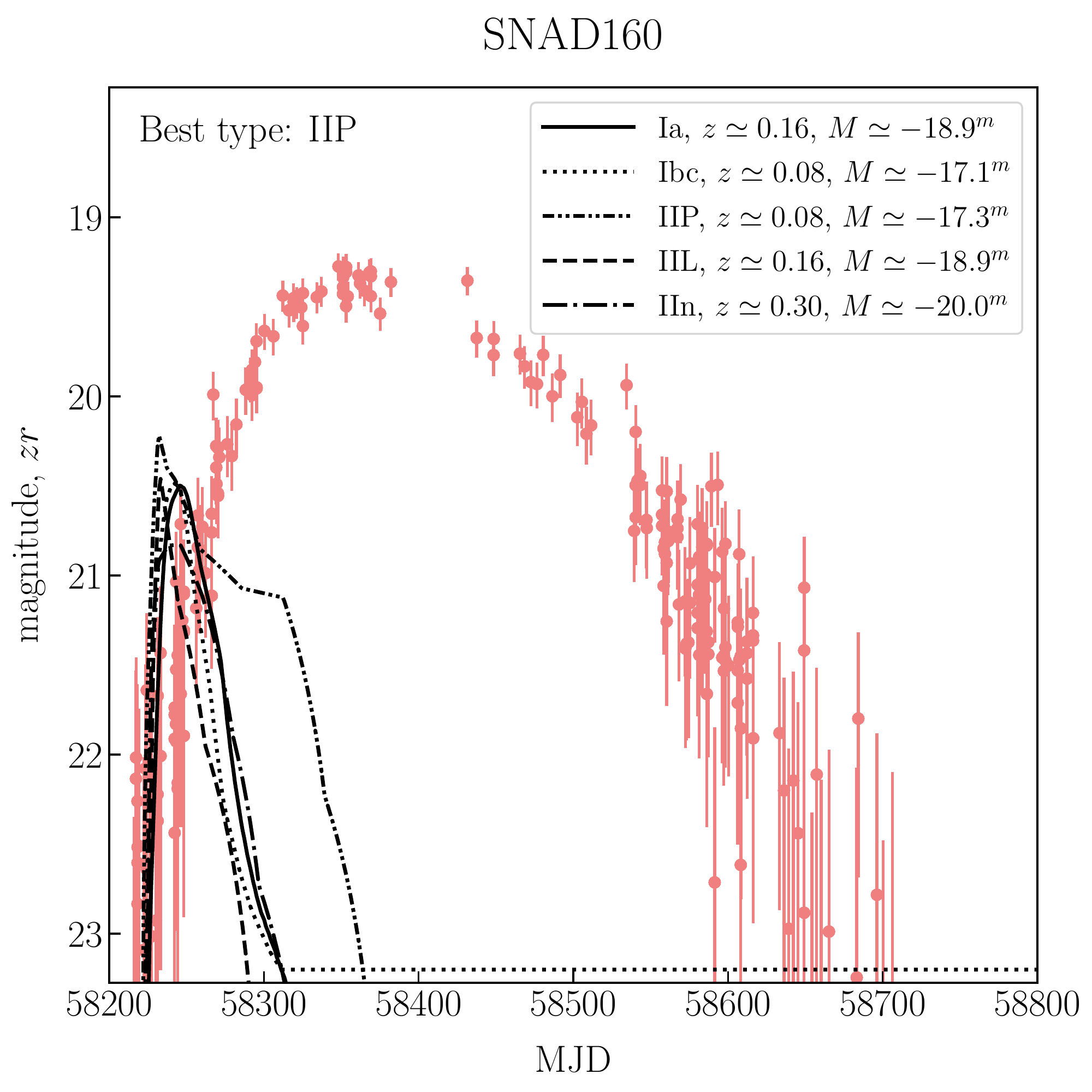}\\c) \\
    \end{minipage}
    \hfill
    \begin{minipage}{0.49\linewidth}
        \centering
        \includegraphics[scale=0.35]{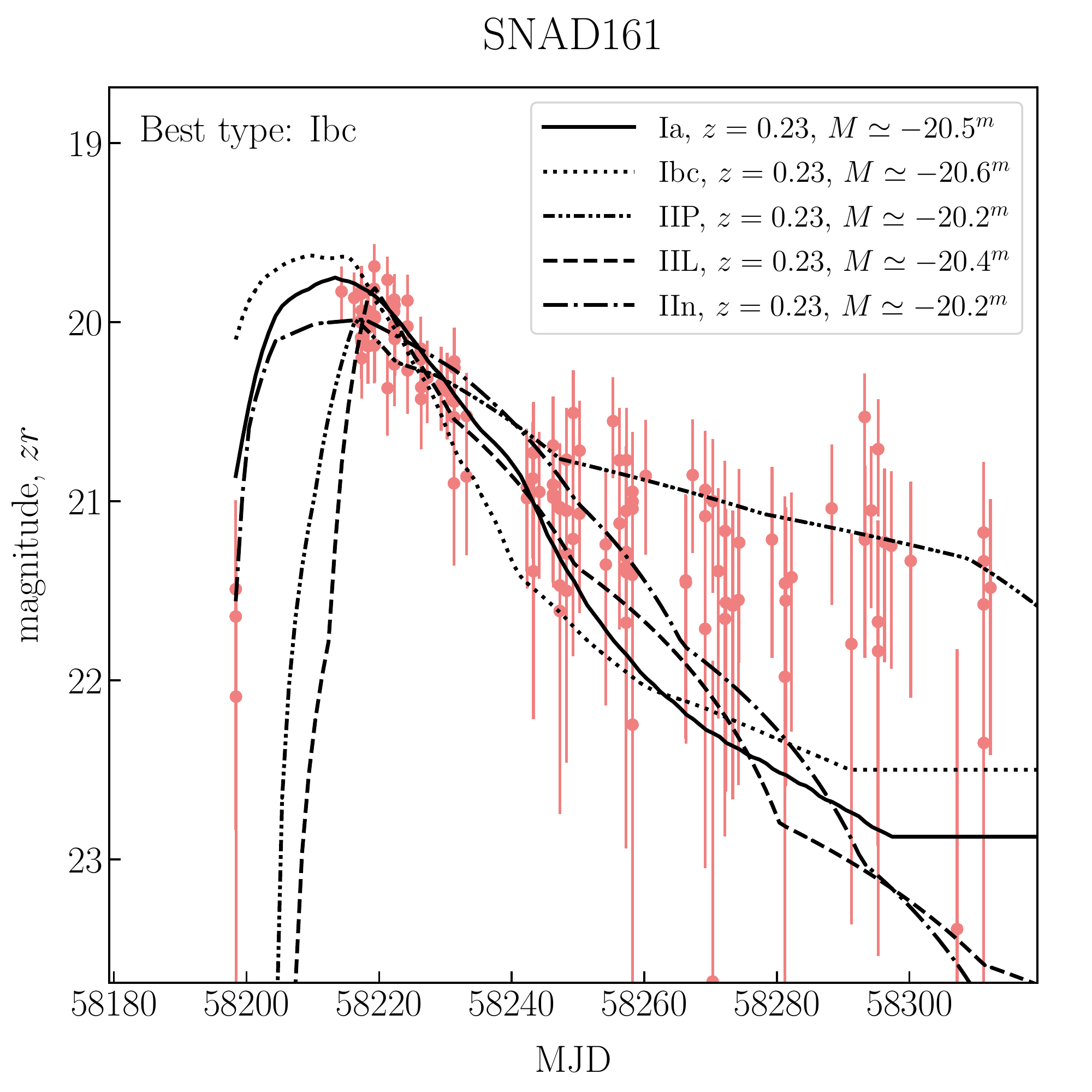}\\d) \\
    \end{minipage} 
    \vfill
    \begin{minipage}{0.49\linewidth}
        \centering
        \includegraphics[scale=0.35]{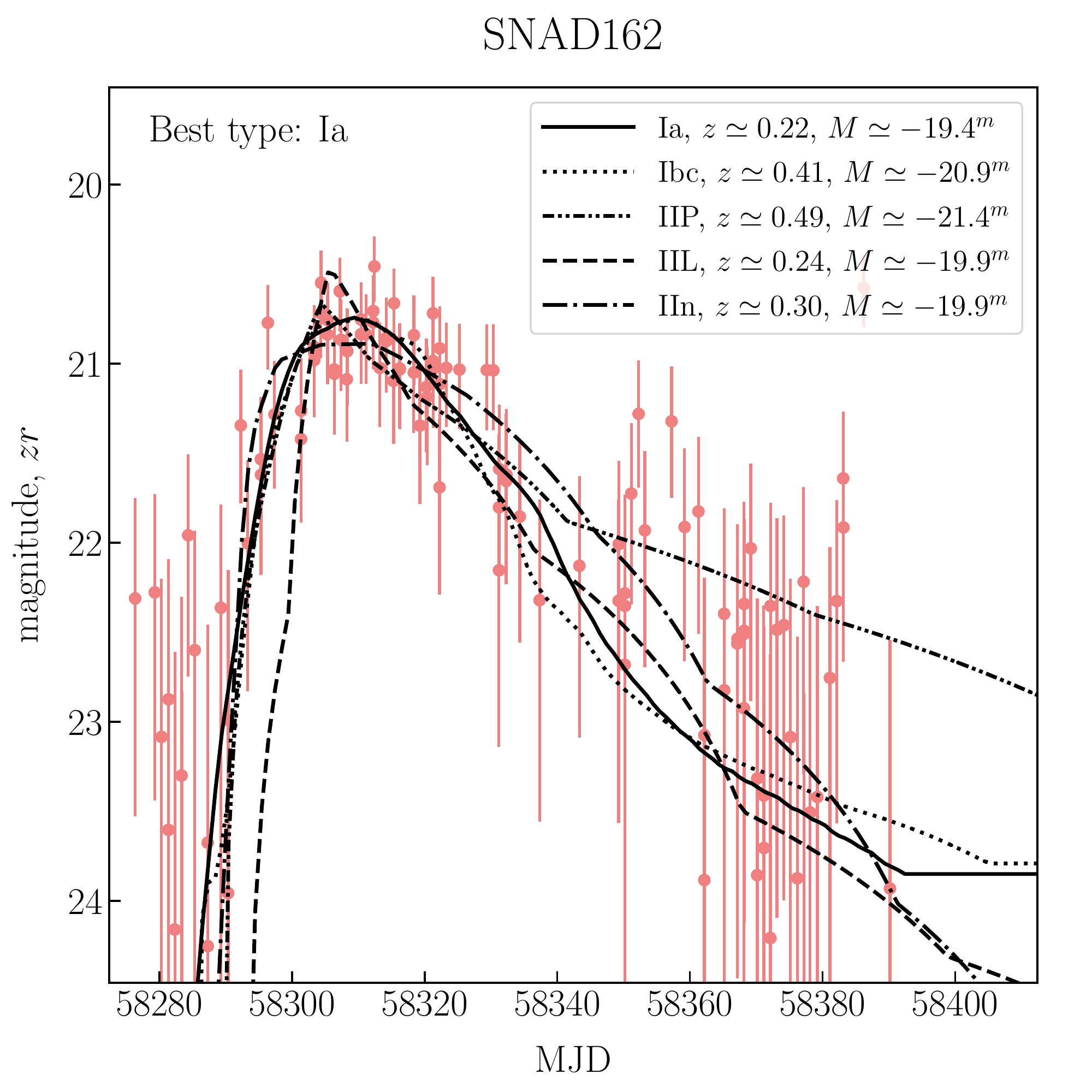}\\e) \\
    \end{minipage}
    \hfill
    \begin{minipage}{0.49\linewidth}
        \centering
        \includegraphics[scale=0.35]{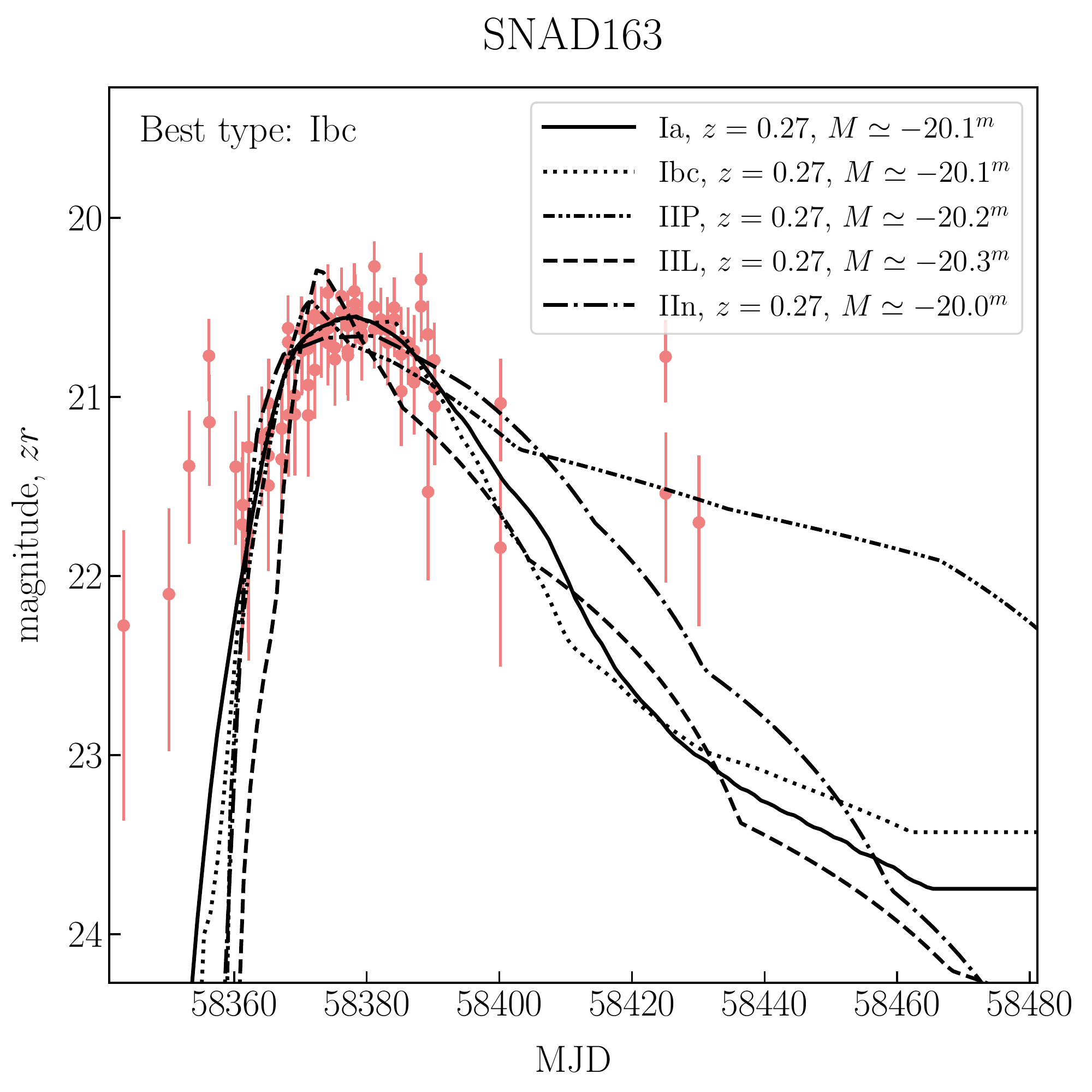}\\f) \\
    \end{minipage} 
    \caption{Light curves of SNAD supernova candidates in $zr$-band and the results of their fit by Nugent's supernova models.}
        \label{fig:snad_LC8}
\end{figure*}

\begin{figure*}
    \begin{minipage}{0.49\linewidth}
        \centering
        \includegraphics[scale=0.35]{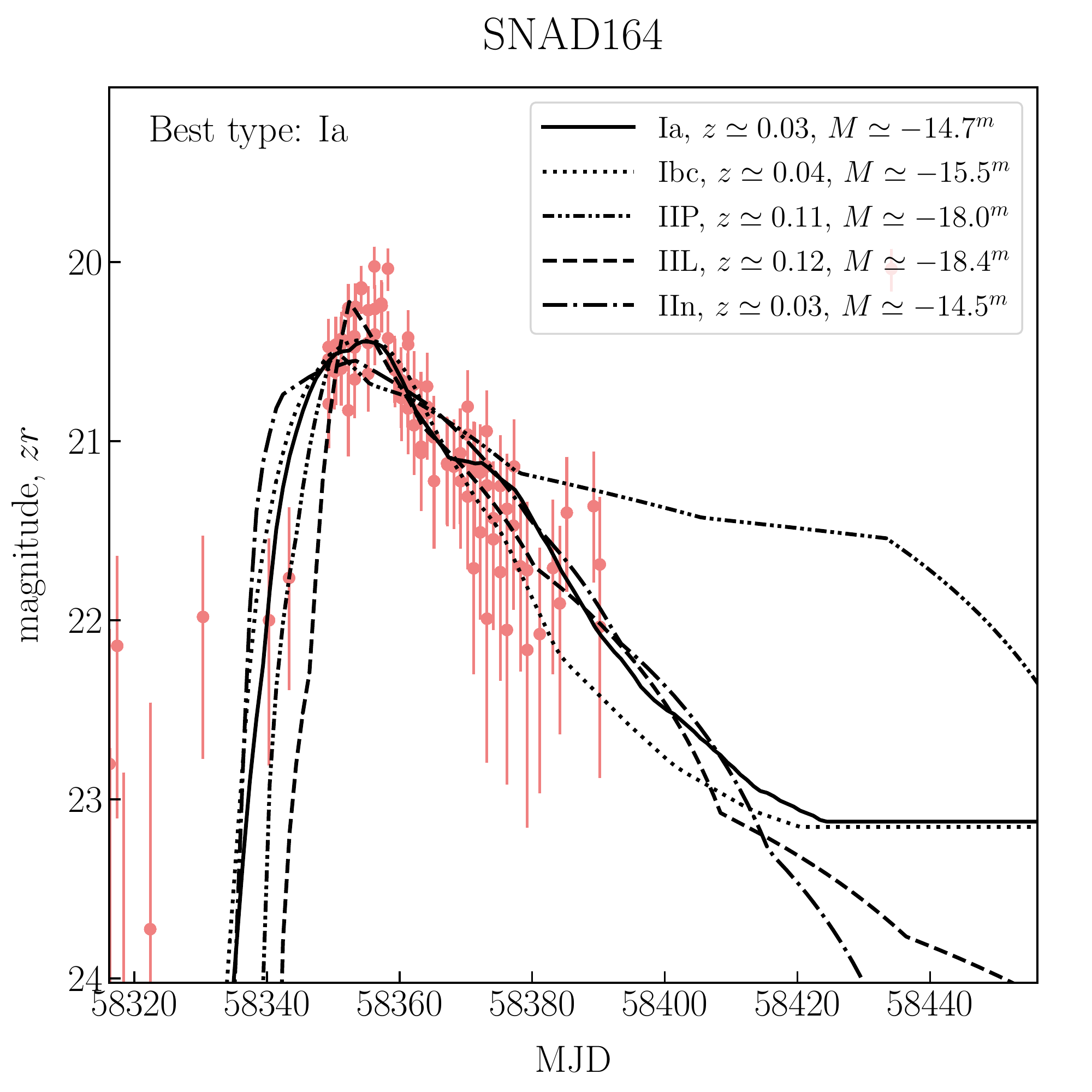}\\a) \\
    \end{minipage}
    \hfill
    \begin{minipage}{0.49\linewidth}
        \centering
        \includegraphics[scale=0.35]{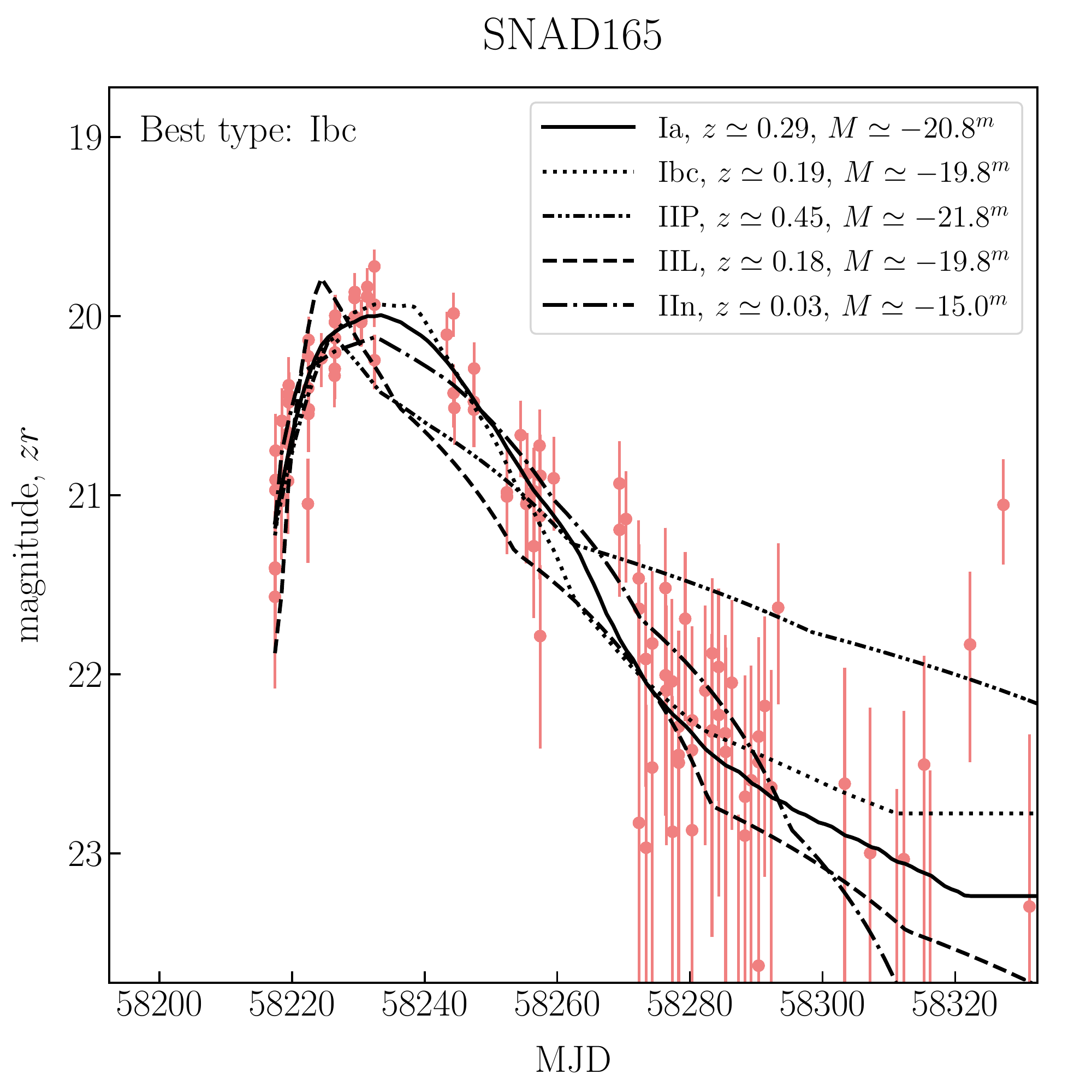}\\b) \\
    \end{minipage}  
    \vfill
    \begin{minipage}{0.49\linewidth}
        \centering
        \includegraphics[scale=0.35]{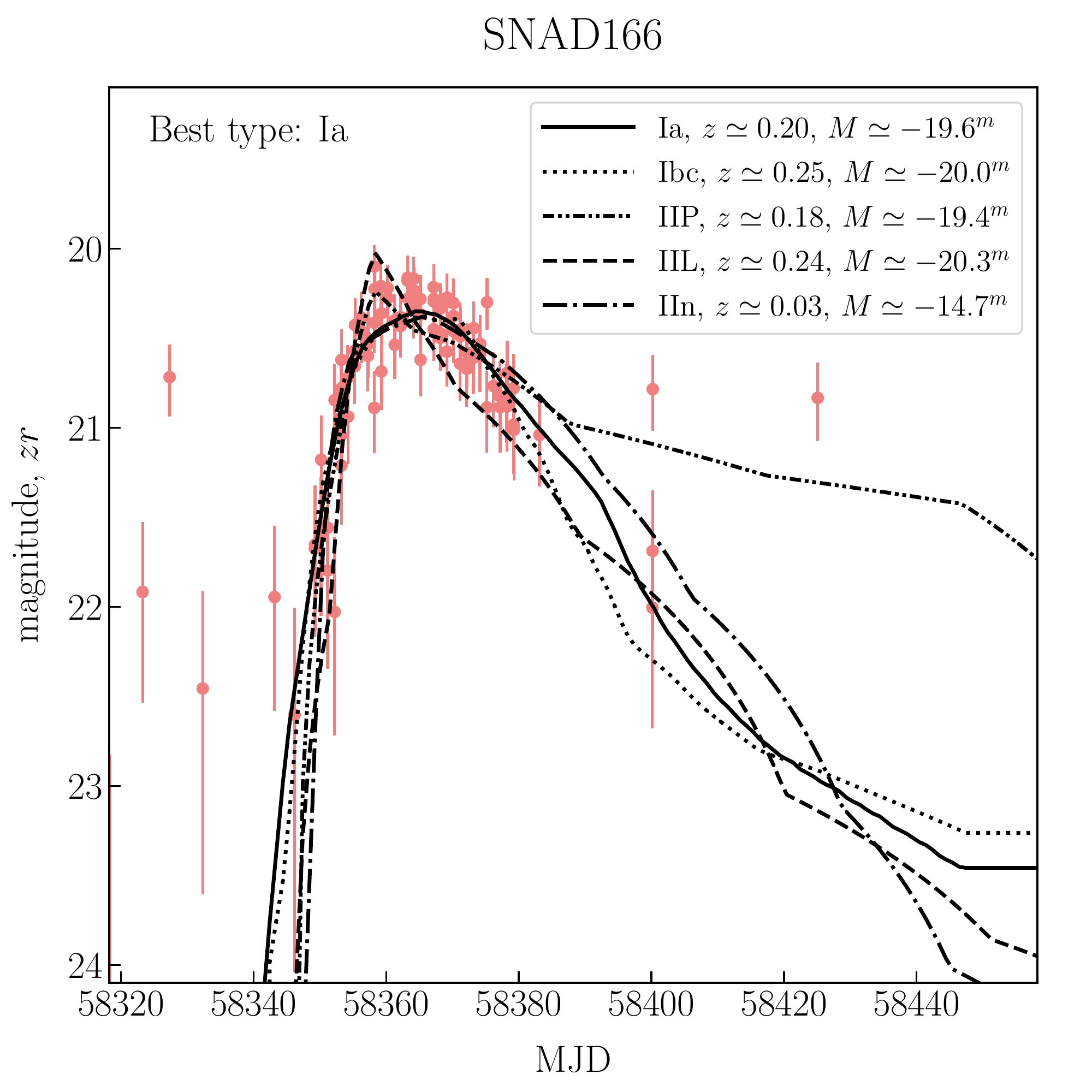}\\c) \\
    \end{minipage}  
    \hfill
    \begin{minipage}{0.49\linewidth}
        \centering
        \includegraphics[scale=0.35]{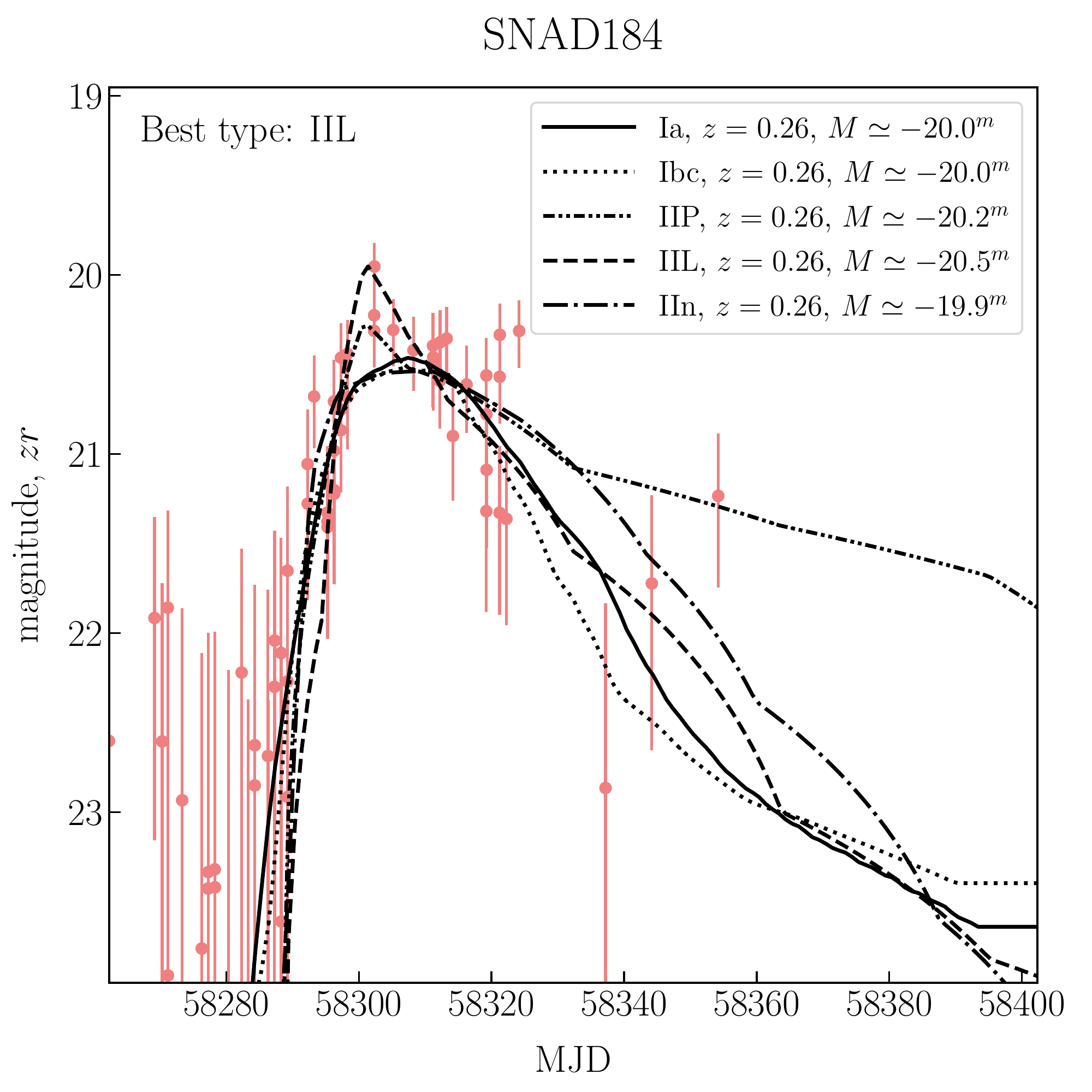}\\d) \\
    \end{minipage}  
    \vfill
    \begin{minipage}{0.49\linewidth}
        \centering
        \includegraphics[scale=0.35]{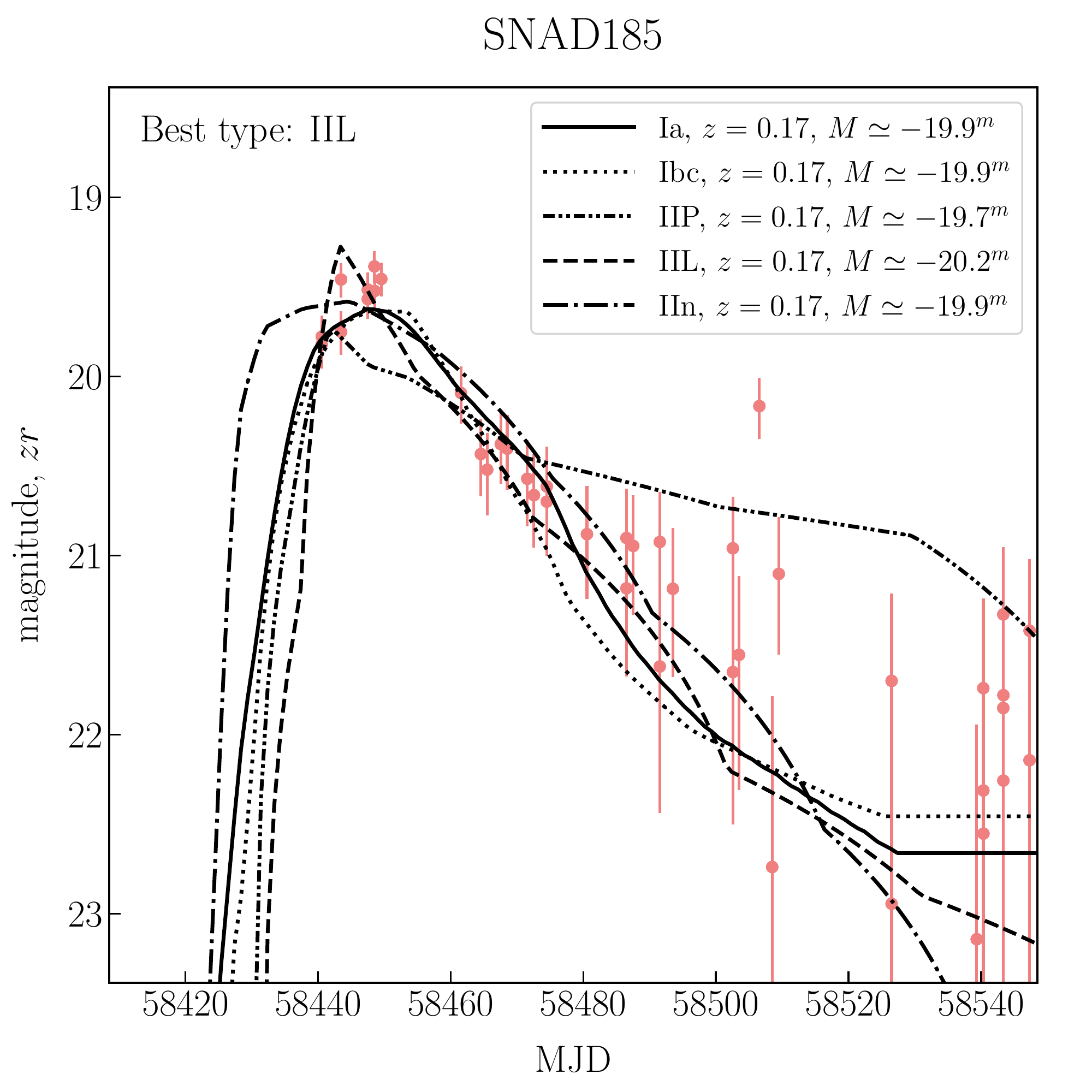}\\e) \\
    \end{minipage}  
    \hfill
    \begin{minipage}{0.49\linewidth}
        \centering
        \includegraphics[scale=0.35]{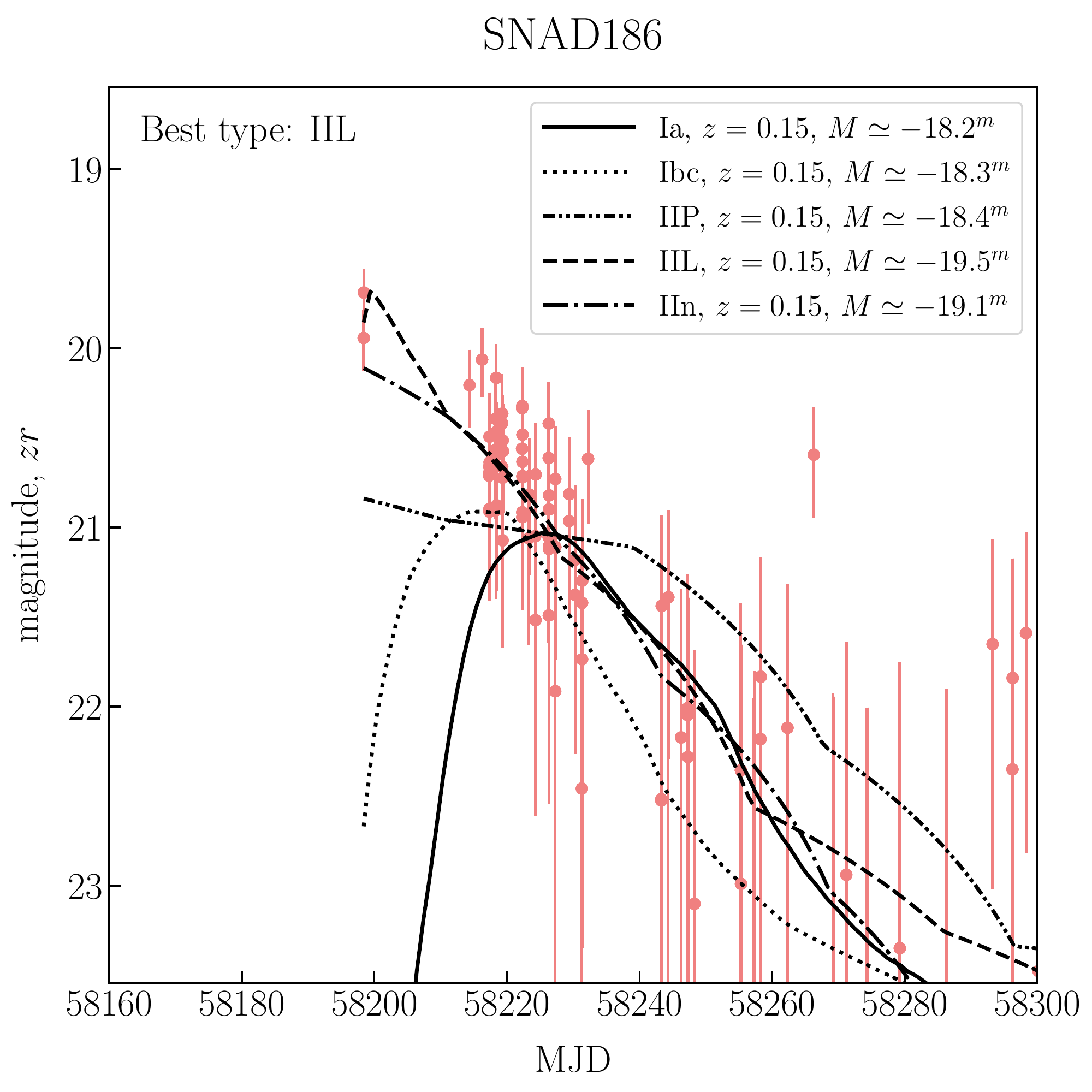}\\f) \\
    \end{minipage}
    \caption{Light curves of SNAD supernova candidates in $zr$-band and the results of their fit by Nugent's supernova models.}
        \label{fig:snad_LC9}
\end{figure*}

\begin{figure*}
    \begin{minipage}{0.49\linewidth}
        \centering
        \includegraphics[scale=0.35]{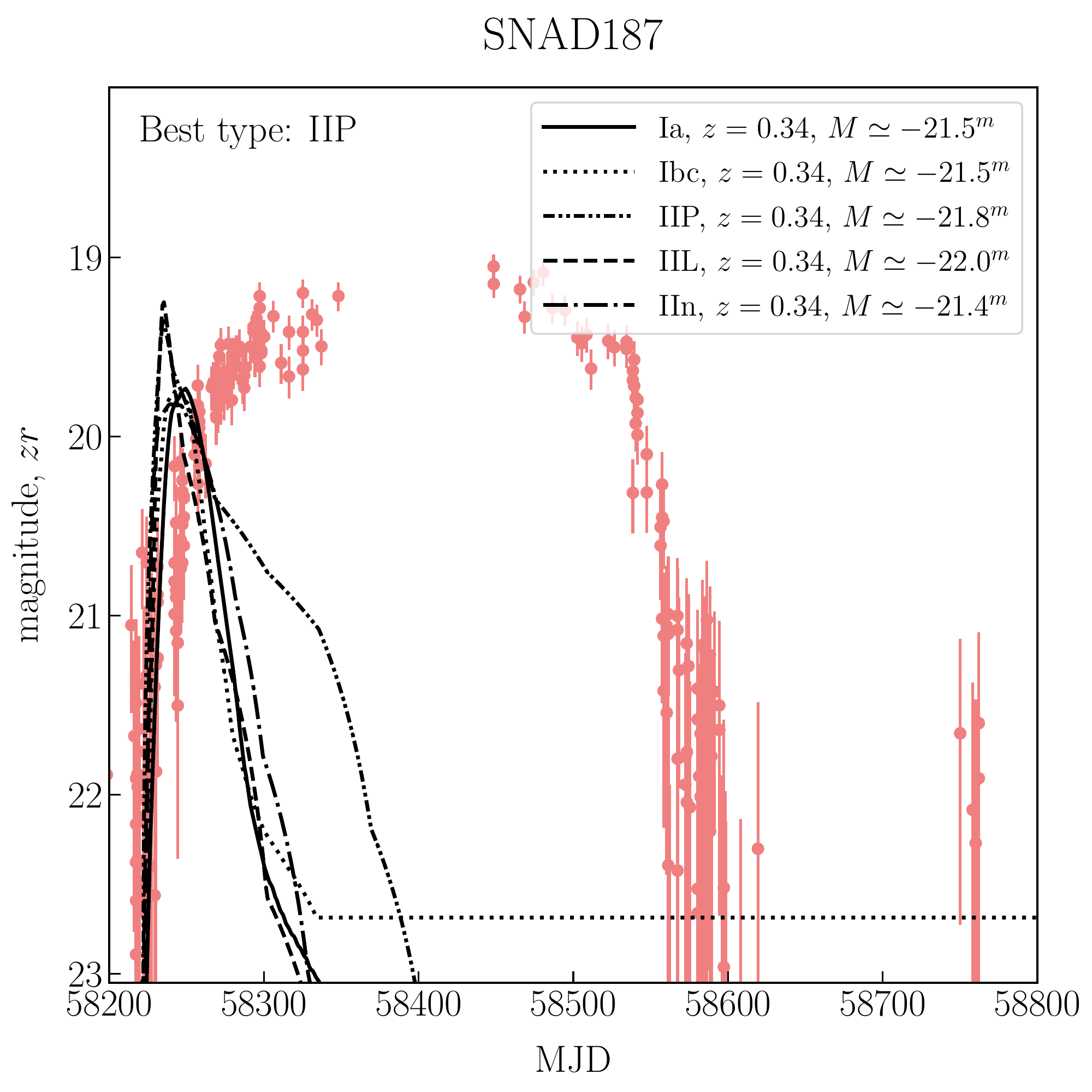}\\a) \\
    \end{minipage}
    \hfill
    \begin{minipage}{0.49\linewidth}
        \centering
        \includegraphics[scale=0.35]{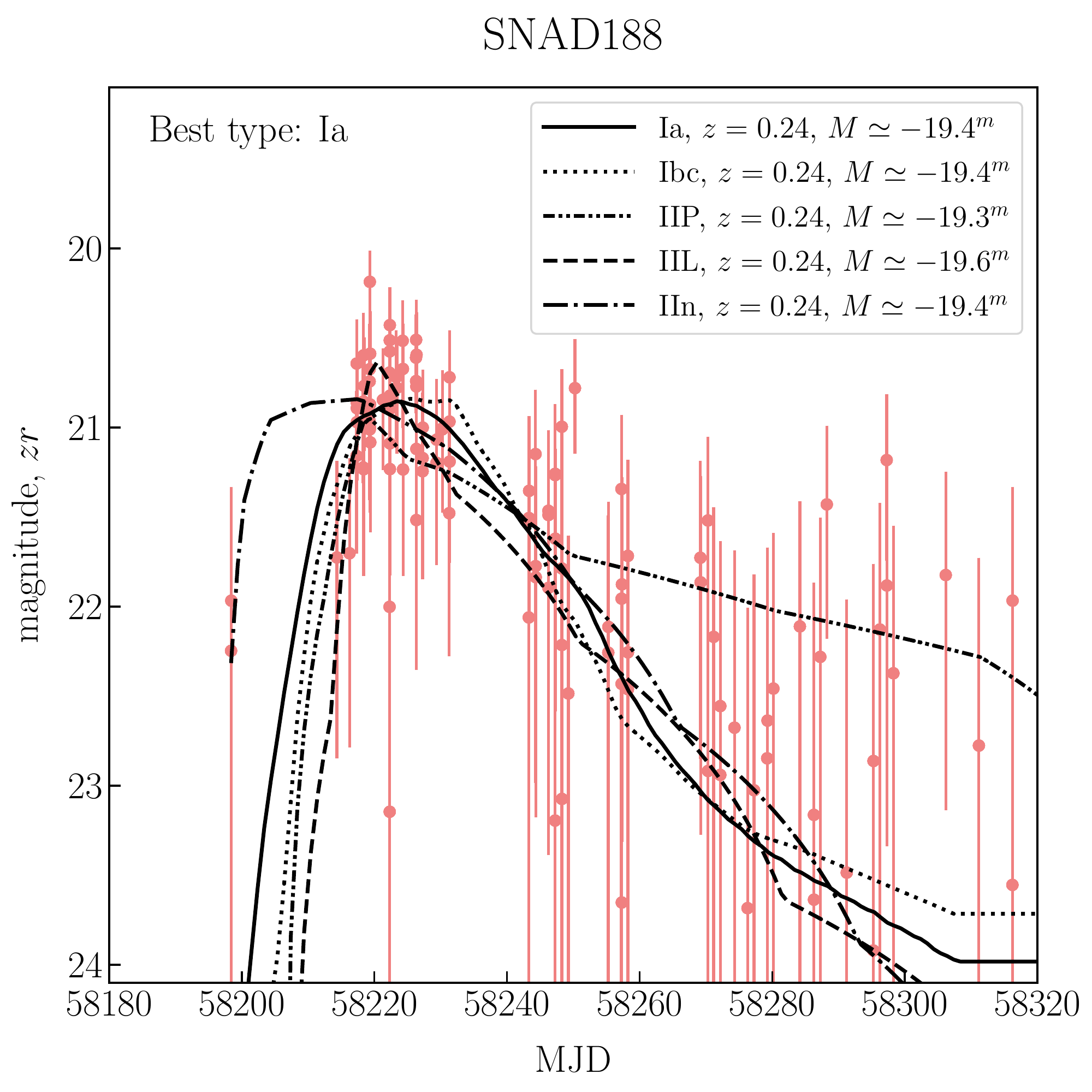}\\b) \\
    \end{minipage}  
    \vfill
    \begin{minipage}{0.49\linewidth}
        \centering
        \includegraphics[scale=0.35]{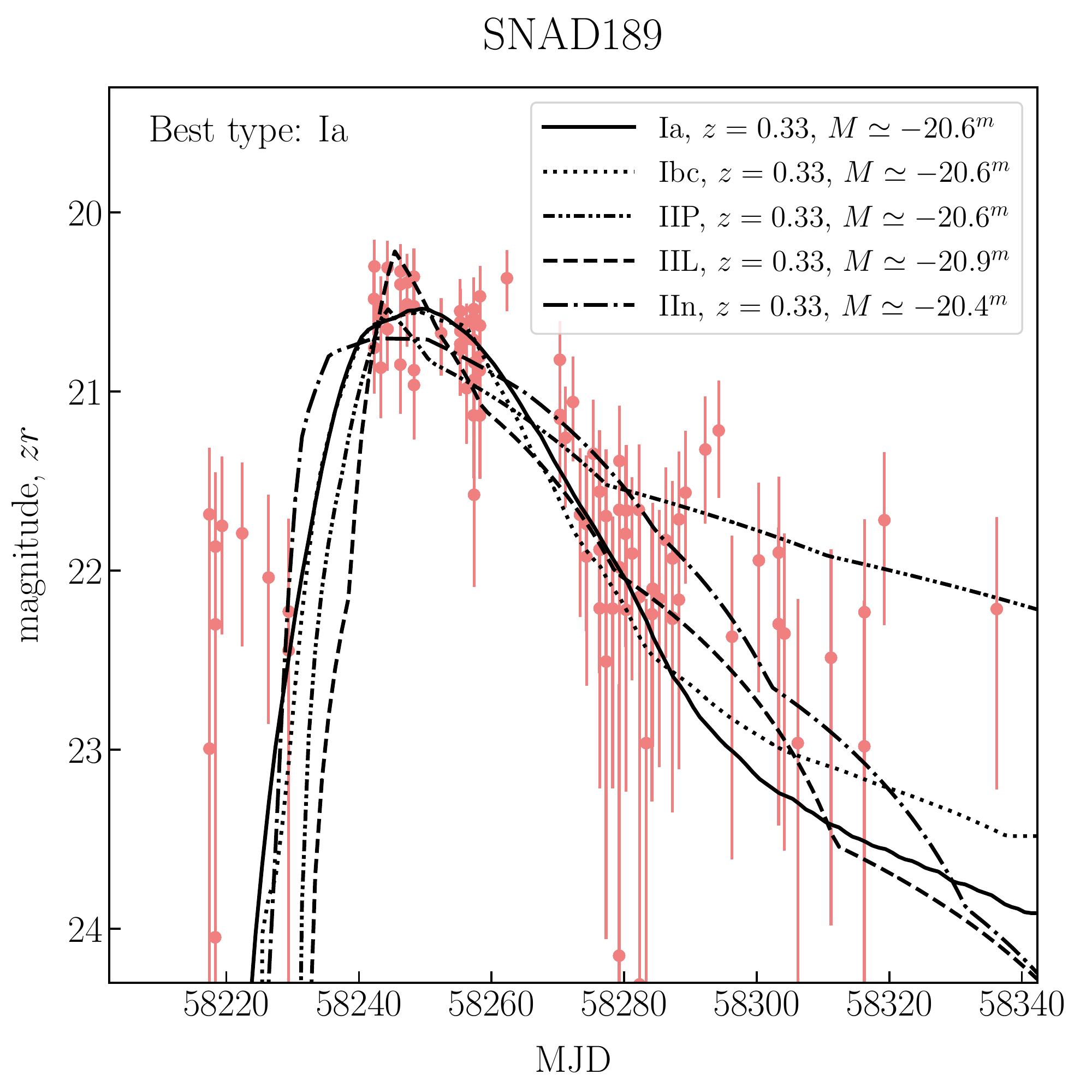}\\c) \\
    \end{minipage}  
    \caption{Light curves of SNAD supernova candidates in $zr$-band and the results of their fit by Nugent's supernova models.}
        \label{fig:snad_LC10}
\end{figure*}

% \newpage
% \bigskip
\section{SNAD ZTF viewer}
\label{ap:viewer}

We present below a glimpse on the expert's  session of the SNAD viewer\footnote{\url{https://ztf.snad.space/}} \maria{\citep{2021ascl.soft06034M,2022arXiv221107605M}}.

\begin{figure}
    \centering
    \includegraphics[width=\textwidth]{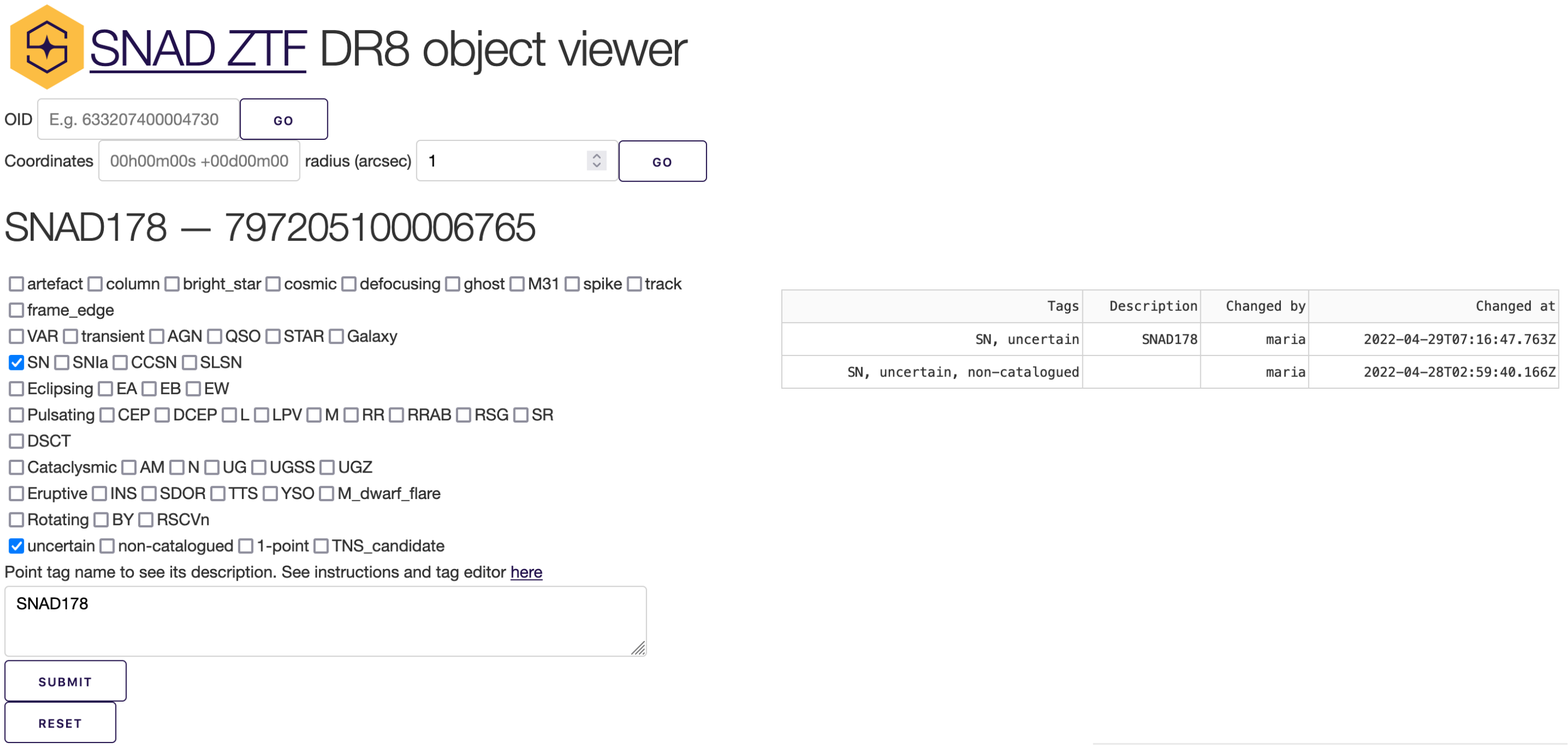}
    \caption{SNAD ZTF viewer tags and logs on the example of SNAD178.}
    \label{viewer}
\end{figure}
\end{appendix}

\end{document}